\documentclass[ba]{imsart}\usepackage[]{graphicx}\usepackage[]{color}
\makeatletter
\def\maxwidth{ %
  \ifdim\Gin@nat@width>\linewidth
    \linewidth
  \else
    \Gin@nat@width
  \fi
}
\makeatother

\definecolor{fgcolor}{rgb}{0.345, 0.345, 0.345}

\usepackage{framed}
\makeatletter
 {\par\unskip\endMakeFramed%
 \at@end@of@kframe}
\makeatother

\definecolor{shadecolor}{rgb}{.97, .97, .97}
\definecolor{messagecolor}{rgb}{0, 0, 0}
\definecolor{warningcolor}{rgb}{1, 0, 1}
\definecolor{errorcolor}{rgb}{1, 0, 0}

\usepackage{alltt}

\pubyear{2021}
\volume{TBA}
\issue{TBA}
\firstpage{1}
\lastpage{1}

\usepackage[toc,page]{appendix}
\newcommand{\stoptocwriting}{%
  \addtocontents{toc}{\protect\setcounter{tocdepth}{-5}}}
\newcommand{\resumetocwriting}{%
  \addtocontents{toc}{\protect\setcounter{tocdepth}{\arabic{tocdepth}}}}

\usepackage[notocbib]{apacite}

\usepackage{dblfloatfix} 

\DeclareFontFamily{U}{cbgreek}{}
\DeclareFontShape{U}{cbgreek}{m}{n}{
        <-6>    grmn0500
        <6-7>   grmn0600
        <7-8>   grmn0700
        <8-9>   grmn0800
        <9-10>  grmn0900
        <10-12> grmn1000
        <12-17> grmn1200
        <17->   grmn1728
      }{}
\DeclareFontShape{U}{cbgreek}{bx}{n}{
        <-6>    grxn0500
        <6-7>   grxn0600
        <7-8>   grxn0700
        <8-9>   grxn0800
        <9-10>  grxn0900
        <10-12> grxn1000
        <12-17> grxn1200
        <17->   grxn1728
      }{}

\makeatletter
\newcommand{\normalorbold}{%
  \ifnum\pdf@strcmp{\math@version}{bold}=\z@ bx\else m\fi
}
\makeatother

\usepackage{expl3}

\usepackage{multirow}
\usepackage[flushleft]{threeparttable}
\usepackage{todonotes}
\usepackage{subcaption}
\usepackage{natbib}
\usepackage[colorlinks,citecolor=blue,urlcolor=blue,filecolor=blue,backref=page]{hyperref}
\usepackage{graphicx}
\RequirePackage{amsthm,amsmath,amsfonts,amssymb,graphicx,url,placeins,multirow,booktabs,dsfont,bbm,lineno,makecell,enumitem,placeins,url,comment,longtable,footnote,array,mathtools,tabulary,colortbl,siunitx}
\startlocaldefs
\numberwithin{equation}{section}
\theoremstyle{plain}

\usepackage{listings}
\definecolor{codegreen}{rgb}{0,0.6,0}
\definecolor{codegray}{rgb}{0.5,0.5,0.5}
\definecolor{codepurple}{rgb}{0.58,0,0.82}
\definecolor{backcolour}{rgb}{0.95,0.95,0.92}

\lstdefinestyle{mystyle}{
    backgroundcolor=\color{backcolour},   
    commentstyle=\color{codegreen},
    keywordstyle=\color{magenta},
    numberstyle=\tiny\color{codegray},
    stringstyle=\color{codepurple},
    basicstyle=\ttfamily\footnotesize,
    breakatwhitespace=false,         
    breaklines=true,                 
    captionpos=b,                    
    keepspaces=true,                 
    numbers=left,                    
    numbersep=5pt,                  
    showspaces=false,                
    showstringspaces=false,
    showtabs=false,                  
    tabsize=2
}

\usepackage{xr}
\usepackage{cleveref}
\makeatletter
\newcommand*{\addFileDependency}[1]{
  \typeout{(#1)}
  \@addtofilelist{#1}
  \IfFileExists{#1}{}{\typeout{No file #1.}}
}

\newcommand{\includeCountrySummaryIfexists}[3]{
    \begin{figure}[!hb]
    \centering
    \includegraphics[width = 1\textwidth]{results/\JOBIDBSGP/\JOBIDBSGP-panel_plot_1_#2.png}
    \caption{\textbf{State summary for #3.} \textbf{(A)} Weekly COVID-19 attributable deaths reported by the CDC stacked by age groups (color). \textbf{(B)} Posterior mean of the weekly COVID-19 attributable deaths stacked by age, estimated with a regularised B-splines projected GP prior. Three weeks in the observation period are selected and designated by a shape. The black line shows the all-age weekly COVID-19 attributable deaths reported by JHU, that were used to adjust the CDC data for reporting delays~\eqref{eq:rescaledeaths}. \textbf{(C)} Posterior median (black line) and 95\% credible interval (ribbon) of the age-specific contribution to COVID-19 attributable deaths in three weeks.}
     \label{fig:summary_#2}
    \end{figure}
}

\endlocaldefs

\newcommand{\JOBIDBSGP}{211201a-2047}
\newcommand{\JOBIDGP}{211125a-11969}
\IfFileExists{upquote.sty}{\usepackage{upquote}}{}
\begin{document}

\begin{frontmatter}
\title{Regularised B-splines projected Gaussian Process priors to estimate time-trends of age-specific COVID-19 deaths related to vaccine roll-out}
\runtitle{Projected GP priors to model trends in COVID-19 deaths by age}

\begin{aug}
\author{\fnms{Mélodie} \snm{Monod}\thanksref{t1,a1,addr1}\ead[label=e1]{melodie.monod18@imperial.ac.uk}},
\author{\fnms{Alexandra} \snm{Blenkinsop}\thanksref{a1}},
\author{\fnms{Andrea} \snm{Brizzi}\thanksref{a1}},
\author{\fnms{Yu} \snm{Chen}\thanksref{a1}},
\author{\fnms{Carlos} \snm{Cardoso Correia Perello}\thanksref{a1}},
\author{\fnms{Vidoushee} \snm{Jogarah}\thanksref{a1}},
\author{\fnms{Yuanrong} \snm{Wang}\thanksref{a1}},
\author{\fnms{Seth} \snm{Flaxman}\thanksref{t2,a1}},
\author{\fnms{Samir} \snm{Bhatt}\thanksref{t2,a2,a3}}
and
\author{\fnms{Oliver} \snm{Ratmann}\thanksref{t2,a1,addr1}\ead[label=e2]{oliver.ratmann@imperial.ac.uk}}

\address[addr1]{Corresponding authors
    \printead{e1} 
    \printead*{e2}
}
\runauthor{M. Monod et al.}

\thankstext{t1}{First supporter of the project}
\thankstext{t2}{Second supporter of the project}
\thankstext{a1}{Department of Mathematics, Imperial College London}
\thankstext{a2}{MRC Centre for Global Infectious Disease Analysis, Jameel Institute, School of Public Health, Imperial College London}
\thankstext{a3}{Section of Epidemiology, Department of Public Health, University of Copenhagen}

\end{aug}


\begin{abstract}

The COVID-19 pandemic has caused severe public health consequences in the United States.
In this study, we use a hierarchical Bayesian model to estimate the age-specific COVID-19 attributable deaths over time in the United States. The model is specified by a novel non-parametric spatial approach, a low-rank Gaussian Process (GP) projected by regularised B-splines. We show that this projection defines a new GP with attractive smoothness and computational efficiency properties, derive its kernel function, and discuss the penalty terms induced by the projected GP. Simulation analyses and benchmark results show that the spatial approach performs better than standard B-splines and Bayesian P-splines and equivalently well as a standard GP, for considerably lower runtimes. The B-splines projected GP priors that we develop are likely an appealing addition to the arsenal of Bayesian regularising priors.
We apply the model to weekly, age-stratified COVID-19 attributable deaths reported by the US Centers for Disease Control, which are subject to censoring and reporting biases. Using the B-splines projected GP, we can estimate longitudinal trends in COVID-19 associated deaths across the US by 1-year age bands. These estimates are instrumental to calculate age-specific mortality rates, describe variation in age-specific deaths across the US, and for fitting epidemic models. Here, we couple the model with age-specific vaccination rates to show that lower vaccination rates in younger adults aged 18-64 are associated with significantly stronger resurgences in COVID-19 deaths, especially in Florida and Texas. These results underscore the critical importance of medically able individuals of all ages to be vaccinated against COVID-19 in order to limit fatal outcomes. 
\end{abstract}


\begin{keyword}[class=MSC]
\kwd[Primary ]{62F15}
\kwd{60G15}
\kwd{65D07}
\kwd[; secondary ]{62P10}
\end{keyword}

\begin{keyword}
\kwd{Bayesian Inference}
\kwd{Gaussian Processes}
\kwd{Splines}
\kwd{Non-parametric estimation}
\kwd{COVID-19}
\end{keyword}

\end{frontmatter}
\stoptocwriting


\section{Introduction}

A new pathogen, Severe Acute Respiratory Syndrome Coronavirus 2 (SARS-CoV-2), emerged in the Wuhan region of China in December 2019 and continues to spread worldwide. The resulting disease, COVID-19, is severe, with overall infection fatality rates (IFRs) between 0.1\% and 1\%~\citep{MeyerowitzKatz2020,Brazeau:2020}, which increase exponentially with age~\citep{Levin2020}. Developed vaccines are highly effective to prevent deaths~\citep{Haas2021,Baden2021}, and initially have been  prioritised to older age groups. In the United States (US), vaccines are now offered to all adults since May 2021, but uptake has been variable across states~\citep{CDC_data_vac_state}. Despite increasing vaccine coverage, several US states, most notably Florida and Texas, are reporting resurgent COVID-19 death waves that are on par or exceed in magnitude the death waves seen in the prevaccination period. A key question is if these death waves since summer 2021 are linked to low vaccination coverage in younger adults, which are known to drive transmission~\citep{Monod2021}.

Here, we provide methods for estimating the age composition of COVID-19 attributable deaths over time, and to characterise the potential shifts in the age composition of deaths. Doing so requires a statistical model because data are partially censored, reported with delays, and reported in age bands that can be inconsistent across locations or do not match those of other data streams such as reported vaccinations. Numerous methods have been developed to interpret the time evolution of overall COVID-19 attributable deaths (e.g., ~\citet{Lavezzo2020,IHME:2020,Chen2020,Blangiardo2020,Zheng2021, Flaxman2020}), but few are suitable to analyze longitudinal data stratified by age brackets or other discrete strata~\citep{Monod2021}, let alone at high resolution such as 1-year age bands. One reason for this paucity of methods is that in SIR-type models, calculating age-specific next generation operators at a time resolution of days becomes computationally slow over observation periods that span years, especially when calculations need to be repeated millions of times within a Bayesian framework~\citep{Wikle2020, Monod2021}. These considerations are prompting us and others to consider non-mechanistic, flexible estimation approaches~\citep{Shah2020,Pokharel2021}. 
As our objective is to reconstruct spatiotemporal trends in age-specific COVID-19 attributable deaths, it does not require invoking many of the assumptions or complexities that underlie mechanistic or semi-mechanistic COVID-19 transmission dynamics models and Bayesian non-parametric models are sufficient.
We present a fully Bayesian, computationally scalable approach to estimate a two-dimensional (2D) surface over ages and weeks that describes the time evolution of COVID-19 deaths by $1$-year age bands at US state level. To impute missing entries on the surface and estimate global trends over ages and weeks, we borrow information across neighboring entries by using a non-parametric smoothing method.

A natural starting point for modelling a surface is a 2D Gaussian process (GP)~\citep{Rasmussen_GP}. However, their computational complexity makes the use of 2D GPs in a fully Bayesian framework difficult when the surface dimension becomes large, even after using a Kronecker decomposition of the kernel function~\citep{Saatci:2011,Wilson:2014}. 
Here we adopt a low-rank approximation via a tensor product of B-splines for which the parameters follow a 2D GP. The resulting approach is equivalent to a 2D GP defined by a low-rank covariance matrix projected by B-splines. B-splines are a popular choice for non-parametric modelling, due to their continuity properties, ensuring smoothness of the fitted surface, and their easy implementation. But choosing the optimal number and position of knots---the defining grid segments where the surface is expected to change its behavior---on the space of ages and weeks is a complex task. Some approaches have focused on adding a penalty to restrict the flexibility of the fitted surface in a frequentist framework~\citep{Sullivan:1986,OSullivan1988,Eilers1996,Eilers2006}.
Following this idea, we regularise the fitted surface by using a kernel function with a free complexity parameter to define the covariance matrix of the low-rank 2D GP. We qualitatively compare the penalty induced by this choice to that of related regularisation methods. We benchmark the proposed regularised B-splines projected GP against several other popular smoothing methods, and we demonstrate that our approach results in substantive computational gains over a standard 2D GP for similar estimation accuracy.


This paper focus on estimating age-specific COVID-19 attributable deaths in the context of vaccine roll-out on publicly available, age-specific COVID-19 death and vaccination data from the $4$ most populated US states, California, Florida, New York (state) and Texas, that are reported by the Centers for Disease Control and Prevention (CDC) (\citeyear{CDC_data_vac_state, CDC_data_website}). It is structured as follows. Section~\ref{sec:CDCdata} introduces the data and their limitations. Section~\ref{sec:methods} describes the proposed methodology to model trends in the age-composition of COVID-19 deaths, including a theoretical characterization of the penalties introduced by B-splines projected GPs. Section~\ref{sec:simulation} presents a comparison of the proposed method to related smoothing approaches on simulated data, and Section~\ref{sec:benchmark} on real world data used for benchmarking. In Section~\ref{sec:realdatanalysis} we estimate the time evolution of age-specific COVID-19 deaths for the four most populated US states. We document the marked variation in the summer 2021 resurgence in age-specific COVID-19 deaths across states, and show strong resurgences in deaths are associated low vaccination coverage in younger adults. Lastly, Section~\ref{sec:discussion} closes with a discussion.

The proposed modelling framework can be further applied to estimate COVID-19 mortality rates in arbitrary age groups, investigate differences in the age composition of deaths across locations or identify time shifts in the age composition of deaths. The code to use our approach and to reproduce the results is available at \sloppy\url{https://github.com/ImperialCollegeLondon/B-SplinesProjectedGPs}. Median and 95\% credible interval of the state- and age-specific COVID-19 attributable deaths predicted by our approach are available in the GitHub repository. Posterior samples are also available upon request. Our results suggest regularised B-splines projected GP priors are likely useful for other surface estimation problems, and we provide templated Stan model files in Supplement~\ref{supp:stan_code}.

\section{COVID-19 deaths and vaccination data} \label{sec:CDCdata}
\subsection{COVID-19 deaths data}
The Center for Disease Control (CDC) and National Center for Health Statistics (NCHS) report each week on the total number of deaths involving COVID-19 that have been received and tabulated through the National Vital Statistics database~\citep{CDC_data_website} for every US state across the age groups
\begin{align} \label{eq:Bagegroup}
    \mathcal{B} &= \big\{0, 1-4, 5-14, 15-24, 25-34, 35-44, 45-54, \\
    &\quad\quad 55-64, 65-74, 75-84, 85+\big\}. \nonumber
\end{align}
We refer to the latter as cumulative reported COVID-19 attributable deaths, $D_{m,b,w}$, for state $m$, in age group $b$ and on week $w$. Historical records are made available by~\cite{CDC_data_repo}.
To simplify the longitudinal dependence and create a time series from the cumulative counts~\citep{King2015}, we obtain the weekly COVID-19 attributable deaths by differencing,
\begin{equation}\label{e:weekly_deaths}
    d_{m,b,w} = D_{m,b,w+1} - D_{m,b,w},
\end{equation}
in each location and age group for all but the last available week (daily deaths, $d$, for state $m$, in age group $b$, on week $w$). We index the weekly deaths by $w=1,\dotsc,W$. 

\begin{figure}[h!]
    \centering
    \includegraphics[width = 0.95\textwidth]{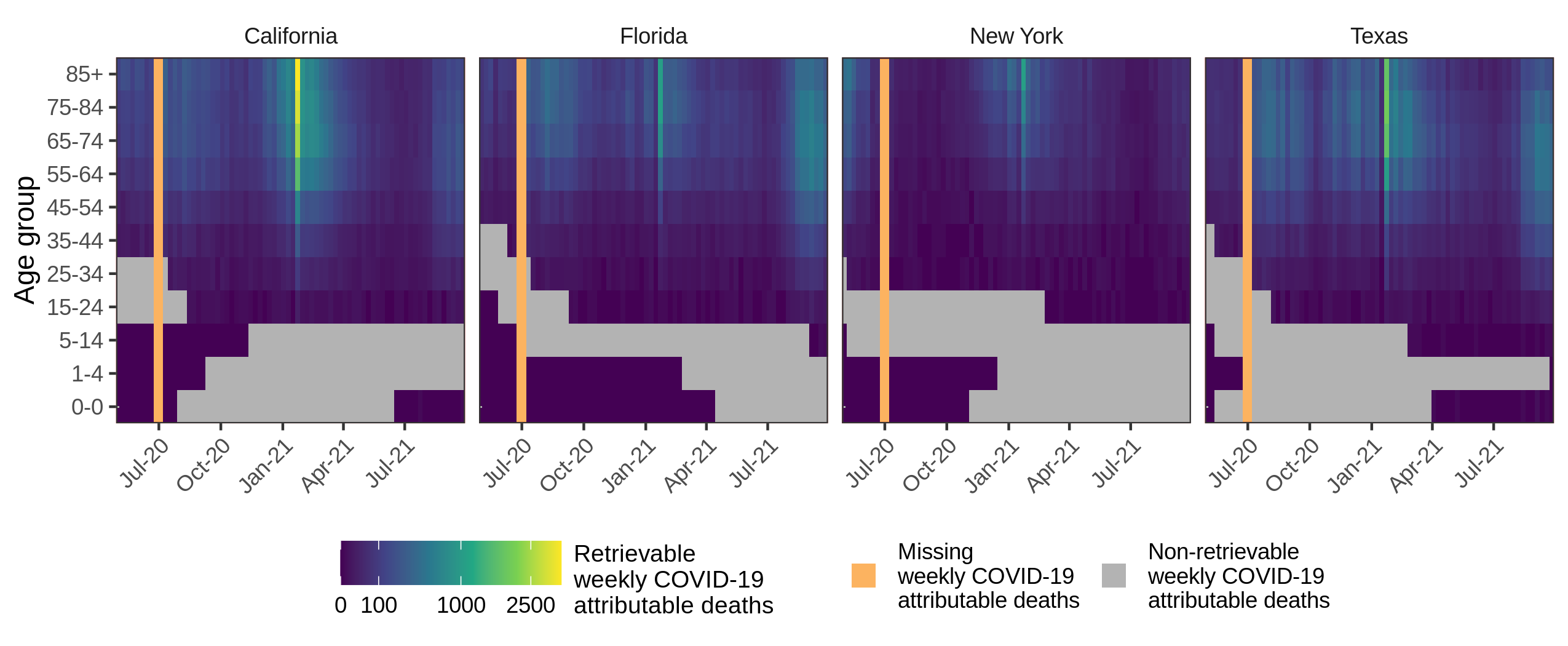}
    \caption{\textbf{Retrievable weekly COVID-19 attributable deaths in four US states.} The retrievable weekly deaths are computed using the first order difference of reported cumulative deaths by the CDC~\citep{CDC_data_website}. Grey cells indicate non-retrievable weekly death counts due to the censoring of cumulative deaths. Orange bars show the missing weekly deaths due to non-reported cumulative deaths.}
    \label{fig:weeklydeathsraw}
\end{figure}

Reflecting the reporting nature of the age-stratified CDC data, the weekly COVID-19 attributable $d_{m,b,w}$ are subject to reporting delays and do not necessarily correspond to the number of individuals who died of COVID-19 in state $m$ and week $w$. The CDC does not report the cumulative deaths count if the count is between $1$ and $9$. Thus, it is not possible to retrieve all weekly deaths, and we refer to the set of weeks that are retrievable in state $m$ for age group $b$ through first order differencing by $\mathcal{W}_{m,b}^{\text{WR}}$, and to the set of weeks that were non-retrievable because of censoring by $\mathcal{W}_{m,b}^{\text{WNR}}$. The censored cumulative deaths are bounded, such that the sum of non-retrievable weekly deaths is also bounded. The exact computations are described in Appendix~\ref{sec:findweeklydeath}. 
In addition, there was no update on July 4, 2020 resulting in missing cumulative deaths in that week. The weekly deaths of that week and the preceding week are declared missing. Note that the missing weekly deaths are not equivalent to the non-retrievable weekly deaths because we cannot bound their sums. 
In the main text, we focus on data reported from May 2, 2020 to September 25, 2021. We thus have $W=74$. Figure~\ref{fig:weeklydeathsraw} show the COVID-19 attributable weekly deaths data for the four most populated US states, California, Florida, New York and Texas. 

\subsection{COVID-19 vaccination data} \label{sec:data_vac}
\begin{figure}[h!]
    \centering
    \includegraphics[width = 0.8\textwidth]{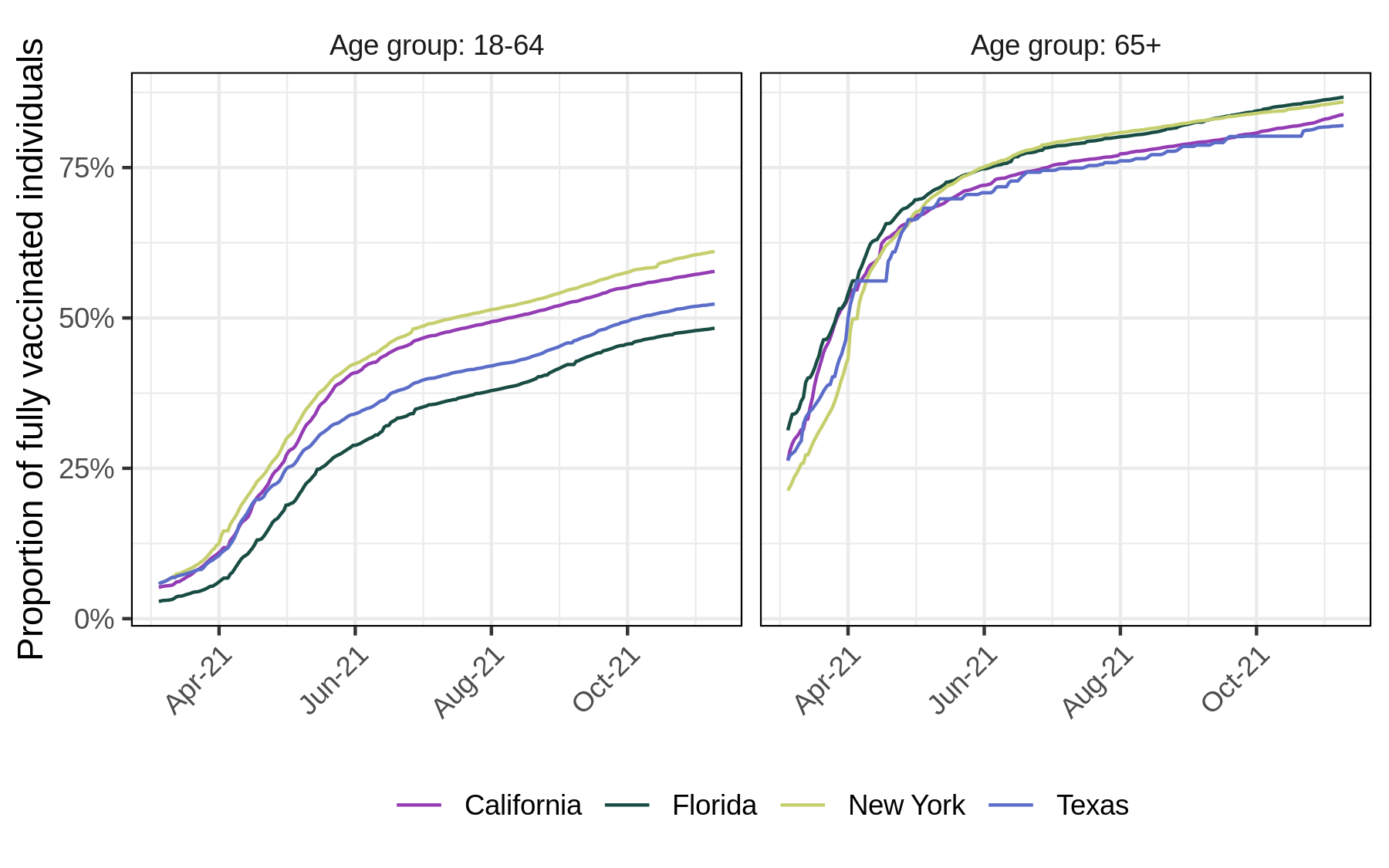}
    \caption{\textbf{Age-specific proportion of fully vaccinated individuals in four US states.} The proportion of fully vaccinated individuals over time were report by the CDC~\citep{CDC_data_vac}. Fully vaccinated individuals are defined as having receive the second dose of a two-dose vaccine or one dose of a single-dose vaccine. Column facets show the proportion of fully vaccinated individuals aged $18$-$64$ and $65+$.}
    \label{fig:vaccinedata}
\end{figure}

The CDC reports weekly time series of the proportions of individuals aged $18$-$64$ and $65+$ who are fully vaccinated~\citep{CDC_data_vac}, referred to as vaccination rate in the following. Fully vaccinated individuals are defined as having received the second dose of a two-dose vaccine or one dose of a single-dose vaccine. We use records from March 02, 2021 to September 25, 2021, and denote by $\text{v}_{m,c,w}$ the vaccination rate for state $m$ in age group $c \in \mathcal{C}$ at the start of week $w$,
\begin{align}
    \mathcal{C} = \{18-64, 65+\}.
\end{align}
Figure~\ref{fig:vaccinedata} presents the vaccination rates among individuals aged $18$-$64$ and $65+$ reported in the four most populated US states, California, Florida, New York and Texas.

\subsection{Other data sets}
We use two additional data sets.
Firstly, we retrieved the weekly COVID-19 attributable deaths regardless of age from John Hopkins University (JHU) for all US states from May 2, 2020 to September 25, 2021~\citep{jhu_datasource}, which we denote by $d^{\text{JHU}}_{m,w}$ for state $m$ and week $w$. In these data, reporting-delayed deaths are back-distributed where possible, and for this reason we use the overall deaths as a reference to mitigate the reporting delays in the age-stratified CDC data~\eqref{e:weekly_deaths}. 
Secondly, age-specific weekly COVID-19 attributable deaths are also directly reported by state Departments of Health (DoH) on their websites, or data repositories. For California, Florida and Texas, data records are available up to April 1, 2021 at~\url{https://github.com/ImperialCollegeLondon/US-covid19-agespecific-mortality-data}. The DoH do not censor low death counts, and the all-ages sum of the DoH data correspond well to JHU data. This prompts us to use the DoH data as an independent data set to assess the accuracy of our model estimates that are derived from the CDC data.

\section{Methods} \label{sec:methods}

\subsection{Modelling weekly COVID-19 attributable deaths by age} \label{sec:model}
For simplicity we suppress the state index in what follows, with all equations being analogous. Our aim is to estimate the weekly deaths by $1$-year age bands, $a\in\mathcal{A} = \{0, 1, \dots, 104, 105 \}$ in week $w\in \mathcal{W}=\{1,\dotsc,W\}$, and we denote their expectation by $\mu_{a,w}$. We first decompose $\mu_{a,w}$ as the product of the weekly deaths for all ages $\lambda_w$ with the relative contribution $\pi_{a,w}$ of age $a$ to weekly deaths, where $\sum_a \pi_{a,w}=1$ for all $w$. Second, we model the longitudinal age composition of deaths $\pi_{a,w}$ as a bivariate random function $(a,w) \to f(a,w)\in\mathbb{R}$, that is exponentiated, normalised, and evaluated on the 2D grid $\mathcal{A}\times \mathcal{W}$. Our basic latent model structure is thus
\begin{align}\label{eq:fa}
    \begin{split}
    \mu_{a,w} & = \lambda_w \pi_{a,w} \\
    \pi_{a,w} &= \text{softmax}\Big([f(a,w)]_{a \in \mathcal{A}}\Big)\\
    &= \bigg( \frac{\exp f(a,w)}{\sum_{\tilde{a}\in\mathcal{A}} \exp f(\tilde{a},w)}\bigg). 
    \end{split}
\end{align}
To link the expected weekly deaths by $1$-year age band, $\mu_{a,w}$, to the data \eqref{e:weekly_deaths}, we aggregate them over the age groups specified by the CDC, $\mu_{b,w} = \sum_{a \in b} \mu_{a,w}$, for all $b \in \mathcal{B}$. Then, we model the observed, weekly deaths in age $b$ and week $w$ for all ages and weeks through Negative Binomial distributions in the shape-scale parameterisation,
\begin{subequations}\label{e:age_composition_model}
\begin{align}
    d_{b,w} \:| \:\alpha_{b,w}, \theta & \sim \text{NegBin}(\alpha_{b,w}, \theta)\label{eq:shapescape} \\
    \mu_{b,w} & =\alpha_{b,w} \frac{\theta}{1 - \theta}\\
    \sigma_{b,w}^2 & =\alpha_{b,w} \frac{\theta}{(1 - \theta)^2} = \mu_{b,w} (1 - \theta)^{-1} = \mu_{b,w} \: (1 + \nu),\label{eq:varweeklydeaths}
\end{align}
\end{subequations}
with mean $\mu_{b,w}$ and variance $\sigma_{b,w}^2$, and where $\nu>0$ is interpreted as an overdispersion parameter. The purpose of the shape-scale parameterisation with identical $\theta$~\eqref{eq:shapescape} is that then, the weekly deaths conditional on their total follow a Dirichlet-Multinomial distribution with parameters $\alpha_{b,w}$, resulting in the succinct identity $\alpha_{b,w}=\sum_{a\in b}\alpha_{a,w}$~\citep{townes2020review}. For implementation, notice that the shape and scale parameters can be rewritten as $\alpha_{b,w} = \mu_{b,w} / \nu$ and $\theta = \nu/(1 + \nu)$.

Our Bayesian model comprises as log likelihood the sum of the log Negative Binomial densities~\eqref{eq:shapescape} over retrievable weekly deaths parameterised by $\mu_{b,w}$ and $\nu$. Considering data censoring, similar terms involving log cumulative density functions of the Negative Binomial that bound sums of the non-retrievable weekly deaths are added in the log likelihood. Taken together, for the collection of weekly deaths $\mathbf{d}=\big(d_{b,w}, b\in\mathcal{B}, w\in\{\mathcal{W}_b^{\text{WR}}, \mathcal{W}_b^{\text{WNR}}\}\big)$, our log posterior density is
\begin{subequations}
\begin{align}
    \log p(\boldsymbol{\lambda}, \boldsymbol{f}, \boldsymbol{\phi}, \nu \: | \: \mathbf{d} ) &\propto \sum_{b \in \mathcal{B}}  \sum_{w \in \mathcal{W}_b^{\text{WR}}} \log \text{NegBin}(d_{b,w} \: | \:  \mu_{b,w}/\nu, \frac{\nu}{1+\nu})  \\ 
    &\quad \quad\quad\quad+ \:\log p\big(\{d_{b,w}\}_{w \in \mathcal{W}_b^{\text{WNR}}}  \: | \: \dotsc \big)   \label{eq:priornrw} \\
    &\quad+ \:
    \sum_{w\in\mathcal{W}} \log\text{Gamma}(\lambda_w \: | \: T_w, T_w/(2\eta))  \label{eq:priorlambda}\\
    &\quad+ \: \log\mathcal{N}^+(\nu^{-1 / 2} \: | \: 0,  1) \: + \: \log p(\boldsymbol{f} \: | \: \boldsymbol{\phi}) \: + \: \log p(\boldsymbol{\phi}), \label{eq:loglikf}
\end{align}
\end{subequations}
where for brevity the term $p\big(\{d_{b,w}\}_{w \in \mathcal{W}_b^{\text{WNR}}} | \dotsc \big)$ in~\eqref{eq:priornrw} is detailed in Appendix~\ref{sup:likelihoodform}. 
The priors are specified as follows.
First, the prior on the total number of weekly deaths $\lambda_w$, in~\eqref{eq:priorlambda}, is specified in the mean-standard deviation parametrisation. The prior expectation on the total deaths is found by summing the retrievable deaths, such that $T_w = \sum_{b\in\mathcal{B},w \in \mathcal{W}_b^{\text{WR}}} d_{b,w}$. Next, we assume that the standard deviation is equal to twice the first order difference in the total deaths. We find the empirical ratio of the total death relative to their first order difference $\eta$, using ordinary least square method without intercept, and specify the standard deviation accordingly. 
Second, the inverse of the square root of the overdispersion parameter is given a standard normal distribution truncated on the positive support, the recommended prior in this context.
Lastly, the model is completed with a prior on the random function $\boldsymbol{f}$ and its hyperparameters $\boldsymbol{\phi}$, which we investigate in detail in the next section.

Considering reporting delays in the data, we explicitly allow for a rescaling of the sum of deaths across age groups in the model according to other curated data sets, which effectively re-distributes reporting delayed deaths in the CDC data set to earlier dates. Specifically, we require that our posterior predictions of the COVID-19 attributable deaths in age group $a$ and week $w$, $d^{\star}_{a,w}$, sum to $\sum_a d^{\star}_{a,w} = d_w^{\text{JHU}}$. We achieve this by exploiting the probabilistic relationship between Negative Binomial distributions in the shape-scale parameterisation and the Dirichlet-Multinomial, through 
\begin{equation} \label{eq:rescaledeaths}
\begin{split}
    & p(d^{\star}_{\cdot,w} \: | \: \text{\textbf{d}}, d_w^{\text{JHU}}) =  \\  
    & \quad\quad \int \text{DirMult}\big(d^{\star}_{\cdot,w} \: | \: d^{\text{JHU}}_w, (\alpha_{a,w})_{a\in\mathcal{A}} \big) \: p(\boldsymbol{\lambda},\nu,\boldsymbol{f},\boldsymbol{\phi} \: | \: \text{\textbf{d}} ) \: d (\boldsymbol{\lambda},\nu,\boldsymbol{f},\boldsymbol{\phi}), 
\end{split}
\end{equation}
where as before $\alpha_{a,w}=\mu_{a,w}/\nu$. We denote by $\mu^{\star}_{a,w}$ the expected predictive COVID-19 attributable deaths in age group $a$ at week $w$,
\begin{align} \label{eq:mean_predicted_deaths}
    \mu^{\star}_{a,w} \: | \: \text{\textbf{d}}, d_w^{\text{JHU}} =  \mathbb{E}\big[d^{\star}_{a,w} \: | \:  \text{\textbf{d}}, d_w^{\text{JHU}} \big].
\end{align}

\subsection{Modelling age-specific contributions to COVID-19 weekly deaths} \label{sec:f}
In~\eqref{eq:fa}, we introduced a 2D function $f(a,w)$, for which we shall find a prior in this section.
Let the number of points of the age axis be $n =|\mathcal{A}|$ and on the week axis be $m = |\mathcal{W}|$.
The total number of points on the grid is $N = n \times m$. The ensemble of pairs of points is $\boldsymbol{X} = \big(\boldsymbol{x}_1, \dots, \boldsymbol{x}_N\big) = \mathcal{A} \times \mathcal{W}$. We now investigate different modelling approaches for the function $f(a,w)$. 

\subsubsection{Two-dimensional Gaussian Process}\label{sec:2DGP}
Given observations $(\boldsymbol{X}, \boldsymbol{f}) = \big\{\big(\boldsymbol{x}_1, f(\boldsymbol{x}_1) \big), \dots, \big(\boldsymbol{x}_N, f(\boldsymbol{x}_N) \big)\big\}$, we start by considering as model of $\boldsymbol{f}$ a zero-mean 2D GP, 
\begin{align} \label{eq:2DGP}
   \boldsymbol{f} \: | \: \boldsymbol{\phi} & \sim \mathcal{GP}\big(\boldsymbol{0}, \boldsymbol{K}\big).
\end{align}
The covariance matrix $\boldsymbol{K}$ is evaluated at all pairs of points in $\boldsymbol{X}$ and has entries $\boldsymbol{K}_{\boldsymbol{x}, \boldsymbol{x}'} = \text{Cov}\big(f(\boldsymbol{x}), f(\boldsymbol{x}')\big) = k(\boldsymbol{x},\boldsymbol{x}')$ with $\boldsymbol{x}, \boldsymbol{x}' \in \boldsymbol{X}$, where $k(.,.)$ is a kernel function. 
For computational efficiency and because our output is on a multidimensional grid, we decompose the kernel function,
\begin{align} \label{eq:kernel_decomposition}
    k\Big((a,w), (a',w')\Big)  = k^1(a,a') \: k^2(w,w')
\end{align}
where $k^1(.,.)$ and $k^2(.,.)$ are kernel functions over ages and weeks, respectively. The corresponding covariance matrix $\boldsymbol{K}$ is calculated with the Kronecker product $\boldsymbol{K} = \boldsymbol{K}^2 \otimes \boldsymbol{K}^1$, and the number of operations to evaluate the covariance matrix reduces from $\mathcal{O}(N^3)$ to $\mathcal{O}(2N^{3/2})$~\citep{JMLR:v12:gonen11a,Saatci:2011}. For implementation, we note that the order of the product follows from the fact that the matrix's entries are stacked columnwise. We rely on the efficient Kronecker product implementation via matrix-vector products proposed in the Supplementary Material, Section 4 of~\cite{Wilson:2014}. In our applications, we set $k^1$ and $k^2$ to be squared exponential kernel functions with variance scale $\zeta^2$ and specific lengthscales for the rows and the columns, $\gamma_1$ and $\gamma_2$. Our priors on the length scales are independent $\gamma_i\sim\text{Inv-Gamma}(5,5)$ for $i=1,2$, and $\zeta\sim\text{Cauchy}^+(0, 1)$. For brevity, we denote $\boldsymbol{\phi} = \{\zeta, \gamma_1, \gamma_2\}$.

\subsubsection{B-splines surface} \label{sec:2DBS}
B-splines basis functions - or, more simply, B-splines - are constructed from polynomial pieces that are joined at certain values over the input space, called knots, and defined by a polynomial degree, $d$, and a non-decreasing sequence of knots, $(t_1 \leq t_2 \leq \dots \leq t_{K})$, where $K$ is the number of knots. Given those, the total number of B-splines is $I = K + d - 1$. We show how a B-spline is constructed in Appendix~\ref{sup:bsplineconstruction}. In this paper, we use cubic B-splines, with $d=3$. Moreover, we use equally spaced knots on both dimensions, such that the only tuning parameter is the number of knots. A fundamental property of B-splines, that constitutes most of their attractiveness, is that they are smooth. More rigorously, a cubic B-spline defined on strictly increasing knots is piecewise infinitely differentiable between the knots, and of continuity $C^{2}$ on the knots (\cite{goldman:2002}, chap. 7). 
$\boldsymbol{f}$ can be modelled with a tensor product of B-splines given by
\begin{align}\label{eq:2DBS}
    f(a, w) = \sum_{i = 1}^{I} \sum_{j = 1}^{J} \beta_{i, j} \: B^1_{i}(a) \: B^2_{j}(w),
\end{align}
where $B^1_{i}(.)$ is the $i$th B-spline defined on the knot vector $(t^1_{1} \leq t^1_{2} \leq \dots \leq t^1_{K_1})$ over the space $\mathcal{A}$ with $K_1$ being the number of knots, $i \in \mathcal{I} = \{1, \dots, I\}$ and $I = K_1 + 2$. Similarly, $B^2_{j}(.)$ is the $j$th B-spline defined on the knot vector $(t^2_{1} \leq t^2_{2} \leq \dots \leq t^2_{K_2})$ over the space $\mathcal{W}$ with $K_2$ being the number of knots, $j \in \mathcal{J} = \{1, \dots, J\}$ and $J = K_2 + 2$.
The ensemble of B-splines form a matrix basis denoted by $\boldsymbol{B}^1$ and $\boldsymbol{B}^2$ of size $I \times n$ and $J \times m$, respectively. The pairs of B-splines indices is denoted by $\boldsymbol{U} = \big(\boldsymbol{u}_1, \dots, \boldsymbol{u}_{M}\big) = \mathcal{I} \times \mathcal{J}$ with $M = I \times J$. $[\beta_{\boldsymbol{u}}]_{u \in \boldsymbol{U}}$ is a rectangular set of coefficients. 
In our applications, we obtain standard B-splines surfaces by placing independent standard normal priors on all B-splines coefficients $\beta_{\boldsymbol{u}}$. To keep notation streamlined, notice that there are no hyperparameters $\boldsymbol{\phi}$. 

\subsubsection{Regularised B-splines projected Gaussian Process}
Given the B-splines indices and corresponding coefficients $(\boldsymbol{U}, \boldsymbol{\beta}) = \big\{(\boldsymbol{u}_1, {\beta}_{\boldsymbol{u}_1}), \dots, (\boldsymbol{u}_{M}, {\beta}_{\boldsymbol{u}_{M}})\big\}$, we place a 2D GP on the coefficients, 
\begin{align}\label{eq:beta_GP}
    \boldsymbol{\beta} \: | \: \boldsymbol{\phi} \sim \mathcal{GP}(0, \boldsymbol{K}_{\beta}).
\end{align}
The covariance matrix has entries $(\boldsymbol{K}_\beta)_{\boldsymbol{u}, \boldsymbol{u}'} = \text{Cov}\big(\beta_{\boldsymbol{u}}, \beta_{\boldsymbol{u}'}\big) = k_{\beta}\big(\boldsymbol{u}, \boldsymbol{u}'\big)$ where $k_{\beta}(.,.)$ is a kernel function depending on unknown hyperparameters $\boldsymbol{\phi}$. This approach is equivalent to directly placing a GP on $\boldsymbol{f}$ with a covariance matrix projected by B-splines. To show this, we  rewrite~\eqref{eq:2DBS} as a matrix calculation, $\boldsymbol{f}= (\boldsymbol{B}^1)^T \boldsymbol{\beta} \boldsymbol{B}^2$, and find its vectorized form
\begin{align}\label{eq:affine}
    \text{vec}(\boldsymbol{f}) &= \text{vec}\Big((\boldsymbol{B}^1)^T \boldsymbol{\beta} \boldsymbol{B}^2\Big)
    = (\boldsymbol{B}^2 \otimes \boldsymbol{B}^1)^T \text{vec}(\boldsymbol{\beta}). 
\end{align}
The linear operator produces functions from $\mathbb{R}^{IJ}$ to $\mathbb{R}^{nm}$. GPs are closed under linear operations (\cite{Papoulis:2002}, chap.10), and so $\boldsymbol{f}$ specified by (\ref{eq:2DBS}-\ref{eq:beta_GP}) is the GP 
\begin{align} \label{eq:GPview}
    \boldsymbol{f} \: | \: \boldsymbol{\phi}\sim \mathcal{GP}\Big(0, \big(\boldsymbol{B}^2 \otimes \boldsymbol{B}^1\big)^T \boldsymbol{K}_{\beta} \big(\boldsymbol{B}^2 \otimes \boldsymbol{B}^1\big) \Big).
\end{align}
In our applications, we decompose the kernel function $k_{\beta}$ into two kernel functions over the rows and columns of the B-spline coefficients matrix as in~\eqref{eq:kernel_decomposition}. We use squared exponential kernel functions with variance scale $\zeta^2$ and specific lengthscales for the rows and the columns, $\gamma_1$ and $\gamma_2$. Our priors on the length scale priors are independent $\gamma_i\sim\text{Inv-Gamma}(5,5)$ for $i=1,2$, and $\zeta\sim\text{Cauchy}^+(0, 1)$. Under this prior, $\boldsymbol{\phi} = \{\zeta, \gamma_1, \gamma_2\}$.

The kernel function obtained by projecting a base kernel function with cubic B-splines is $C^2$ as it can be represented as a linear combination of $C^{2}$ functions (proof in Appendix~\ref{supp:cinfprop}). Therefore the surface obtained when modelling a surface with the B-spline tensor product~\eqref{eq:GPview}, unlike a standard 2D GP~\eqref{eq:2DGP}, has inherent smoothness properties not matter the base kernel functions $k_{\beta}$ used.
We can also interpret~\eqref{eq:GPview} as a regularised splines method. Indeed, existing regularised spline methods such as smoothing splines~\citep{Sullivan:1986,OSullivan1988} and P-splines~\citep{Eilers1996,Eilers2006} aim to minimise in a frequentist setting the loss function 
\begin{align}
    \underbrace{\sum_{a \in \mathcal{A}} \sum_{w \in \mathcal{W}} \Big( f(a, w) - \sum_{i = 1}^{I} \sum_{j = 1}^{J} \hat{\beta}_{i, j} B^1_{i}(a) B^2_{j}(w) \Big)^2}_{\text{data fit}} + \underbrace{P}_{\text{penalty}}, \label{eq:lossfunction}
\end{align}
where $\hat{\boldsymbol{\beta}}$ are the estimated coefficients and $P$ is a penalty applied on the second derivative of the fitted curve for smoothing splines and on finite differences of adjacent B-splines coefficients for P-splines.
Now adopting a Bayesian approach, our aim is to maximise the probability of the posterior parameter conditional upon the data, which is proportional to the likelihood multiplied by the prior. By placing a GP on the B-splines coefficients, the prior of $\boldsymbol{f}$ modelled with the B-spline tensor product~\eqref{eq:GPview} can be decomposed in the same way as~\eqref{eq:lossfunction}, 
\begin{align}
\begin{split}
    \text{log p} (\text{vec}(\boldsymbol{f}) \: | \: \boldsymbol{\phi}) & \propto -\frac{1}{2} \Big( \text{vec}(\boldsymbol{f})^T \big(\boldsymbol{B}^2 \otimes \boldsymbol{B}^1\big)^T \boldsymbol{K}_{\beta} \big(\boldsymbol{B}^2 \otimes \boldsymbol{B}^1\big) \text{vec}(\boldsymbol{f})  \\
    &\quad\quad\quad+ \text{log} \:  \Big|\big(\boldsymbol{B}^2 \otimes \boldsymbol{B}^1\big)^T \boldsymbol{K}_{\beta} \big(\boldsymbol{B}^2 \otimes \boldsymbol{B}^1\big)\Big|  \Big) \\
   & \propto -\frac{1}{2} \Big( \underbrace{\text{vec}(\boldsymbol{f})^T \big(\boldsymbol{B}^2 \otimes \boldsymbol{B}^1\big)^T \boldsymbol{K}_{\beta} \big(\boldsymbol{B}^2 \otimes \boldsymbol{B}^1\big) \text{vec}(\boldsymbol{f})}_{\text{data fit}} + \underbrace{\text{log} \: | \boldsymbol{K}_{\beta} |}_{\text{penalty}}  \Big), 
   \end{split}
\end{align}
where $|\boldsymbol{A}|$ denotes the determinant of matrix $\boldsymbol{A}$. Thus, the log determinant of the covariance matrix in~\eqref{eq:beta_GP} can be interpreted as a complexity penalty, or regulariser, which comes into play if the kernel has a free parameter to control the model complexity (\cite{MacKay:2003}, chap. 28). 

\subsubsection{Bayesian P-splines}
Bayesian P-splines have been developed as an extension of the frequentist P-splines, in~\citet{Lang2004, Brezger2006}, and impose a spatial dependency on the B-splines coefficients to avoid overfitting. Bayesian P-splines are obtained by placing the Gaussian Markov Random Field prior on the B-splines coefficients,
\begin{align} \label{eq:CAR}
    \beta_{i,j} \: | \: \boldsymbol{\beta}_{-(i,j)}, \boldsymbol{\phi} \sim \mathcal{N} \Big(  \sum_{i' = 1}^{I} \sum_{j' = 1}^{J} \frac{ \delta_{(i,j),(i',j')} \: \beta_{i',j'}}{\delta_{(i,j)}^+}, \frac{\tau^2}{\delta_{(i,j)}^+} \Big),
\end{align}
where $\boldsymbol{\beta}_{-(i,j)}$ is the matrix $\boldsymbol{\beta}$ without entry $(i,j)$, $\delta_{(i,j),(i',j')}$ evaluates to one if $(i,j)$ and $(i',j')$ are neighboring coordinates vertically or horizontally, and otherwise to zero, and $\delta_{(i,j)}^+ = \sum_{i' = 1}^{I} \sum_{j' = 1}^{J} \delta_{(i,j),(i',j')}$. Here, $\tau$ is a spatially varying precision parameter. This model implies that each $\beta_{i,j}$ is normally distributed with a mean equal to the average of its neighbors. 
We use the Bayesian P-splines prior in benchmark comparisons, due to their close structural relationship with B-splines projected GP. 
In our applications, we accelerated the evaluation of the likelihood by simplifying the joint likelihood to the pairwise difference formulation and we placed a sum-to-zero constrain on the parameters to ensure the identifiability of the pairwise differences as proposed by~\citep{Morris2019}. We used the prior $\tau\sim\text{Cauchy}^+(0, 1)$. To keep notation streamlined, we write $\boldsymbol{\phi} = \{\tau\}$. 

\subsection{Numerical inference}
All inferences using model~(\ref{eq:fa}-\ref{e:age_composition_model}) coupled with different priors on the random surface $\boldsymbol{f}$ were fitted with RStan version 2.21.0, using an adaptive Hamiltonian Monte Carlo (HMC) sampler~\citep{rstan}. 8 HMC chains were run in parallel for 1,500 iterations, of which the first 500 iterations were specified as warm-up. 
Inferences in simulation analysis and benchmarking were performed similarly.

\section{Simulation results} \label{sec:simulation}
Prior to application to the CDC data, we compared the performance of the spatial models defined in Section~\ref{sec:f} at retrieving the mean surface of 2D count data. 
In the simulations, we considered observations on a 2D grid $\{0, 0.02 \dots, 0.98, 1\} \times \{0, 0.02 \dots, 0.98, 1\}$, such that there are $50\times50$ entries. The observations are generated through a Negative Binomial model using as mean the exponential of a 2D GP with a squared exponential kernel. The variance scale of the 2D GP is fixed to $1$ and the length scale is varied ($\gamma \in \{0.05, 0.25, 1\}$) to generate weakly, mildly and strongly correlated observations. 
The training set includes 40\% uniformly sampled observations (i.e., number of observations in the training set $N= 640$). Here, we compare the performance of four models with different prior specifications on the mean surface, a standard 2D GP, a standard B-splines surface, Bayesian P-splines and regularised B-splines projected 2D GP.
For the methods using B-splines, the same number of knots are placed along the two axes. 

\begin{figure*}[t]
\centering
    \centering
       \includegraphics[width=0.95\linewidth]{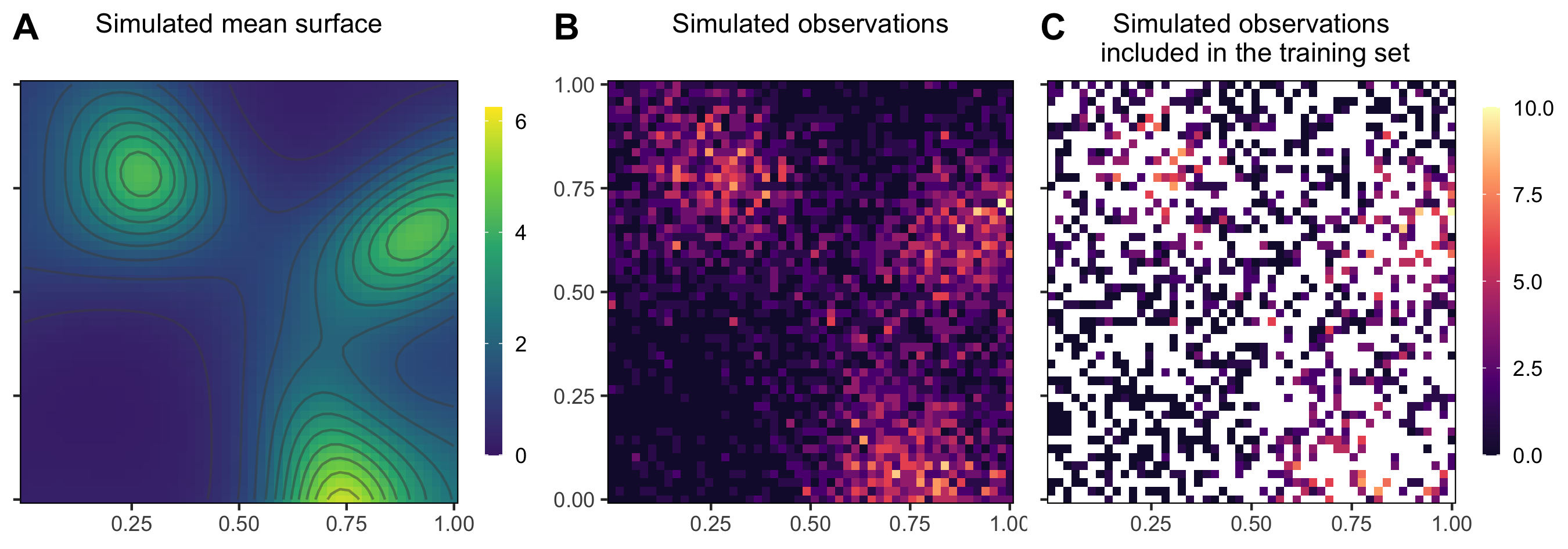}
      \caption{\textbf{Midly correlated 2D simulated observations.} \textbf{(A)} The mean surface is defined by the exponential of a 2D GP with a squared exponential kernel implying mild correlations ($\gamma = 0.25$). \textbf{(B)} Count data were simulated from a negative Binomial model. \textbf{(C)} 40\% of the simulated observations were included in the training set. }
    \label{fig:simu_l21}
\end{figure*}

Figure~\ref{fig:simu_l21}A shows the simulated mean surface in the mildly correlation scenario obtained with $\gamma=0.25$, Figure~\ref{fig:simu_l21}B the simulated observations and Figure~\ref{fig:simu_l21}C the simulated observations included in the training set. The mean surface estimated by the standard 2D GP model is shown in Supplementary Figure~\ref{fig:simu_gp}. Figure~\ref{fig:simu_l22} presents the estimated mean surface when a standard B-spline surface, Bayesian P-splines, and regularised B-splines projected 2D GP priors are used for inference with different numbers of knots.

We find that the standard 2D GP and the regularised B-splines projected 2D GP recover well the simulated true mean surface (Figures~\ref{fig:simu_l22} and Supplementary Figure~\ref{fig:simu_gp}). Moreover, the regularised B-splines projected 2D GP obtains equivalent or better predictive performance as the standard 2D GP for a shorter running time, 73.95\% faster on average for $30$ knots (Table~\ref{tab:LOO_simu} and Supplementary Table~\ref{tab:CI_simu}). In contrast, the standard B-splines and the Bayesian P-splines approaches overfit the data when the number of knots increase (Figures~\ref{fig:simu_l22} and Table~\ref{tab:LOO_simu}). Similar results were obtained for the other correlation scenarios (Figures~\ref{fig:simu_data_1}-\ref{fig:simu_bsplines_3}, Table~\ref{tab:LOO_simu} and Supplementary Table~\ref{tab:CI_simu}).

\begin{figure}[p]
    \centering
    \includegraphics[width=0.95\linewidth]{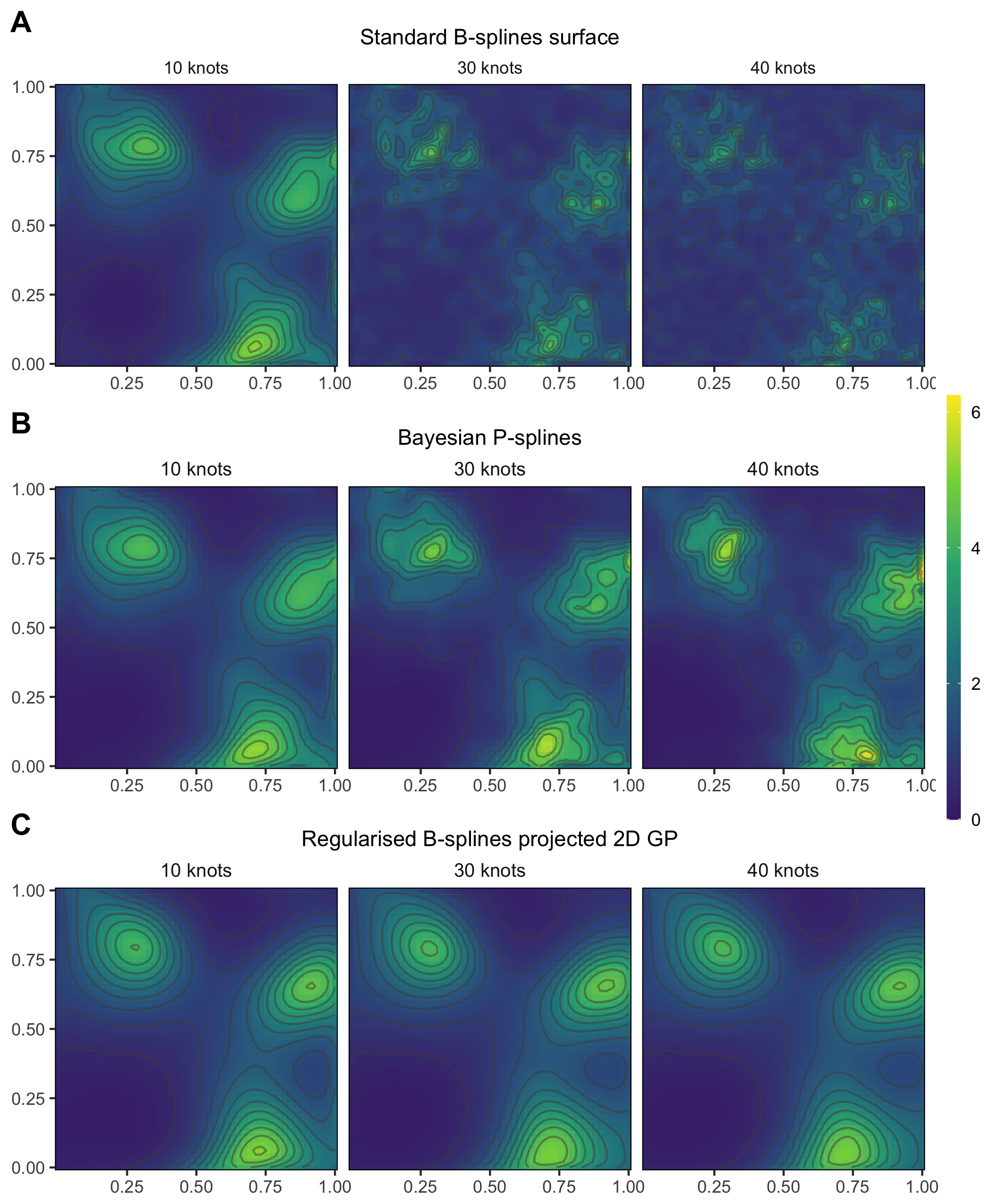}
    \caption{\textbf{Estimated mean surfaces using non-regularized and regularized B-splines spatial models.} Posterior median of the mean surface estimated by \textbf{(A)} standard B-splines, \textbf{(B)} Bayesian P-splines and \textbf{(C)} regularised B-splines projected 2D GP with different number of knots given mildly correlated simulated data ($\gamma = 0.25$). The simulated data are illustrated in Figure~\ref{fig:simu_l21}.}
    \label{fig:simu_l22}
\end{figure}

\begin{table*}[!p]
\small
\begin{threeparttable}
 \centering
  \centering
      \begin{tabular}{l l | r r r}
      \toprule
      \hline
       \multicolumn{2}{c|}{\multirow{2}{*}{Method}}  & \multicolumn{3}{c}{Simulation scenarios} \\[0.05cm]
      && Weakly correlated & Mildly correlated & Strongly correlated \\[0.05cm]
      \hline
      &&&&\\[-0.35cm]
      && \multicolumn{3}{c}{Mean squared error of the mean surface} \\[0.05cm]
      \cline{3-5} 
        &&&&\\[-0.3cm]

        \multicolumn{2}{l|}{\textbf{Standard 2D GP}} \\
        &&0.12  & 0.04\tnote{$\star$} & 0.05\tnote{$\star$}   \\[0.3cm]
        
       \multicolumn{5}{l}{\textbf{Standard B-splines surface}}  \\
        &Number of knots  & & \\
        & $10$ & 0.11   & 0.15  & 0.25  \\[0.05cm]
        & $30$ & 0.46  & 0.61  & 0.91  \\[0.05cm]
         & $40$ &  0.67   & 0.93  & 1.35  \\[0.2cm]
         
         \multicolumn{5}{l}{\textbf{Bayesian P-splines surface}}  \\
        &Number of knots  & & \\
        & $10$ & 0.11  & 0.09  & 0.10  \\[0.05cm]
        & $30$ & 0.14   & 0.13  & 0.12  \\[0.05cm]
         & $40$ &  0.15  & 0.13  & 0.12 \\[0.2cm]
        
      \multicolumn{5}{l}{\textbf{Regularised B-splines projected 2D GP}} \\
        &Number of knots & & \\
        & $10$ & 0.10   & 0.06  & 0.06 \\[0.05cm]
        & $30$ & 0.09   & 0.05  & 0.06 \\[0.05cm]
         & $40$ &  0.08\tnote{$\star$}  & 0.05  & 0.06 \\[0.05cm]
          
        \hline
         &&&&\\[-0.35cm]
        
        & &  runtime in minutes &  runtime in minutes &  runtime in minutes \\[0.02cm]
        & &  (-\%longest runtime) &  (-\%longest runtime) &  (-\%longest runtime) \\[0.05cm]
            \cline{3-5} 
        &&&&\\[-0.3cm]
        
    \multicolumn{2}{l|}{\textbf{Standard 2D GP}} \\
         && 24 (-0.00\%) &  47 (-0.00\%) &  21 (-0.00\%)
         
         \\[0.2cm]
         
    \multicolumn{5}{l}{\textbf{Standard B-splines surface}}  \\
        &Number of knots  & & \\
        & $10$ & 1 (-97.23\%) &   1 (-98.24\%) &  1 (-96.96\%) \\[0.05cm]
        & $30$ &  1 (-93.92\%) & 1 (-96.98\%) & 1 (-93.22\%) \\[0.05cm]
         & $40$ & 2 (-90.97\%) &  2 (-95.46\%) & 2 (-88.18\%) \\[0.2cm]
         
    \multicolumn{5}{l}{\textbf{Bayesian P-splines}}  \\
        &Number of knots  & & \\
        & $10$ & 7 (-69.18\%) &   6 (-86.44\%) &  4 (-78.81\%) \\[0.05cm]
        & $30$ &  3 (-85.95\%) & 3 (-93.04\%) & 4 (-81.05\%) \\[0.05cm]
         & $40$ & 3 (-86.95\%) &  4 (-90.49\%) & 4 (-82.95\%) \\[0.2cm]
         
     \multicolumn{5}{l}{\textbf{Regularised B-splines projected 2D GP}} \\
        &Number of knots & & \\
        & $10$ & 5 (-80.58\%) & 3 (-93.26\%) & 4 (-79.78\%) \\[0.05cm]
         &$30$& 6 (-76.60\%) & 8 (-83.36\%) & 8 (-61.89\%) \\[0.05cm]
         &$40$ & 6 (-76.61\%) & 8 (-82.00\%) & 8 (-60.02\%) \\
          
\bottomrule
 \end{tabular}
    \begin{tablenotes}
        \item[$\star$] Best predictive performance.
    \end{tablenotes}
    \caption{{\bf Performance of four spatial models on simulated data.} \textbf{(Top)} Means and standard deviations of the squared errors between the posterior median estimates of the mean surface and the true values, evaluated on the test data withheld from fitting. \textbf{(Bottom)} Proportional decreases in runtimes compared to the slowest model (for which the proportional decrease is null). }
    \label{tab:LOO_simu} 
    \end{threeparttable}
\end{table*}


\section{Benchmark results} \label{sec:benchmark}
We next benchmarked the performance of the spatial models defined in Section~\ref{sec:f} on data unrelated to COVID-19, which has previously been used to benchmark spatial models. The data set consists of more than 83,000 locations across East Africa with measured deviations in land surface temperatures~\citep{Ton2018}. We use the same training data as in~\cite{mishra:2020}, consisting of 6,000 uniformly sampled locations, and fit them using a Gaussian likelihood with mean surface specified by the various spatial models, and a free observation variance parameter. We compared standard 2D B-splines, Bayesian P-splines, regularised B-splines projected 2D GPs, low rank Gaussian Markov random fields, neural network models, and the $\pi$VAE model~\citep{mishra:2020} as a prior for the mean surface. For the spline methods, $125$ equidistant knots were placed along both axes. The results are summarised in Supplementary Table~\ref{tab:benchmarkMSE} and Figure~\ref{fig:benchmark}. We find that the regularised B-splines projected 2D GP model (testing MSE: 2.96) obtained better predictive performance than the standard B-splines surface model (testing MSE: 4.45), a low rank Gaussian Markov random field (testing MSE: 4.36), and a neural network model (testing MSE: 14.94) and similar predictive performance than Bayesian P-splines (testing MSE: 2.57) and a standard 2D GP (testing MSE: 2.47). It had a worse predictive performance than the $\pi$VAE model (testing MSE: 0.38), but training $\pi$VAE is not trivial and requires considerable computational runtimes. 

\section{Time trends of age-specific COVID-19 deaths in the US}\label{sec:realdatanalysis}

Based on the simulation and benchmark results, we used the regularised B-splines projected 2D GP model to reconstruct the time trends in age-specific COVID-19 attributable deaths across the US. We placed $12$ equidistant knots over the age axis and $10$ over the week axis. This choice was determined such that the predictive performance did not significantly increase with more knots. Overall, fitting the regularised B-splines projected 2D GP model took 48 hours on a high performance computing environment, with typical effective sample sizes between 342 to 22127, and there were no reported divergences in Stan's Hamiltonian Monte Carlo algorithm. 
Here we focus on studying the recovered trends in age-specific COVID-19 deaths during the summer 2021 resurgence in relation to vaccine coverage. Vaccinate coverage is reported by age strata that do not match the age strata in which the weekly deaths are reported by the CDC, which renders our Bayesian semi-parametric modelling approach on 1-year age groups advantageous. 

\begin{figure}[!p]
    \centering
    \includegraphics[width = 0.95\textwidth]{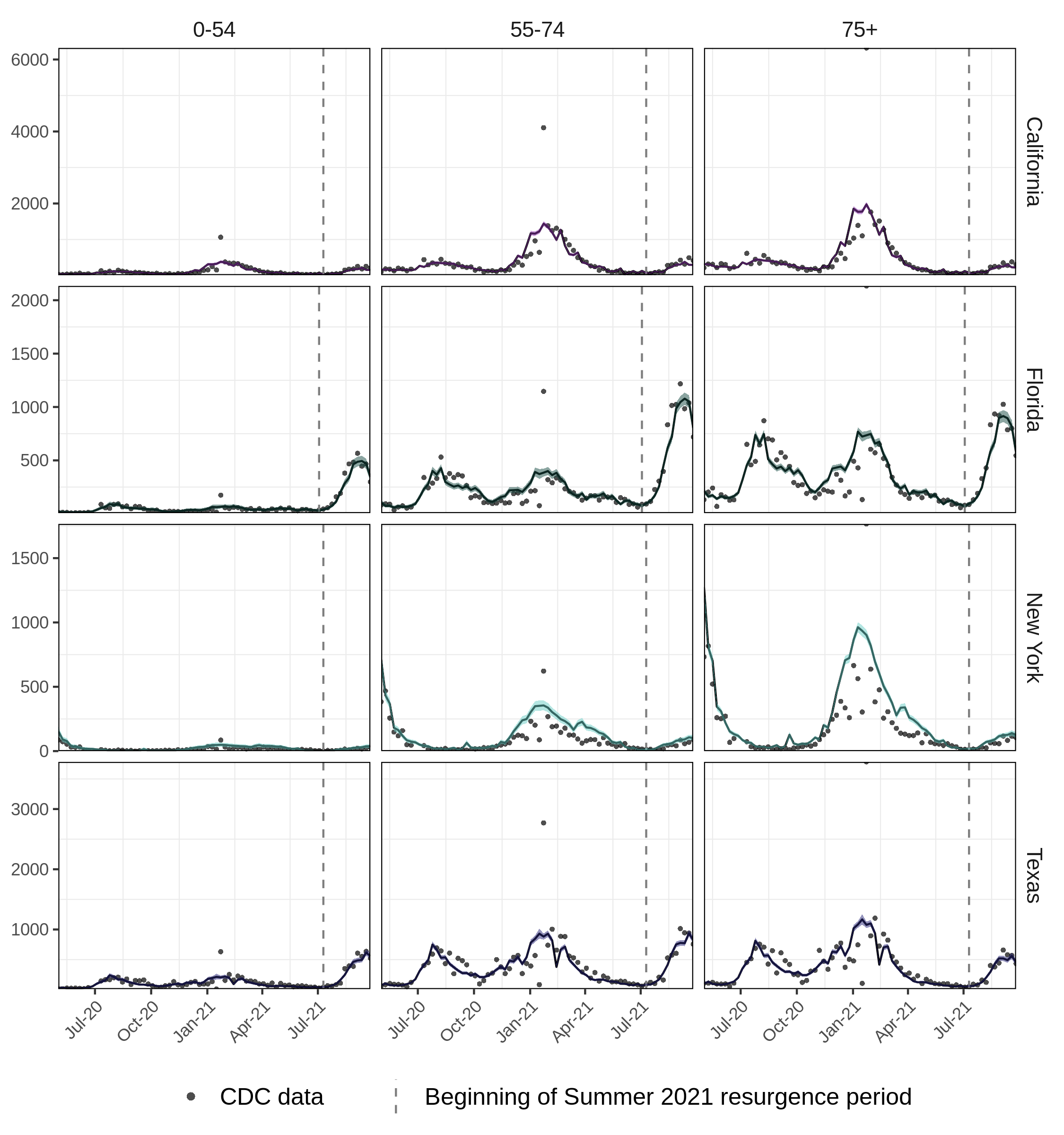}
    \caption{\textbf{Predicted age-specific COVID-19 attributable deaths in the four most populated US states.} Shown are the posterior median (line) and 95\% credible intervals (ribbon) of the predicted weekly COVID-19 attributable deaths obtained with~\eqref{eq:rescaledeaths}. The CDC data are shown with dots. The start of the 2021 summer resurgence period is indicated as dashed vertical line.}
    \label{fig:predicteddeath}
\end{figure}

\subsection{Smooth estimates of age-specific COVID-19 attributable deaths over time}

Figure~\ref{fig:predicteddeath} shows the predicted weekly attributable COVID-19 deaths over time that occurred in the four most populated US states. The predictions for each $1$-year age are aggregated into the age bands $0$-$54$, $55$-$74$ and $75+$ for direct comparison to the CDC data, shown as black dots. The data and estimates clearly reflect the multiple COVID-19 waves across the four states, with high numbers of deaths in the elderly.

The CDC data contain many anomalies, including censored and delayed entries, such as the unrealistically high number of deaths reported for the week starting on January 23, 2021 (Figure~\ref{fig:predicteddeath}). We used the curated all-ages weekly COVID-19 deaths records reported by JHU to re-distribute the delayed deaths to earlier weeks~\eqref{eq:relative_deaths}. The posterior median and 95\% credible intervals in Figure~\ref{fig:predicteddeath} illustrate how the reporting delays apparent in the CDC data are adjusted by using the JHU data as external calibration, and how the discretized CDC data informs our estimates of the age composition of deaths. Using our model in conjunction with the curated JHU data on all COVID-19 deaths, we can estimate cumulative mortality for any age band. We find that tragically, as of September 25, 2021, the cumulative mortality rates in individuals aged 85+ now exceed 3\% in Texas and New York (Supplementary Figure~\ref{fig:mortalityrateall}). 

We validated our estimates of the age profile of COVID-19 attributable deaths using an external data set, i.e. that was not used to inform the model. The external data set includes age-specific COVID-19 deaths reported directly by US states DoH (presented in Section~\ref{sec:CDCdata}). We computed the empirical age-specific contribution to weekly COVID-19 attributable deaths from the DoH data and checked if the latter lied inside the 95\% posterior credible interval estimated by our model. We find a 93.82\% prediction using the regularised B-splines projected 2D GP model across the four states, California, Florida, New York and Texas. State stratification of the prediction performance are presented in Supplementary Table~\ref{tab:comparison_estimate_cum_deaths_age}.

For completeness, we also fitted the model using alternative smoothing priors. Supplementary Section~\ref{supp:realdatanalysis_one_state} shows the estimated age-specific composition of the COVID-19 attributable deaths in three weeks in Florida, as obtained with a standard 2D GP smoothing prior, standard 2D B-splines, Bayesian P-splines and the regularised B-splines projected 2D GP. Moreover, Supplementary Table~\ref{tab:comparison_estimate_cum_deaths_age} presents the accuracy of their estimates compared to the external data set, ranging from 94.31\% to 95.29\%. Under the model specified with a Bayesian P-splines and standard 2D B-Splines surface, the estimated 2D surface over ages and weeks was wiggly and had a large uncertainty. Moreover, better predictive performances, quantified by the expected log pointwise predictive density~\citet{Verity2020}, were obtained for the B-Splines projected 2D GP compared to all other methods (Supplementary Table~\ref{tab:ELPD_comp}). Those results motivate the use of our proposed prior, the B-Splines projected 2D GP.

\subsection{Strong summer 2021 resurgence in age-specific COVID-19 deaths are associated with limited vaccine coverage}

Since July 2021, COVID-19 cases started to increase substantially~\citep{jhu_datasource} and with them COVID-19 deaths in all age groups (Figure~\ref{fig:predicteddeath}). We define the summer 2021 resurgence as the upward trend in COVID-19 attributable deaths spanning from July 03, 2021 to September 25, 2021 across the United States, and denote by $w^{\text{start-r}}_m$ the week index corresponding to the first week included in the summer 2021 resurgence period for state $m$. The week index $w^{\text{start-r}}_m$ is defined as the first week from July 01 2021 for which a 4-week central moving average on the all-age weekly COVID-19 deaths was increasing. The state-specific start of the resurgence period is indicated by a dashed vertical line in Figure~\ref{fig:predicteddeath}. 
As is evident from the empirical all-age data and confirmed in our reporting-adjusted age-specific estimates, the increase in COVID-19 deaths was highly heterogeneous across US states, with small increases seen in California and New York, and large increases seen in Florida and Texas, relative to the magnitude of COVID-19 deaths in the previous waves.

We hypothesised that the variation in the magnitudes of the summer 2021 resurgence in COVID-19 deaths could be associated with differences in how comprehensively age-specific populations in each state took up the COVID-19 vaccine offer. To test this hypothesis, we attached as transformed parameter to the model presented in Section~\ref{sec:model}, age-specific predicted COVID-19 deaths relative to previous waves over the summer 2021. The relative COVID-19 deaths for state $m$, in age group $c$, at week $w$ in summer 2021, denoted by $r_{m,c,w}$, are defined as the ratio of the expected posterior predictive weekly deaths in week $w$ to the maximum of the expected posterior predictive weekly deaths attained before the summer 2021 resurgence period in the same state and age group,
\begin{align} \label{eq:relative_deaths}
 \begin{split} 
   r_{m,c,w}  &= \mu^{\star}_{m,c,w} \: \big/ \: \mu^{\star, \text{max-pre-r}}_{m,c} 
 \end{split}
\end{align}
for $c \in \mathcal{C}$ and $w \geq w^{\text{start-r}}_m$, where $\mu^{\star}_{m,c,w}$ are the expected posterior predictive weekly deaths for state $m$, in age group $c$, at week $w$ defined in~\eqref{eq:mean_predicted_deaths}, and $\mu^{\star, \text{max-pre-r}}_{m,c} = \text{max}\big(\{\mu^{\star}_{m,c,w}\}_{w < w^{\text{start-r}}_m} \}\big)$. For simplicity we here suppress in the notation the conditional dependence on the data. By dividing the mean posterior predictive weekly deaths by their maximum value attained on previous waves, the resulting relative COVID-19 deaths are adjusted for state-specific factors that are known to influence the magnitude of SARS-CoV-2 infections such as population density, stringency of non-pharmaceutical interventions, and comprehensiveness of behaviour changes.

We then assessed if the relative resurgences in COVID-19 deaths across age groups, weeks and states could be associated with differences in pre-resurgence vaccination rates, within a random-effects meta-regression model across states. The pre-resurgence vaccination rates are defined as the proportion of fully vaccinated individuals 14 days before the start of the summer 2021 resurgence period and denoted by $\text{v}_{m,c}^{\text{pre-r}}$ for state $m$ and age group $c$. We formulated the model as follows,
\begin{align}
\begin{split}  \label{eq:model_vaccination}
    r_{m,c,w} &\: | \: \xi_{m,c,w}, \kappa_{m} \sim \text{Gamma}\big(\xi_{m,c,w}, \kappa_{m}^2\big) \\
      \xi_{m,c,w} &= \exp\big( \chi_{m,c} + \psi_{m,c} \: (w - w^{\text{start-r}}_m) \big) \\
      \begin{bmatrix}
\chi_{m,18-64}\\
\chi_{m,65+}
\end{bmatrix}
&=
\begin{bmatrix}
\chi_{18-64}^{\text{base}} + \chi_{m,18-64}^{\text{state}} \\
\chi_{65+}^{\text{base}} + \chi_{m,65+}^{\text{state}} \\ 
\end{bmatrix}
+
\begin{bmatrix}
\chi^{\text{vacc}} & \chi^{\text{vacc-cross}}_{18-64}\\
\chi^{\text{vacc-cross}}_{65+} & \chi^{\text{vacc}}\\ 
\end{bmatrix}
\begin{bmatrix}
\text{v}_{m,18-64}^{\text{pre-r}} \\
\text{v}_{m,65+}^{\text{pre-r}} \\ 
\end{bmatrix} \\
      \begin{bmatrix}
\psi_{m,18-64}\\
\psi_{m,65+}
\end{bmatrix}
&=
\begin{bmatrix}
\psi_{18-64}^{\text{base}}  \\
\psi_{65+}^{\text{base}}  \\
\end{bmatrix}
+
\begin{bmatrix}
\psi^{\text{vacc}} & \psi^{\text{vacc-cross}}_{18-64}\\
\psi^{\text{vacc-cross}}_{65+} & \psi^{\text{vacc}}\\ 
\end{bmatrix}
\begin{bmatrix}
\text{v}_{m,18-64}^{\text{pre-r}} \\
\text{v}_{m,65+}^{\text{pre-r}} \\ 
\end{bmatrix}
\end{split} 
\end{align}
for $c \in \{ [18-64], [65+]\}$. The $\chi_{m,c}$ correspond to state- and age-specific baseline terms at the start of the summer resurgence, and the $\psi_{m,c}$ capture state- and age-specific growth rates. Note that the model also allows for indirect effects of vaccination rates in the other age group.
Random effects and fixed effects (in terms of vaccination rate) were specified to the baseline terms, and fixed effects to the growth rates,
\begin{align} 
\begin{split} 
     \chi_c^{\text{base}}, \psi_c^{\text{base}}  &\sim \mathcal{N}(0, 0.5), \\
     \chi_{m,c}^{\text{state}} &\sim \mathcal{N}(0, \sigma_{c,\chi}^2), \\
     \chi^{\text{vacc}}, \psi^{\text{vacc}}  &\sim \mathcal{N}(0, 0.5), \\
     \chi^{\text{vacc-cross}}_c, \psi^{\text{vacc-cross}}_{c}  &\sim \mathcal{N}(0, 0.5), \\
     \sigma_{c,\chi},  \kappa_{m} &\sim \text{Cauchy}^+(0,1).
     \end{split} 
\end{align}
for $c \in \{ [18-64], [65+]\}$.
The model structure was motivated by the fact that by the start of the summer resurgence, the epidemics in each state had widely different magnitudes, likely depending on several state-specific unobserved factors, such as non-pharmaceutical intervention measures, mask use, or mobility, which we aimed to capture with the $\chi_{m,c}^{\text{state}}$ baseline random effects. Next, we hypothesized that after accounting for baseline differences, deaths would grow at smaller rates in populations with higher vaccine coverage at the start of the summer resurgence. To summarize, in the meta-regression model, the growth in deaths during the summer 2021 resurgence is expressed with initial values in terms of baseline epidemic magnitude and initial vaccination rates, which in turn characterise state-level exponential growth rates of this highly transmissible virus over the resurgence period.
We found that model~\eqref{eq:model_vaccination} without state-specific random effects on the growth rate already fitted the data very well (Supplementary Figures~\ref{fig:PPC_preprop_vac1}-\ref{fig:PPC_preprop_vac2}), suggesting a clear negative association of growth rates in deaths and vaccination coverage (after accounting for baseline difference at the start of resurgence, and after accounting for long-term differences in epidemic magnitude across states through $r_{m,c,w}$).

\begin{figure}[t]
    \centering
    \includegraphics[width = 0.95\textwidth]{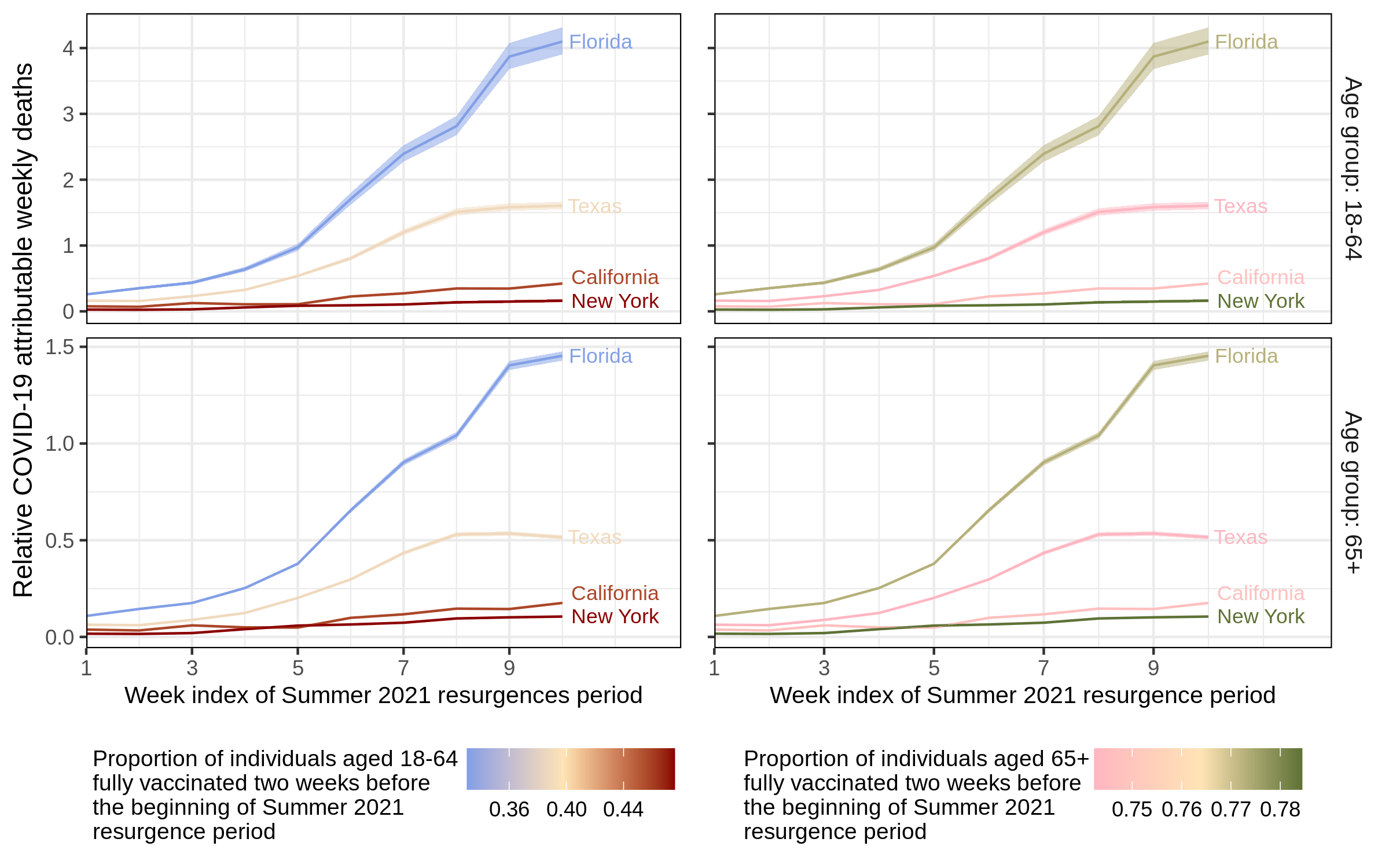}
    \caption{\textbf{Resurgence in COVID-19 deaths and pre-resurgence vaccination coverage during the summer 2021 in four states.}  Posterior median estimates (line) along with 95\% credible intervals (ribbon) of the relative COVID-19 deaths~\eqref{eq:relative_deaths} over the summer 2021 resurgence period are shown on the y-axis for four states. The relative death estimates for each age group were aggregated to the reporting strata of the vaccination data. Week indices of the summer 2021 resurgence are shown on the x-axis, and correspond to different calendar weeks for each state (see text). The pre-resurgence vaccination rate are shown in color. Row facets show resurgences in relative COVID-19 deaths in individuals aged 18-64 and 65+. Column facets show vaccine coverage in individuals aged 18-64 and 65+.}
    \label{fig:relative_deaths_data}
\end{figure}

Figure~\ref{fig:relative_deaths_data} shows on the x-axis the week indices of the summer 2021 resurgence period and in color the state-specific pre-resurgence vaccination rate among individuals aged $18$-$64$ (left) and $65+$ (right). The estimated relative COVID-19 deaths~\eqref{eq:relative_deaths} are shown on the y-axis, now aggregated by the reporting strata of the vaccination data. 
We show results for four states for visual clarity; 
please see Supplementary Figures~\ref{fig:weeklydeathsraw_all}-\ref{fig:relative_deaths_data_all2} for results across the 10 most populated US states California, Florida, Georgia, Illinois, Michigan, New York, North Carolina, Ohio, Pennsylvania and Texas. 
Across US states, we observe that the resurgences in the weekly, relative COVID-19 attributable deaths were stronger in individuals aged $18$-$64$ compared to  individuals aged $65+$, in line with higher vaccination rates in the $65+$ age group compared to the $18$-$64$ age group (note the scales on the y-axes). 

We evaluated model~\eqref{eq:model_vaccination} on the relative COVID-19 deaths of the ten most populated US states, California, Florida, Georgia, Illinois, Michigan, New York, North Carolina, Ohio, Pennsylvania and Texas.
We find that the age-specific vaccination rates in individuals aged $18$-$64$ and $65+$ were statistically significantly associated with a protective effect on averting a strong summer 2021 resurgence in COVID-19 deaths.
These relationships are visualized in Figure~\ref{fig:relative_deaths_data} and further in Supplementary Figure~\ref{fig:params_preprop_vac}, which shows effect sizes as a forest plot, indicating that higher proportions of fully vaccinated individuals prior to resurgence were statistically significantly associated with smaller relative COVID-19 deaths among the same age group. 
Specifically, the baseline effect of vaccination in the same age group ($\chi^{\text{vacc}}$) was strongly negatively associated with resurgent deaths, with a Bayesian p-value of being non-negative of 0.025\%.
Similarly, the effect of vaccination in the same age group on the rate of exponential increase in resurgent deaths ($\psi^{\text{vacc}}$) was strongly negatively associated with resurgent deaths, with a Bayesian p-value of being non-negative of 0\%.

With regard to indirect effects, all the terms $\chi^{\text{vacc-cross}}_{18-64}$, $\chi^{\text{vacc-cross}}_{65+}$, $\psi^{\text{vacc-cross}}_{18-64}$, and $\psi^{\text{vacc-cross}}_{65+}$ were negative but not statistically significantly so in terms of the 95\% posterior credible intervals. Yet, such negative associations are consistent with previous observations that identified young adults as being the main spreaders of COVID-19 across all age groups~\citep{Monod2021,Wikle2020}. In this context, infectious disease theory predicts that reducing infections in the younger adults, has substantial indirect benefits in averting COVID-19 deaths both in younger adults and the elderly. 
This rationale has been widely demonstrated for other highly transmissible pathogens such as seasonal influenza~\citep{Baguelin2013}, which is primarily transmitted through children and their parents, and so achieving high vaccine coverage in these populations is optimal to minimize deaths and hospitalisations, and maximize other cost-benefit metrics. The effect sizes associated with higher vaccine coverage are thus consistent with infectious disease theory, and underscore the tremendous importance that younger adults get vaccinated against COVID-19 if they are medically fit to halt future COVID-19 resurgences among their age group.


\begin{figure}[p]
    \centering
    \includegraphics[width = 0.9\textwidth]{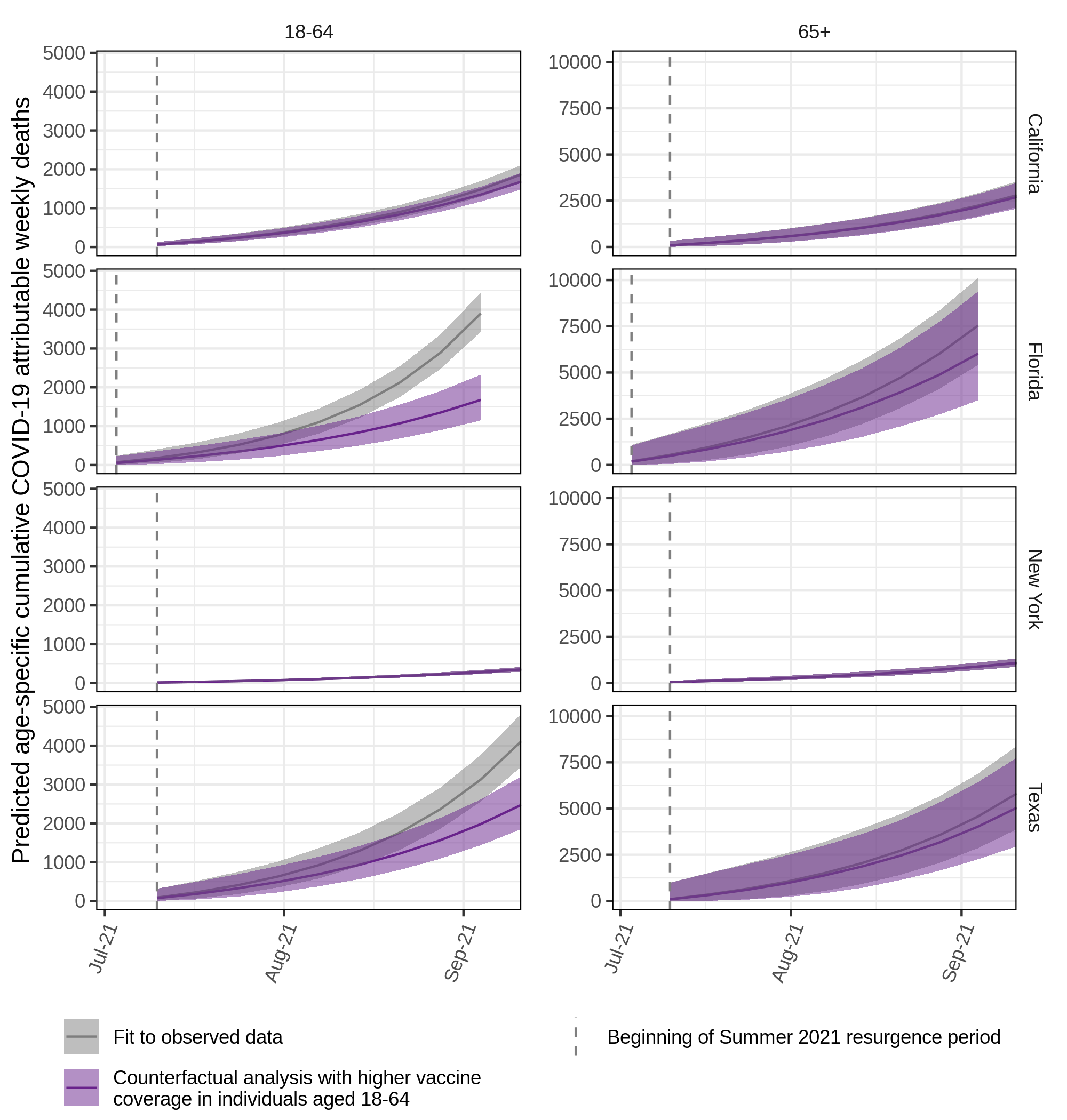}
    \caption{\textbf{Projected avoidable deaths under higher vaccine coverage in individuals aged 18-64 during the summer 2021 resurgence period in four US states.}
    In counterfactual scenarios, we predicted the number of COVID-19 deaths assuming that the vaccine coverage in individuals aged $18$-$64$ in all states had been the same as the maximum vaccine coverage in individuals aged $18$-$64$ across states ($46$\%). 
    Posterior median estimates of the predicted weekly deaths (line) and 95\% credible intervals (ribbon) are shown for the observed COVID-19 deaths (black) and the counterfactual (purple). The start of the 2021 summer resurgence is shown as dashed vertical line. In Florida, an estimated 57.15\% [40.96, 70.47] of deaths in the $18$-$64$ age group, and 20.09\% [-29.54, 54.04] of deaths in the $65+$ age group could have been avoided. In Texas, an estimated 39.80\% [20.81, 54.85] of deaths in the $18$-$64$ age group, and 13.41\% [-41.91, 49.61] of deaths in the $65+$ age group could have been avoided.}
    \label{fig:weeklydeaths_counterfactual}
\end{figure}

\subsection{Projected impact of higher vaccination rates in younger adults}

To illustrate the potential impact of the inferred associations between vaccine coverage in younger adults and resurgent COVID-19 deaths, we considered counterfactual analyses. Specifically, we project the weekly COVID-19 deaths in all age groups assuming that the pre-resurgence vaccine coverage among $18$-$64$ year olds in all states would have been the same as the maximum pre-resurgence vaccine coverage among $18$-$64$ year olds across the ten most populated states, and assuming that the estimated associations are causal. The maximum pre-resurgence vaccine coverage among $18$-$64$ year olds was reported in New York, and was equal to $46$\%. We consider this counterfactual scenario realistic, as it is a retrospective evaluation rather than a forecast, and within the observed range of vaccination rates, i.~e. not an extrapolation. In the counterfactual, we did not consider higher vaccination rates in individuals aged 65+ because the differences in vaccination rates across states were much smaller in this age group, and we here aim to highlight the potential consequences of the more substantial variation in vaccination rates in younger individuals.

Figure~\ref{fig:weeklydeaths_counterfactual} summarises the counterfactual projections for the four most populated states and Supplementary Figures~\ref{fig:weeklydeaths_counterfactual_remaining} for the remaining six of the ten most populated US states. 
We estimate that significantly fewer deaths would have occurred if vaccine coverage in $18$-$64$ year old in all states would have been the same as the maximum vaccine coverage in $18$-$64$ year old across states. 
Overall, from July 03, 2021 to September 25, 2021 and across the ten most populous US states, we project that 41.96\% [31.57, 50.89] of all deaths in individuals aged 1 could have been avoided if pre-resurgence vaccine coverage rates in $18$-$64$ year olds had been $46$\%, and 13.41\% [-41.91, 49.61] of deaths in individuals aged $65+$. Stratified results by US state are shown in Table~\ref{tab:weeklydeaths_counterfactual}.



\section{Discussion} \label{sec:discussion}

We present a Bayesian non-parametric modelling approach to estimate and report longitudinal trends in COVID-19 attributable deaths by 1-year age bands across the United States. The model is informed by COVID-19 attributable weekly deaths reported by the CDC, the official agency providing such data at the finer spatial resolution in the US. In comparison to the crude CDC data, the predictions are for all weeks, and account for missing and censoring present in the crude data. The predictions of the model also adjust for reporting delays as e.~g. seen in November to December 2020 in Florida and for under-reporting as e.~g. seen in December 2020 to March 2021 in New York (Figure~\ref{fig:predicteddeath}), by allowing the use of an external dataset for calibration, for which we used the curated JHU all-age COVID-19 attributable deaths. Importantly, the Bayesian model predicts COVID-19 attributable deaths by $1$-year age band, which for example allows reporting of cumulative mortality rates in any age category of interest. Finally, the model implementation is portable, freely available and computationally efficient through its implementation in Stan and the use of B-splines projected Gaussian process priors. For these reasons, we find that the model is a useful tool to improve the reporting of COVID-19 deaths and facilitate demographic and epidemiologic analyses at finer granularity. 

Our analyses of the US summer 2021 resurgence provide further support for the hypothesis that lower vaccine coverages in several states, most notably Florida and Texas, are associated with larger COVID-19 death counts~\citep{Sah2021,Moghadas2021}. However, there are a large number of behavioural, governmental and state specific differences that modify this effect size. We have here attempted to agnostically account for these via random effect terms, but it should be noted that we do not explicitly model these factors.

We developed B-splines projected Gaussian process priors to describe the time trends in the age profile of COVID-19 deaths through a non-parametric 2D random surface. We placed GP priors on the B-splines coefficients, showed that the model is again a GP, and derived the B-splines projected kernel function of the GP. We also showed that the model can be interpreted as regularised B-splines, and find that the negative determinant of the covariance matrix in the log-likelihood, often referred to as Occam's factor (\cite{MacKay:2003}, chap. 28), plays the same role as the penalty of smoothing splines and P-splines~\citep{Sullivan:1986,Eilers1996}. Compared to a standard 2D GP, improvements in runtimes lie both in the fact that fewer parameters on the 2D surface are estimated (from $n\times m$ to $I \times J$), and that the computational complexity in calculating covariance matrices is reduced (from $\mathcal{O}(2(n \times m)^{3/2})$ to $\mathcal{O}(2(I \times J)^{3/2})$), where $I$ and $J$ are the number of B-splines on the $x$ and $y$ axes, and $n$ and $m$ are the number of points on the $x$ and $y$ axes and with $I < n$ and $J < m$.
On simulations and on real data, we find the model induces better smoothness properties than standard B-splines~\citep{deboor} and Bayesian P-splines~\citep{Lang2004} where these are warranted, and so we find the B-splines projected Gaussian process priors are likely an appealing addition to the arsenal of Bayesian regularising priors.

\FloatBarrier
\newpage
\begin{supplement}
\stitle{\textbf{Supplementary Figures and Tables}}
\sdescription{}
\end{supplement}

\begin{supplement}
\stitle{\textbf{B-splines and properties of the regularised B-splines projected Gaussian Process prior}}
\sdescription{2 parts: (1) Construction of a B-spline curve and a B-spline surface and, (2) Proof that a baseline kernel function projected with cubic B-splines is a $\mathcal{C}^2$ function.}
\end{supplement}

\begin{supplement}
\stitle{\textbf{Modelling COVID-19 weekly deaths}}
\sdescription{2 parts: (1) Finding weekly death count from censored and missing cumulative death count from the CDC data and, (2) Specifying the likelihood form.}
\end{supplement}

\begin{supplement}
\stitle{\textbf{Additional analyses}}
\sdescription{(1) Comparison of random surface priors for one US state and (2) Differences in the pre-vaccination age profile of COVID-19 attributable deaths across states.}
\end{supplement}

\begin{supplement}
\stitle{\textbf{Templated Stan model files}}
\sdescription{Stan model to fit regularised B-splines projected Gaussian Process on the 2D mean surface of count or continuous data.}
\end{supplement}

\begin{supplement}
\stitle{\textbf{State level summary}}
\sdescription{Detailed fit summary for every US state}
\end{supplement}

\bibliographystyle{ba}
\bibliography{ref}

\begin{acks}[Acknowledgments]
We thank Sam Abbott and two anonymous reviewers for their helpful comments.
This work was supported by the the Imperial College Research Computing Service, DOI: 10.14469/hpc/2232, the Imperial College COVID-19 Response Fund and the EPSRC through the EPSRC Centre for Doctoral Training in Modern Statistics and Statistical Machine Learning. S. Bhatt acknowledges The UK Research and Innovation (MR/V038109/1), the Academy of Medical Sciences Springboard Award (SBF004/1080), The MRC (MR/R015600/1), The BMGF (OPP1197730), Imperial College Healthcare NHS Trust-BRC Funding (RDA02), The Novo Nordisk Young Investigator Award (NNF20OC0059309) and The NIHR Health Protection Research Unit in Modelling Methodology. O. Ratmann acknowledges the Bill \& Melinda Gates Foundation (OPP1175094).
\end{acks}

\clearpage
\newpage
\FloatBarrier

\resumetocwriting

\renewcommand{\thefigure}{S\arabic{figure}}
\renewcommand{\thetable}{S\arabic{table}}
\renewcommand{\thesection}{S\arabic{section}}

\begin{frontmatter}
\title{Supplementary Materials}

\end{frontmatter}
Supplementary Materials to Regularised B-splines projected Gaussian Process priors to estimate time-trends of age-specific COVID-19 deaths related to vaccine roll-out by Monod et al.

\tableofcontents

\FloatBarrier
\clearpage
\section{Supplementary Figures and Tables}

\begin{figure*}[h!]
     \centering
     \begin{subfigure}[b]{0.32\textwidth}
         \centering
         \includegraphics[width=\textwidth]{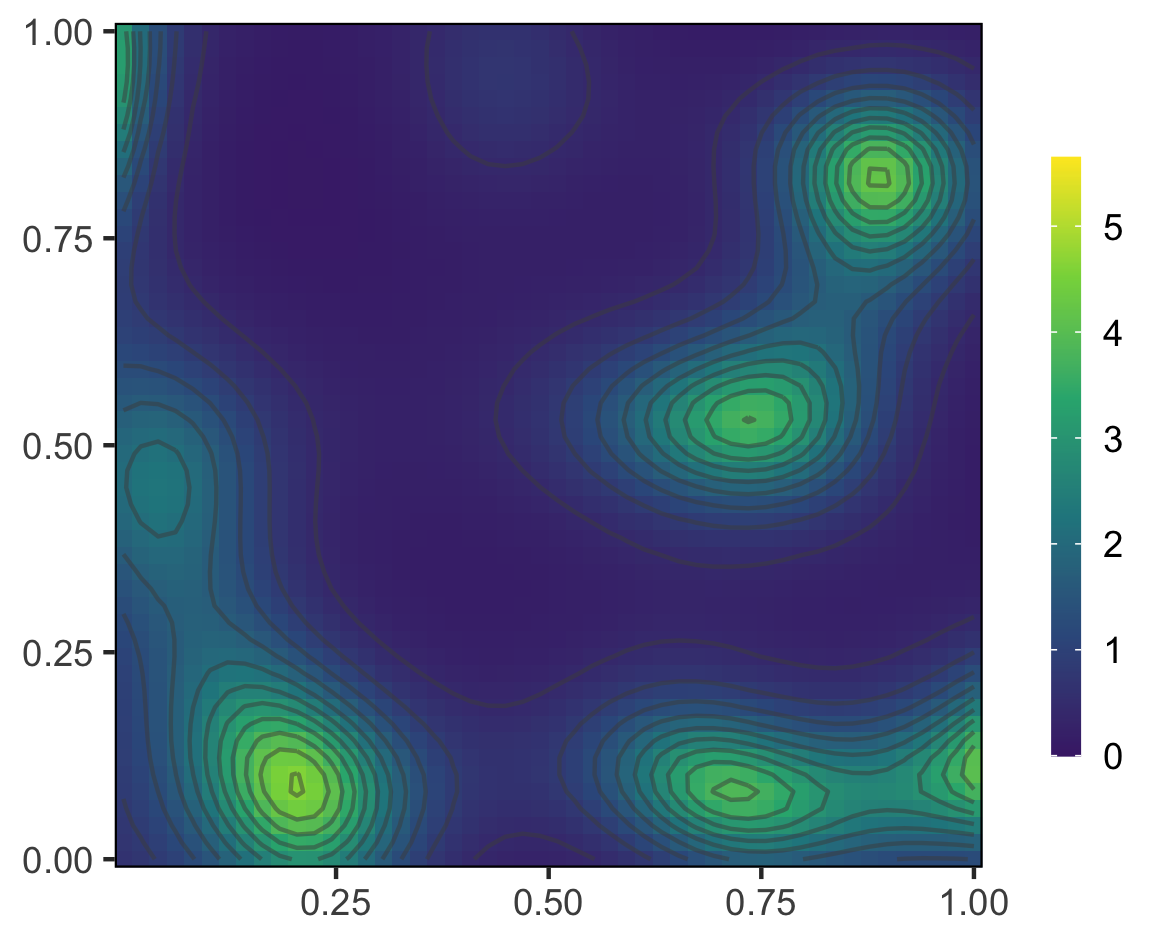}
         \caption{$\gamma = 0.05$}
         \label{fig:y equals x}
     \end{subfigure}
     \hfill
     \begin{subfigure}[b]{0.32\textwidth}
         \centering
         \includegraphics[width=\textwidth]{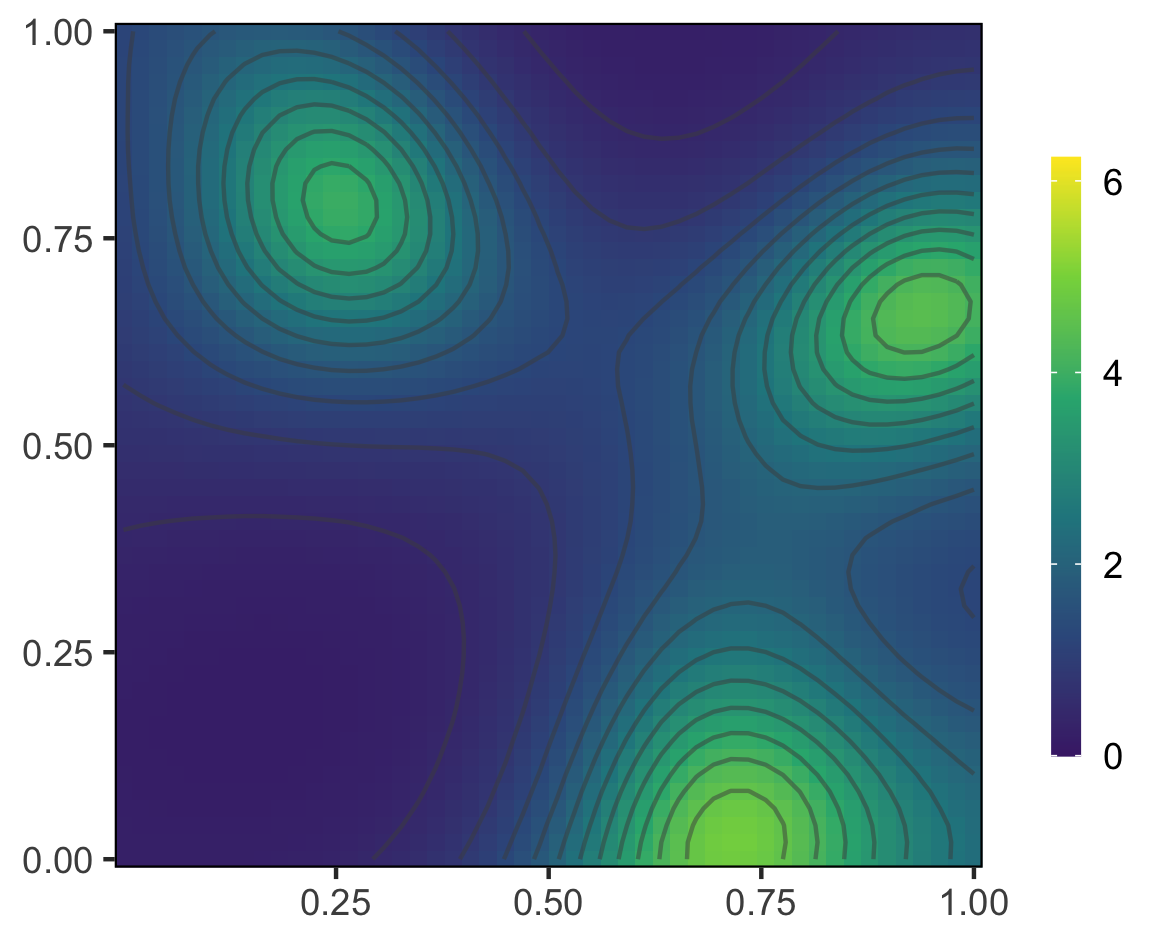}
         \caption{$\gamma= 0.25$}
         \label{fig:five over y}
     \end{subfigure}
     \hfill
     \begin{subfigure}[b]{0.32\textwidth}
         \centering
         \includegraphics[width=\textwidth]{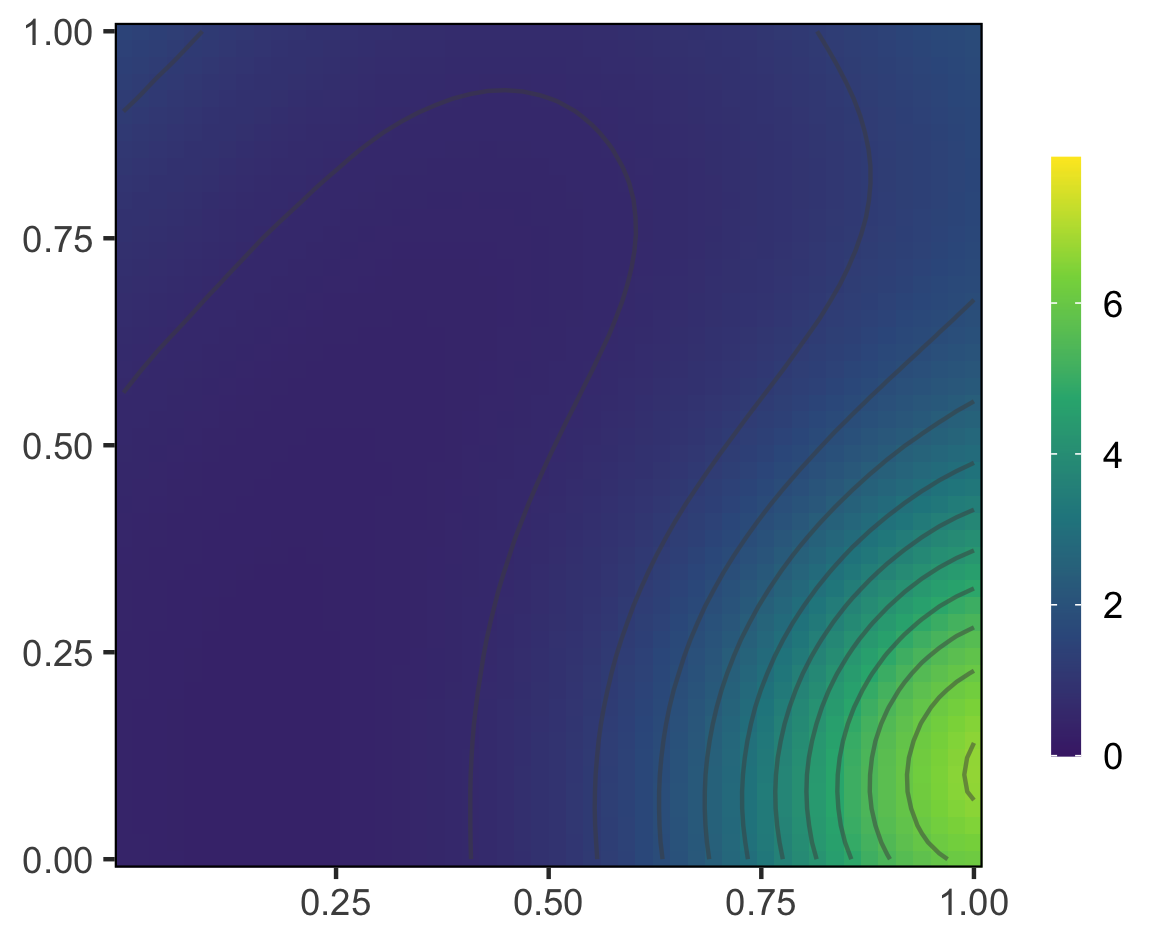}
         \caption{$\gamma= 1$}
         \label{fig:five over x}
     \end{subfigure}
        \caption{\textbf{Estimated mean surfaces using a 2D GP model.} Posterior median of the mean surface estimated with a 2D GP model given weakly correlated ($\gamma = 0.05$), mildly correlated ($\gamma= 0.25$) and strongly correlated ($\gamma= 1$) simulated data.}
        \label{fig:simu_gp}
\end{figure*}

\FloatBarrier
\clearpage
\begin{figure*}[h!]
\centering
    \centering
       \includegraphics[width=0.95\linewidth]{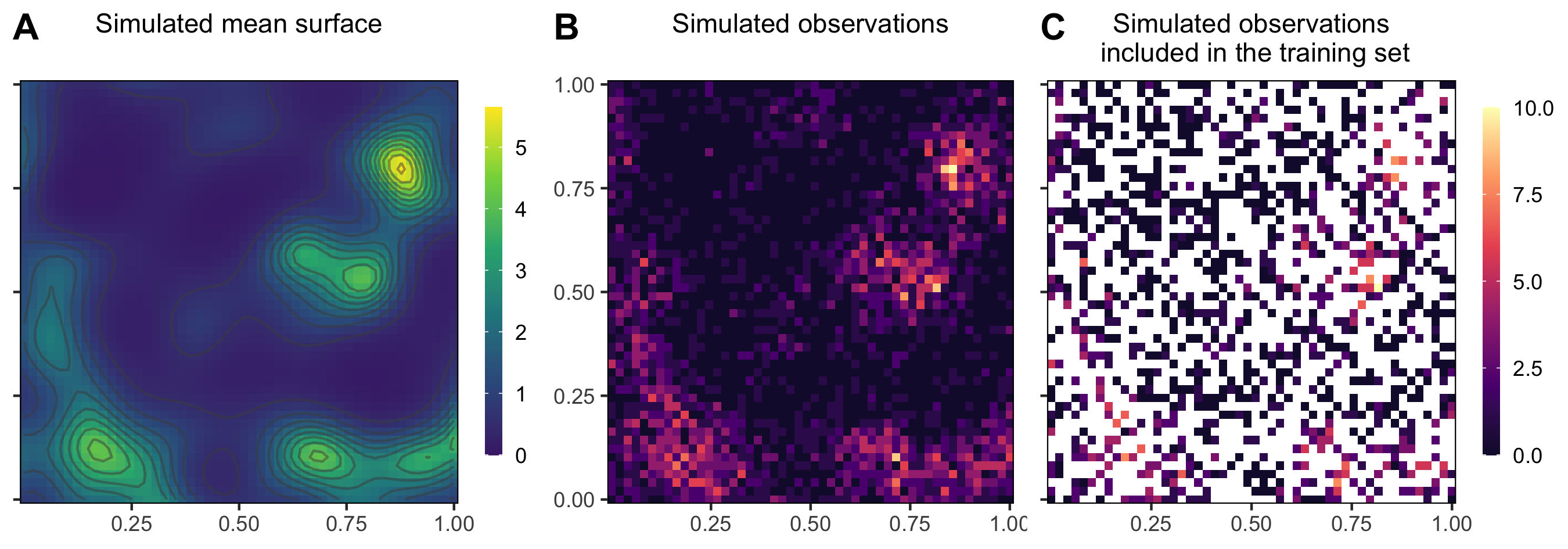}
      \caption{\textbf{Weakly correlated simulated 2D observations.} \textbf{(A)} The mean surface is defined by the exponential of a 2D GP with a squared exponential kernel implying weak correlations ($\gamma = 0.05$). \textbf{(B)} Count data were simulated from a negative Binomial model. \textbf{(C)} 40\% of the simulated observations were included in the training set.  }
    \label{fig:simu_data_1}
\end{figure*}

\begin{figure*}[h!]
\centering
    \centering
       \includegraphics[width=0.95\linewidth]{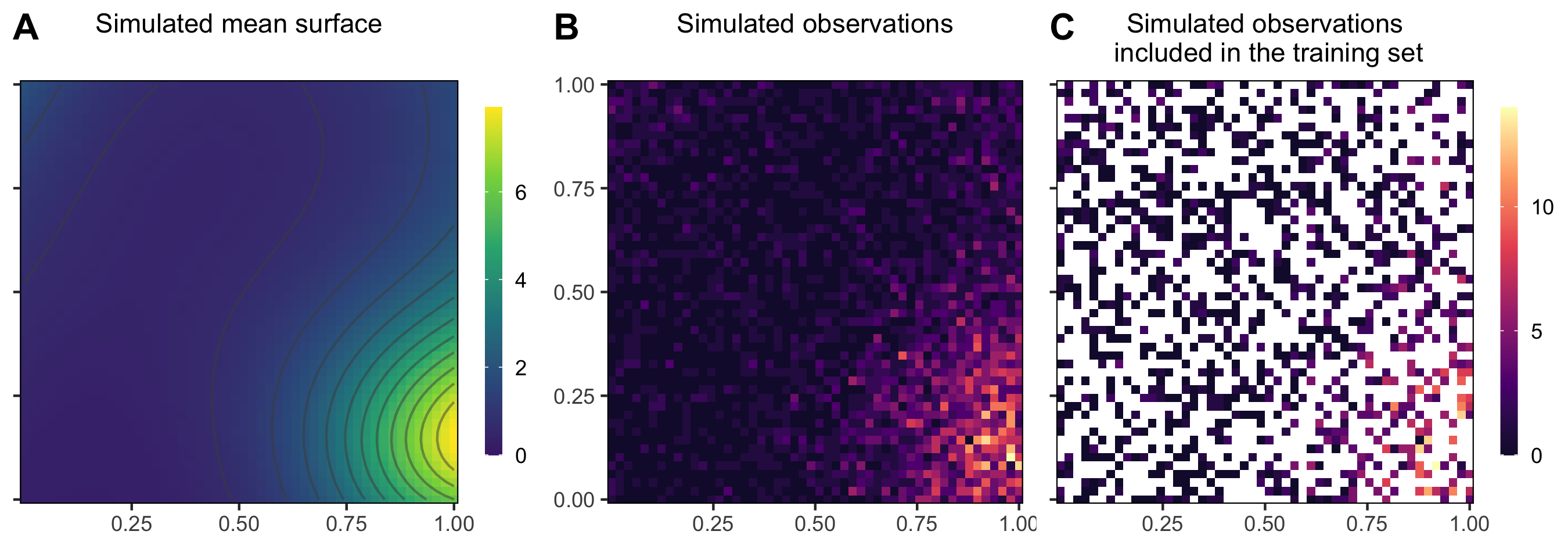}
      \caption{\textbf{Strongly correlated simulated 2D observations.} \textbf{(A)} The mean surface is defined by the exponential of a 2D GP with a squared exponential kernel implying strong correlations ($\gamma = 1$). \textbf{(B)} Count data were simulated from a negative Binomial model. \textbf{(C)} 40\% of the simulated observations were included in the training set. }
    \label{fig:simu_data_3}
\end{figure*}


\begin{figure}[p]
    \centering
    \includegraphics[width=0.95\linewidth]{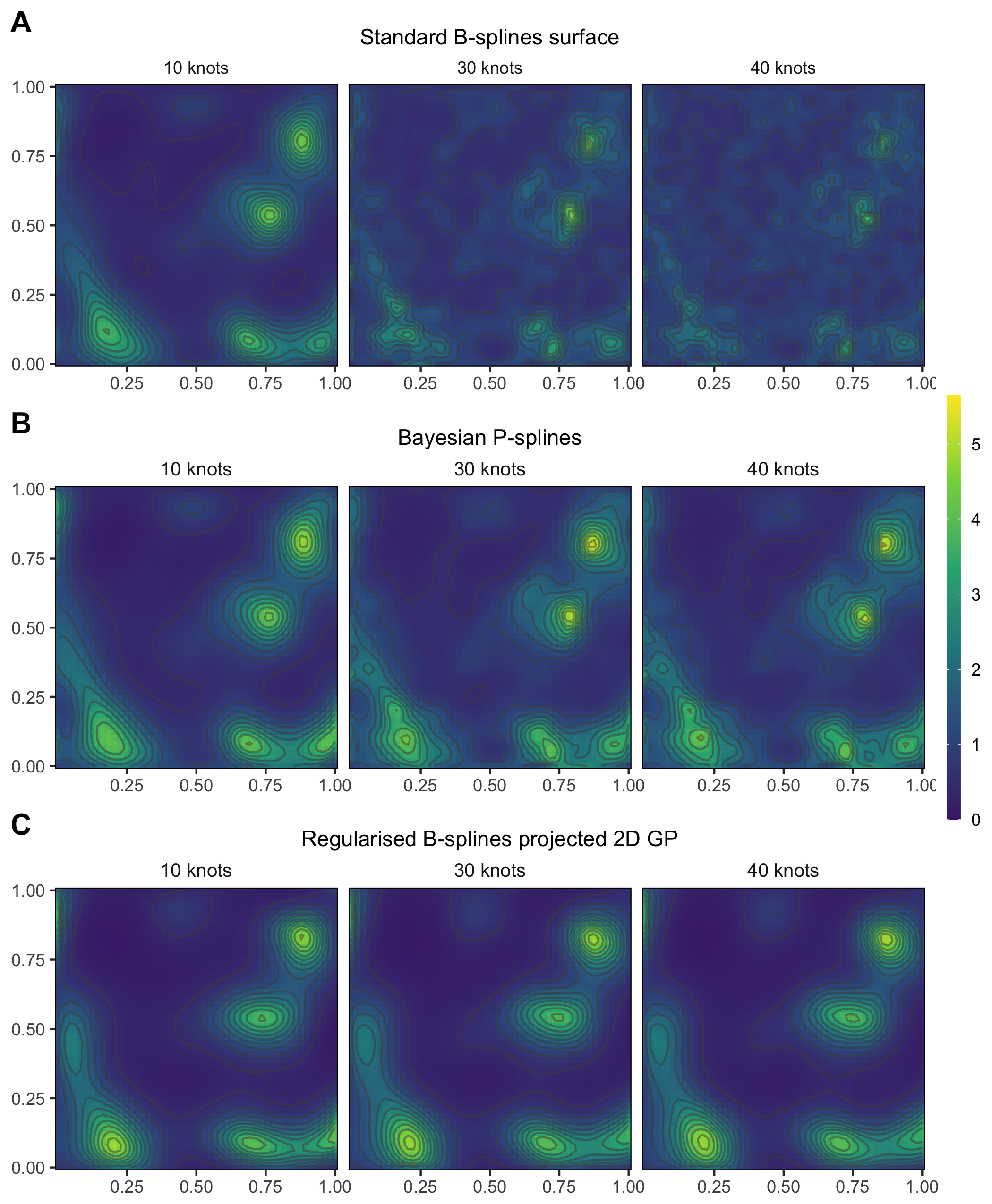}
    \caption{\textbf{Estimated mean surfaces using non-regularized and regularized B-splines spatial models (weak correlation, $\gamma = 0.05$).} Posterior median of the mean surface estimated by \textbf{(A)} standard B-splines, \textbf{(B)} Bayesian P-splines and \textbf{(C)} regularised B-splines projected 2D GP with different number of knots given weakly correlated simulated data ($\gamma = 0.05$). The simulated data are illustrated in Figure~\ref{fig:simu_data_1}.}
    \label{fig:simu_bsplines_1}
\end{figure}

\begin{figure}[p]
    \centering
    \includegraphics[width=0.95\linewidth]{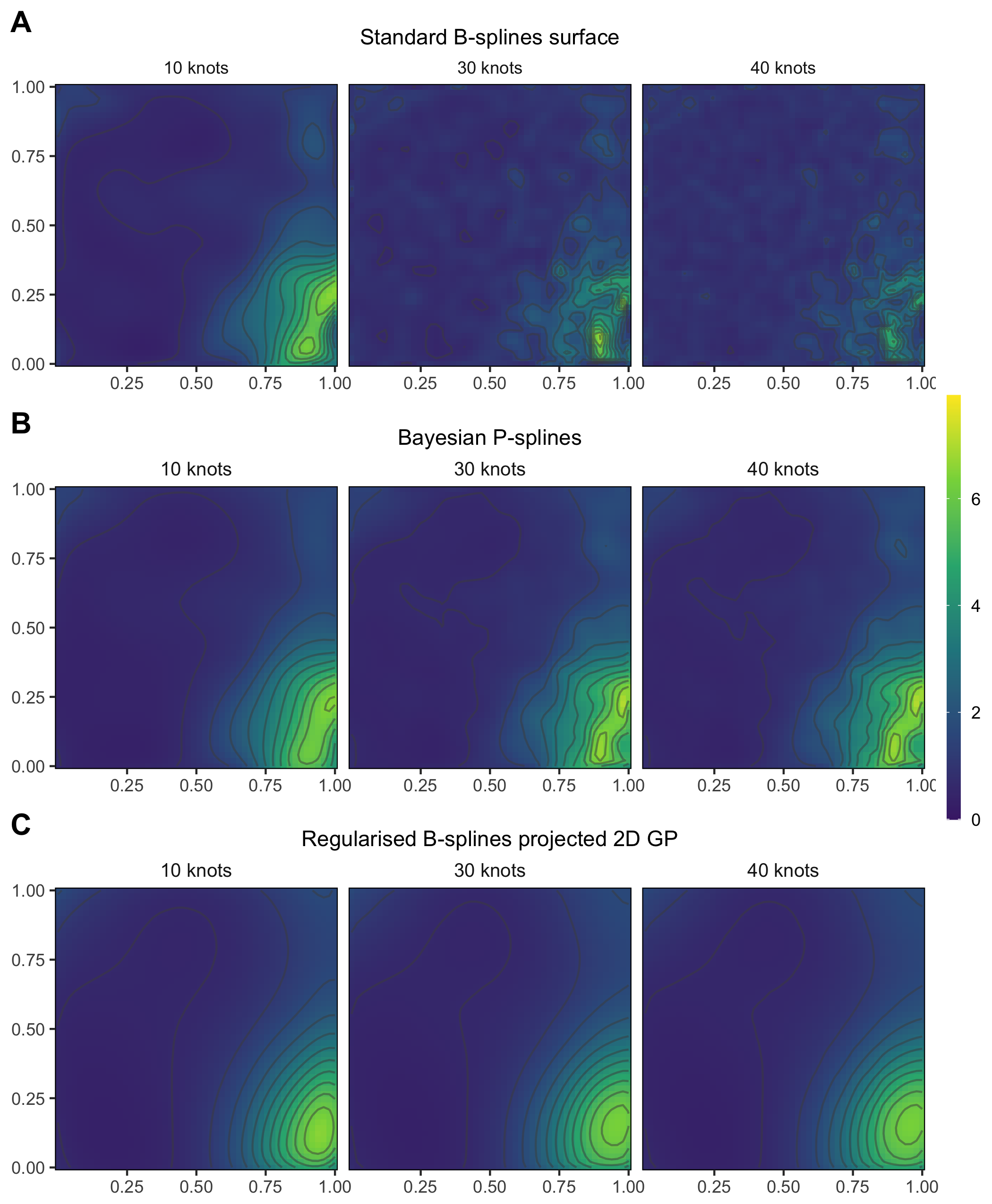}
    \caption{\textbf{Estimated mean surfaces using non-regularized and regularized B-splines spatial models (strong correlation, $\gamma = 1$).} Posterior median of the mean surface estimated by \textbf{(A)} standard B-splines, \textbf{(B)} Bayesian P-splines and \textbf{(C)} regularised B-splines projected 2D GP with different number of knots given strongly correlated simulated data ($\gamma = 1$). The simulated data are illustrated in Figure~\ref{fig:simu_data_3}.}
    \label{fig:simu_bsplines_3}
\end{figure}


\begin{figure}[!ht]
    \centering
    \includegraphics[width = 1\textwidth]{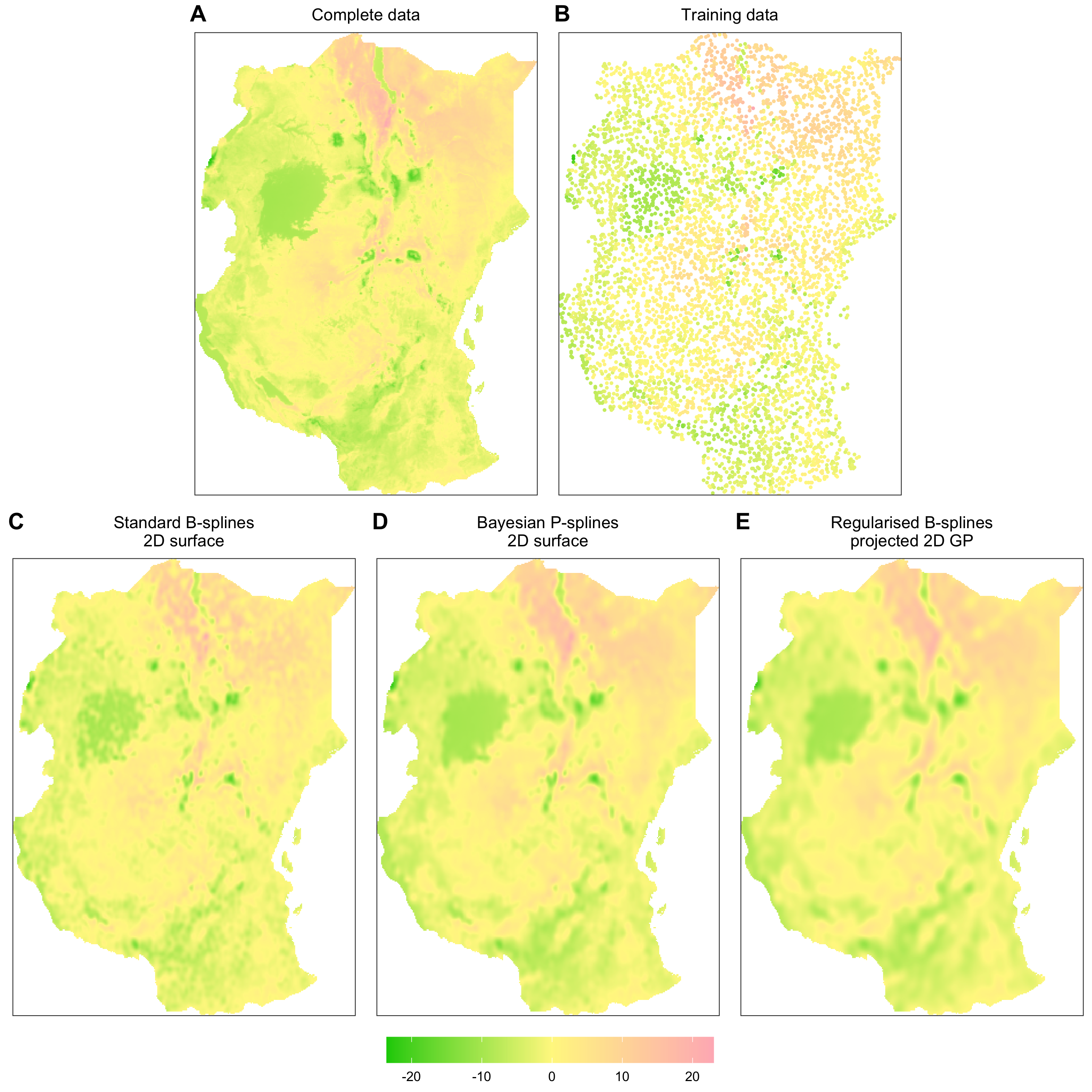}
    \caption{\textbf{Deviation in land surface temperature for East Africa trained on $6000$ random uniformly chosen
locations.} \textbf{(A)} Complete data from~\cite{Ton2018}. \textbf{(B)} Training data, similar as the one used in~\cite{mishra:2020}. The remaining subfigures show the estimated posterior mean surface by \textbf{(C)} a standard B-splines surface, \textbf{(D)} Bayesian P-Splines, and \textbf{(E)} regularised B-splines projected 2D GP.}
    \label{fig:benchmark}
\end{figure}


\clearpage
\newpage

\begin{figure}[!t]
    \centering
    \includegraphics[width = 1\textwidth]{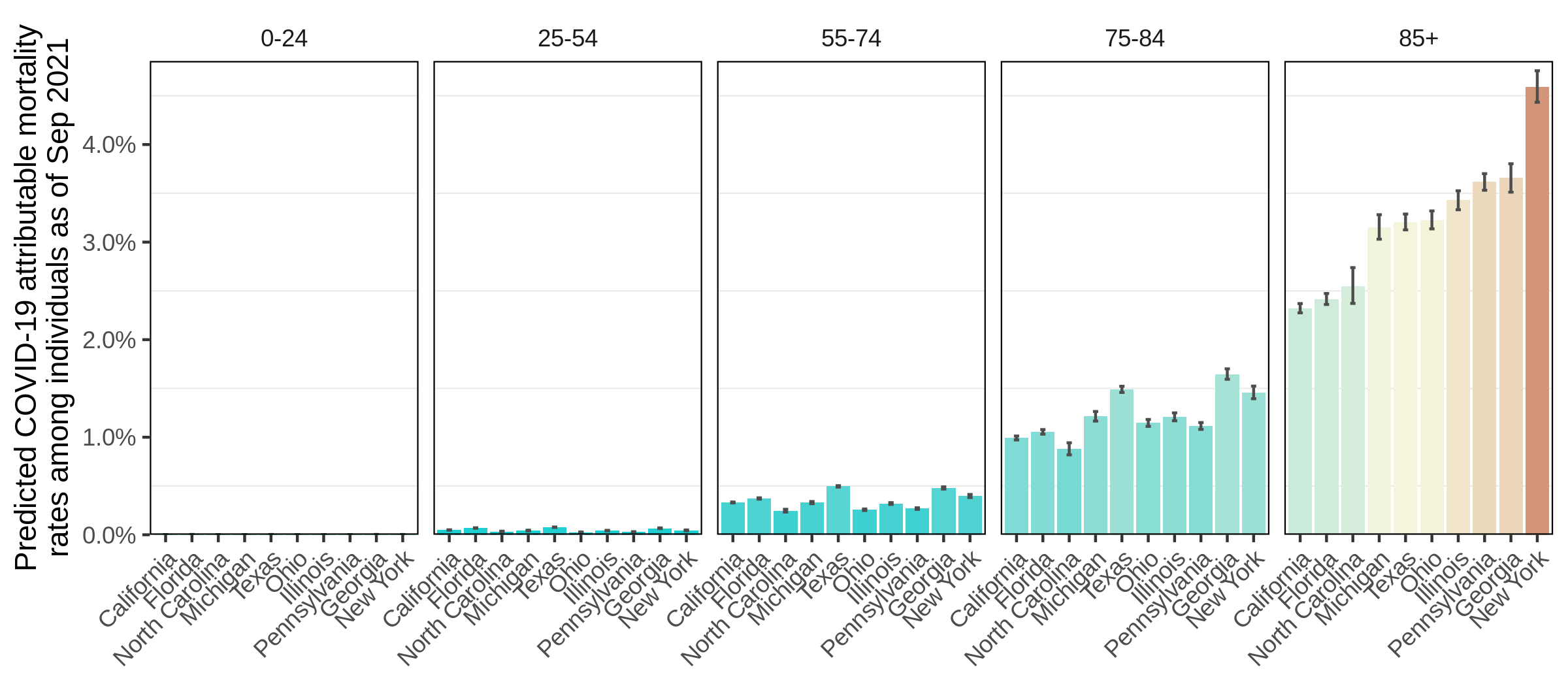}
    \caption{\textbf{Predicted COVID-19 attributable mortality rates.} Posterior median estimates of the mortality rate as of September 25, 2021 (barplot) and 95\% credible intervals (error bars) are presented. Column facets indicate the age group among which the COVID-19 mortality rates are predicted.}
    \label{fig:mortalityrateall}
\end{figure}


\clearpage
\newpage

\begin{figure}[h!]
    \centering
    \includegraphics[width = 0.95\textwidth]{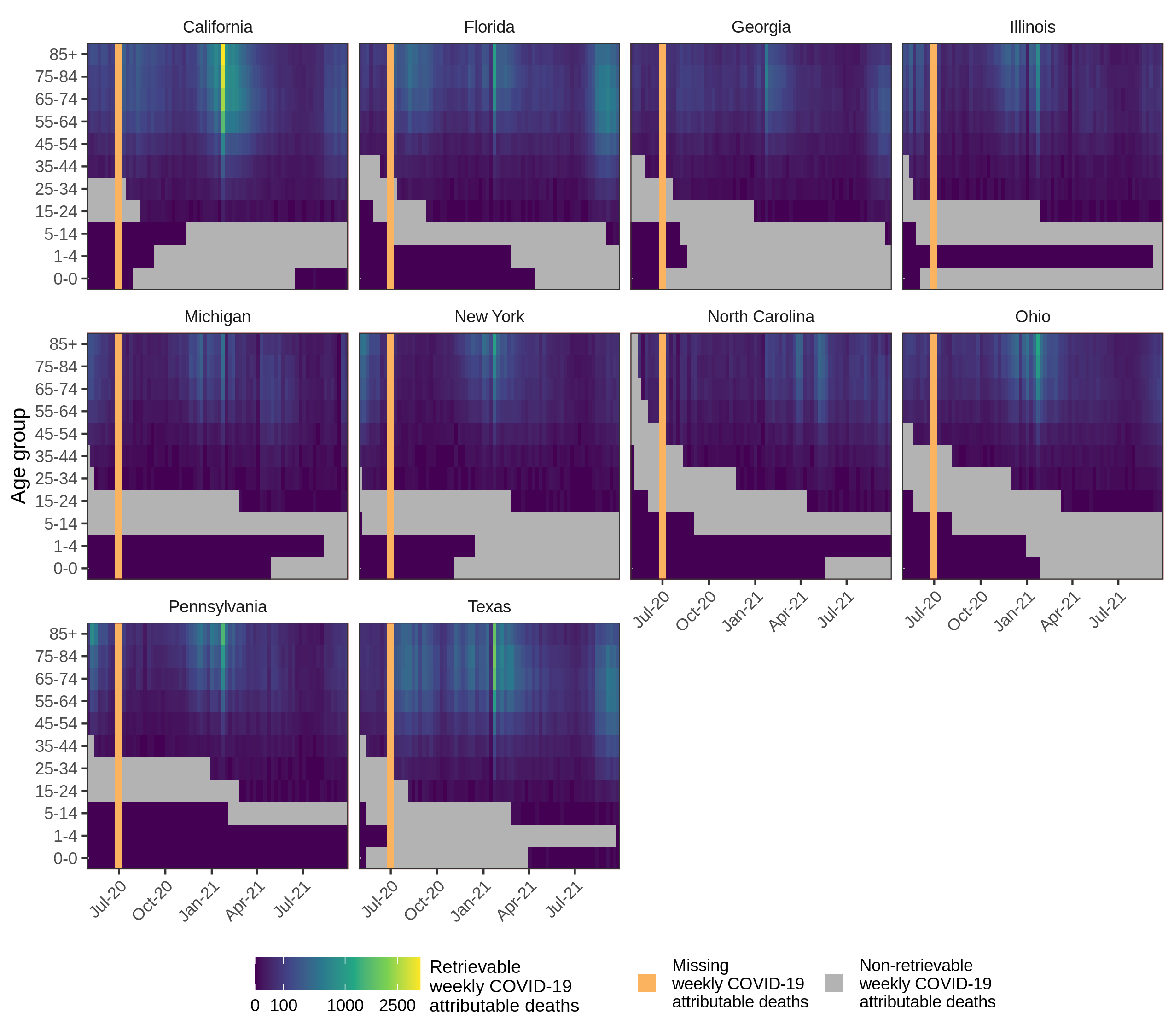}
    \caption{\textbf{Retrievable weekly COVID-19 attributable deaths in ten US states.} The retrievable weekly deaths are computed using the first order difference of reported cumulative deaths by the CDC~\citep{CDC_data_website}. Grey cells indicate non-retrievable weekly death counts due to the censoring of cumulative deaths. Orange bars show the missing weekly deaths due to non-reported cumulative deaths.}
    \label{fig:weeklydeathsraw_all}
\end{figure}

\begin{figure}[t]
    \centering
    \includegraphics[width = 1\textwidth]{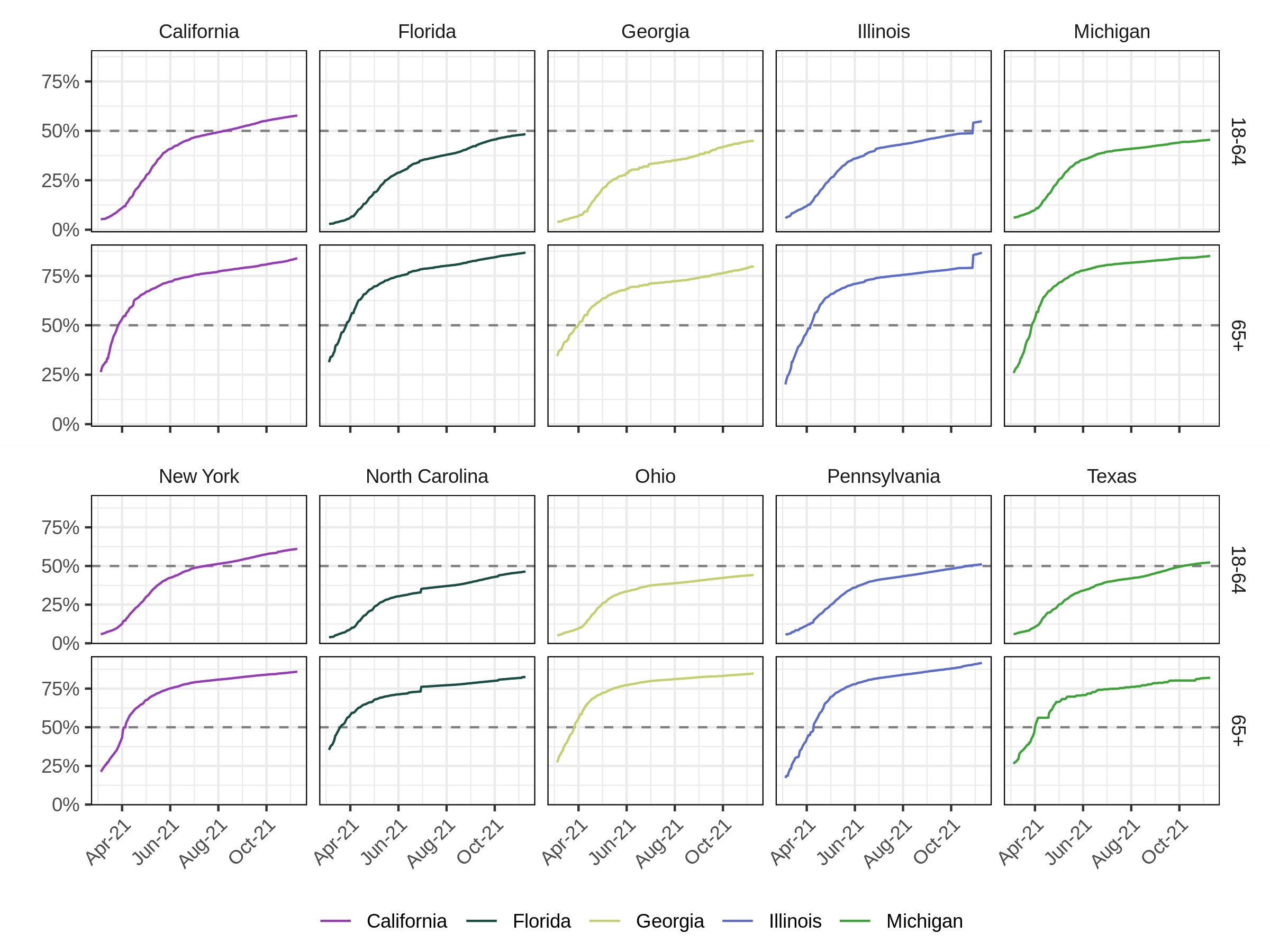}
    \caption{\textbf{Age-specific proportion of fully vaccinated individuals in ten US states.} The proportion of fully vaccinated individuals over time were retrieved from~\cite{CDC_data_vac}. Fully vaccinated individuals are defined as having received the second dose of a two-dose vaccine or one dose of a single-dose vaccine. Row facets show the proportion of fully vaccinated individuals aged $18$-$64$ and $65+$. A vertical line indicates the 50\% threshold.}
    \label{fig:vaccinedata_all}
\end{figure}

\begin{figure}[!p]
    \centering
    \includegraphics[width = 0.95\textwidth]{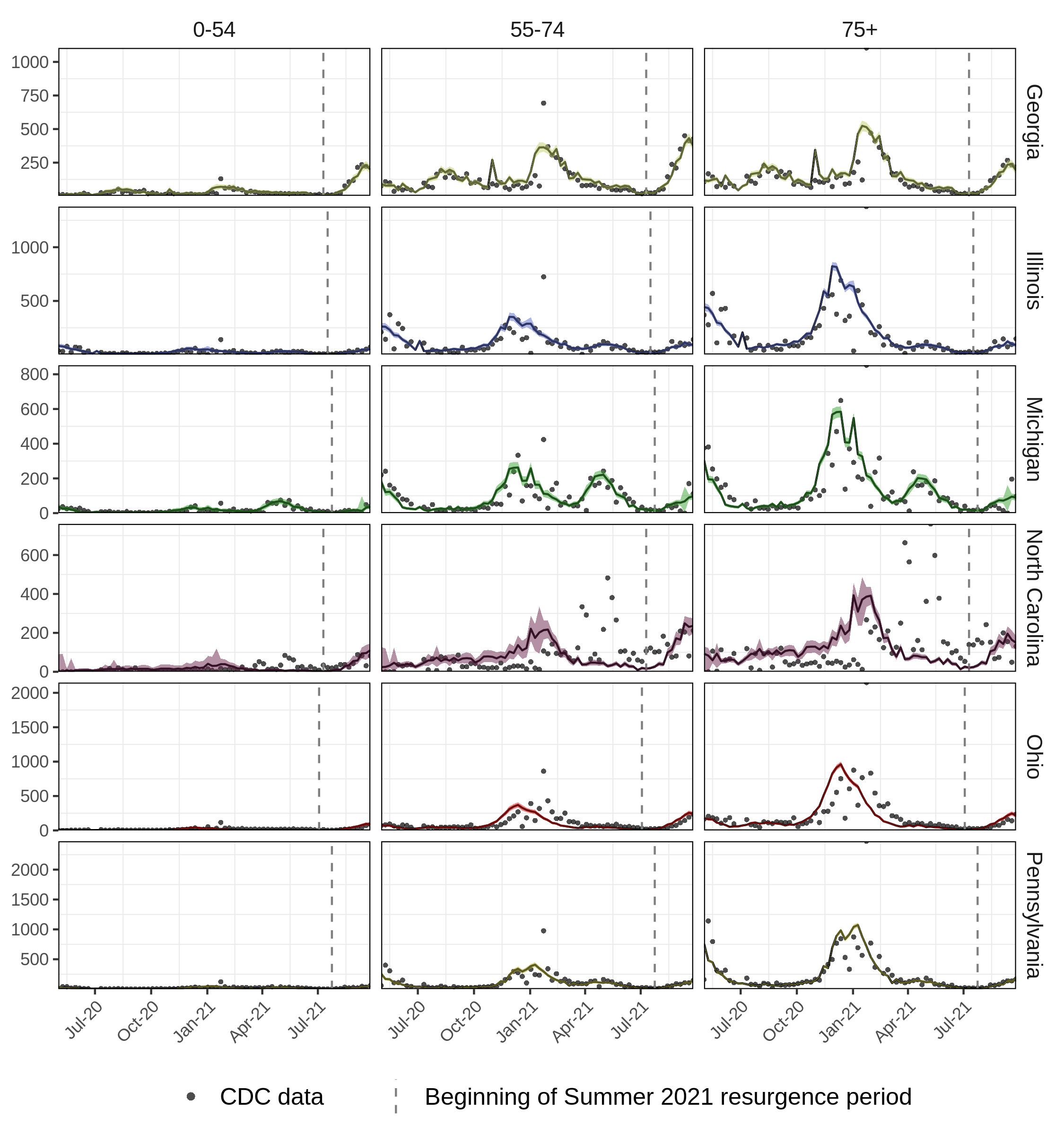}
    \caption{\textbf{Predicted age-specific COVID-19 attributable deaths in the six US states.} Shown are the posterior median (line) and 95\% credible intervals (ribbon) of the predicted weekly COVID-19 attributable deaths obtained with~\eqref{eq:rescaledeaths}. The CDC data are shown with dots. The start of the 2021 summer resurgence period is indicated as dashed vertical line.}
    \label{fig:predicteddeath_remaining}
\end{figure}

\begin{figure}[t]
    \centering
    \includegraphics[width = 0.95\textwidth]{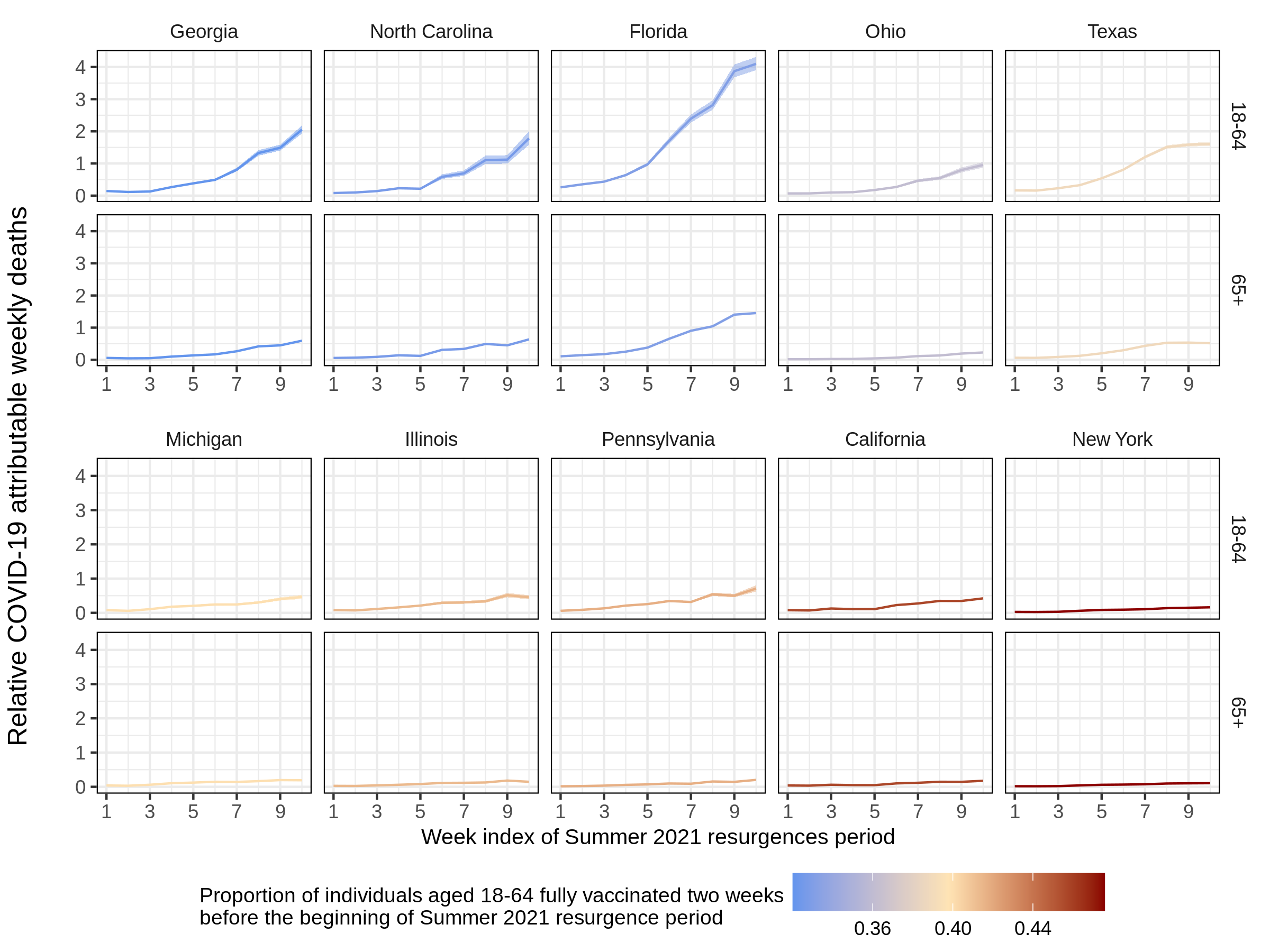}
    \caption{\textbf{Resurgence in COVID-19 deaths and pre-resurgence vaccination coverage during the summer 2021 in the ten most populated states (part 1).}  Posterior median estimates (line) along with 95\% credible intervals (ribbon) of the relative COVID-19 deaths~\eqref{eq:relative_deaths} over the summer 2021 resurgence period are shown on the y-axis for ten states. The relative death estimates for each age group were aggregated to the reporting strata of the vaccination data. Week indices of the summer 2021 resurgence are shown on the x-axis, and correspond to different calendar weeks for each state (see main text). The pre-resurgence vaccination rate among individuals aged 18-64 are shown in color. The states are ordered from left to right by the pre-resurgence vaccination rate among individuals aged 18-64. Row facets show resurgences in relative COVID-19 deaths in individuals aged 18-64 and $65+$.}
    \label{fig:relative_deaths_data_all1}
\end{figure}

\begin{figure}[t]
    \centering
    \includegraphics[width = 0.95\textwidth]{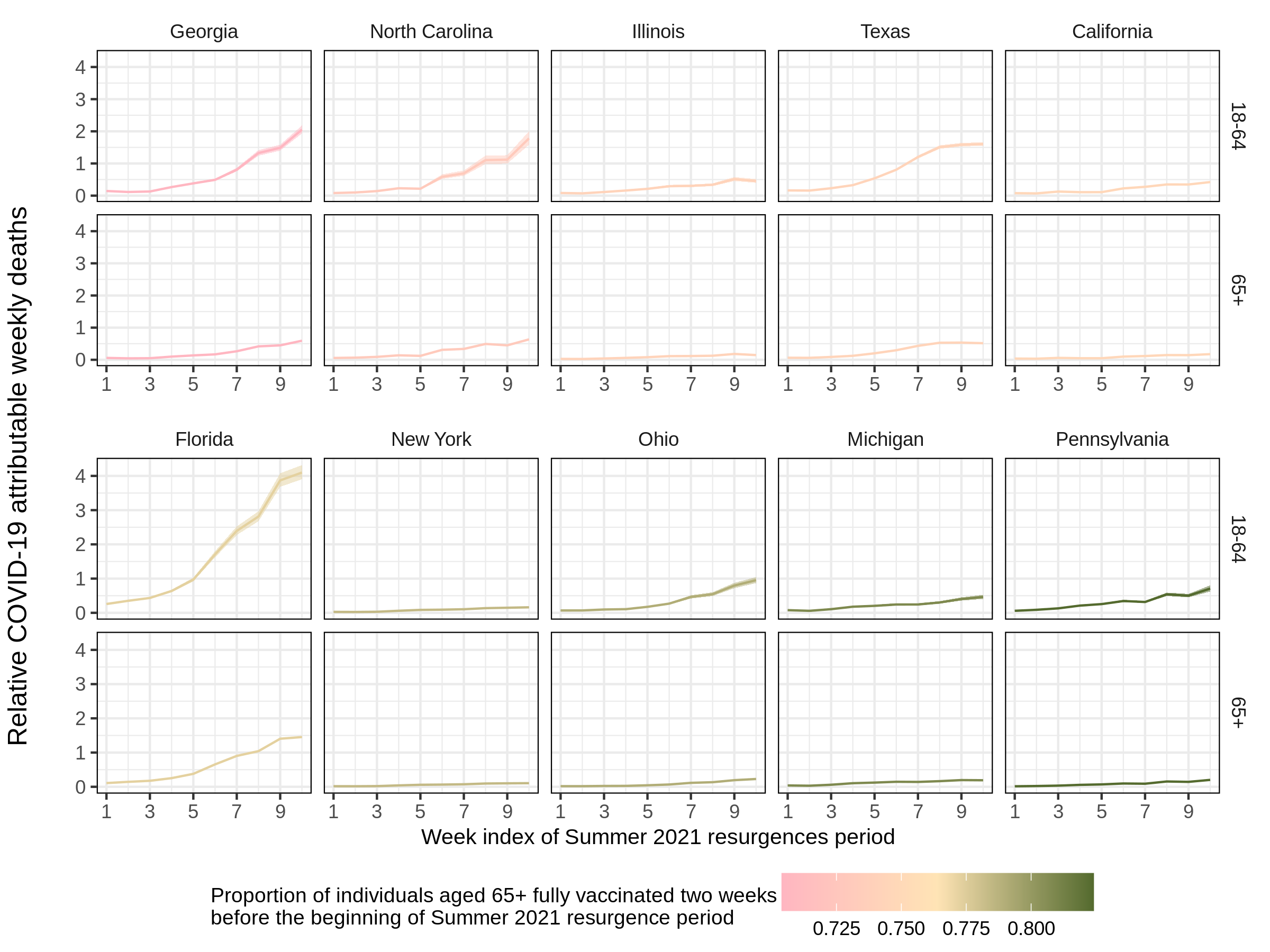}
    \caption{\textbf{Resurgence in COVID-19 deaths and pre-resurgence vaccination coverage during the summer 2021 in the ten most populated states (part 2).}  Posterior median estimates (line) along with 95\% credible intervals (ribbon) of the relative COVID-19 deaths~\eqref{eq:relative_deaths} over the summer 2021 resurgence period are shown on the y-axis for ten states. The relative death estimates for each age group were aggregated to the reporting strata of the vaccination data. Week indices of the summer 2021 resurgence are shown on the x-axis, and correspond to different calendar weeks for each state (see main text). The pre-resurgence vaccination rate among individuals aged $65+$ are shown in color. The states are ordered from left to right by the pre-resurgence vaccination rate among individuals aged $65+$. Row facets show resurgences in relative COVID-19 deaths in individuals aged 18-64 and $65+$. }
    \label{fig:relative_deaths_data_all2}
\end{figure}


%
\begin{figure}[t]
    \centering
    \includegraphics[width = 0.95\textwidth]{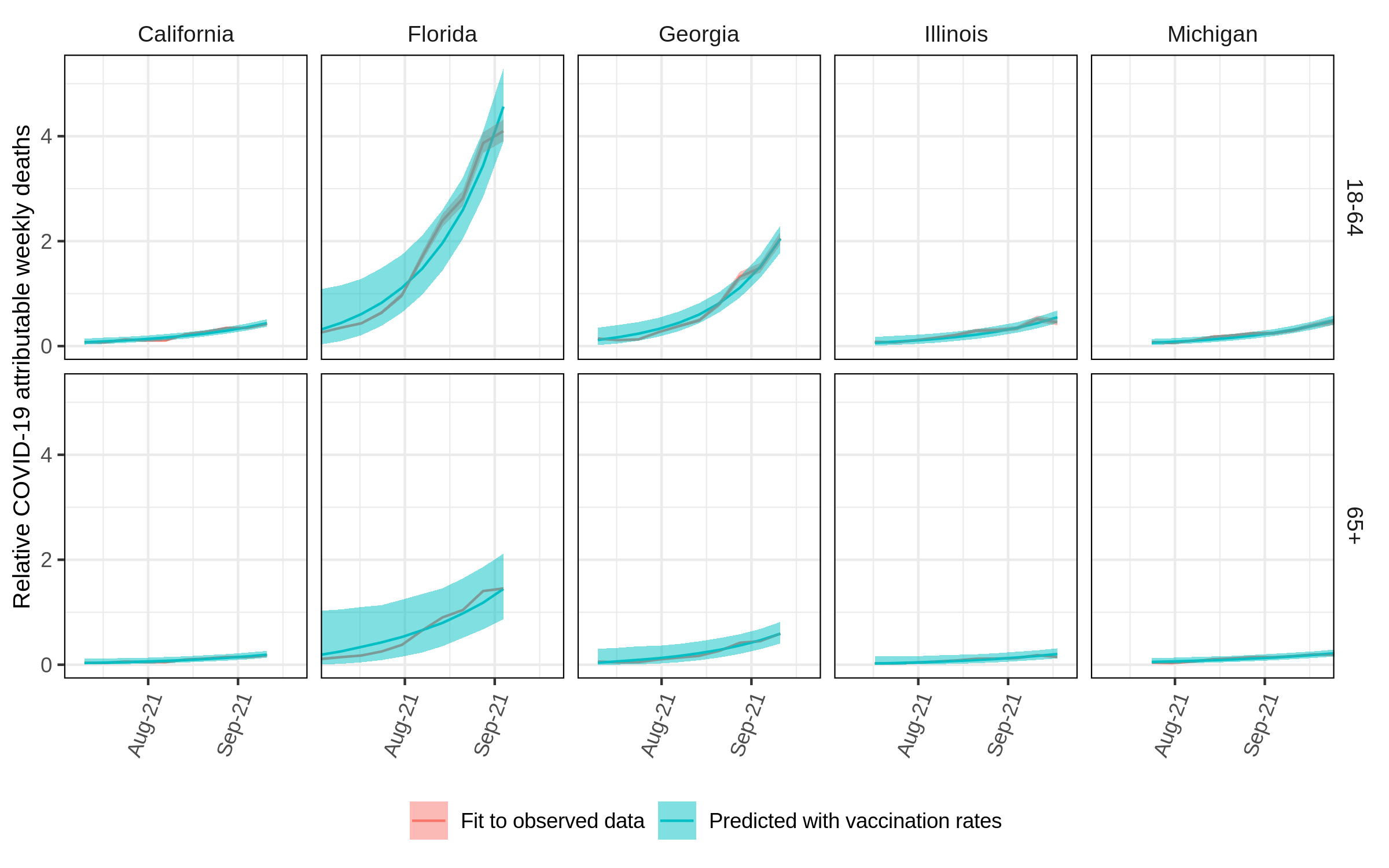}
    \caption{\textbf{Posterior predictive checks of model~\eqref{eq:model_vaccination} (part 1).} Posterior median estimates (line) and 95\% credible intervals (ribbon) of the relative COVID-19 deaths during the summer 2021 resurgence period estimated with pre-resurgence vaccination coverage~\eqref{eq:model_vaccination} are shown in blue. Posterior median estimates (line) and 95\% credible intervals (ribbon) of the relative COVID-19 deaths during the summer 2021 resurgence period estimated with weekly deaths data~\eqref{eq:rescaledeaths}, on which~\eqref{eq:model_vaccination} was fitted, are shown in red. Row facets show resurgences in relative COVID-19 deaths in individuals aged 18-64 and $65+$.}
    \label{fig:PPC_preprop_vac1}
\end{figure}

\begin{figure}[t]
    \centering
    \includegraphics[width = 0.95\textwidth]{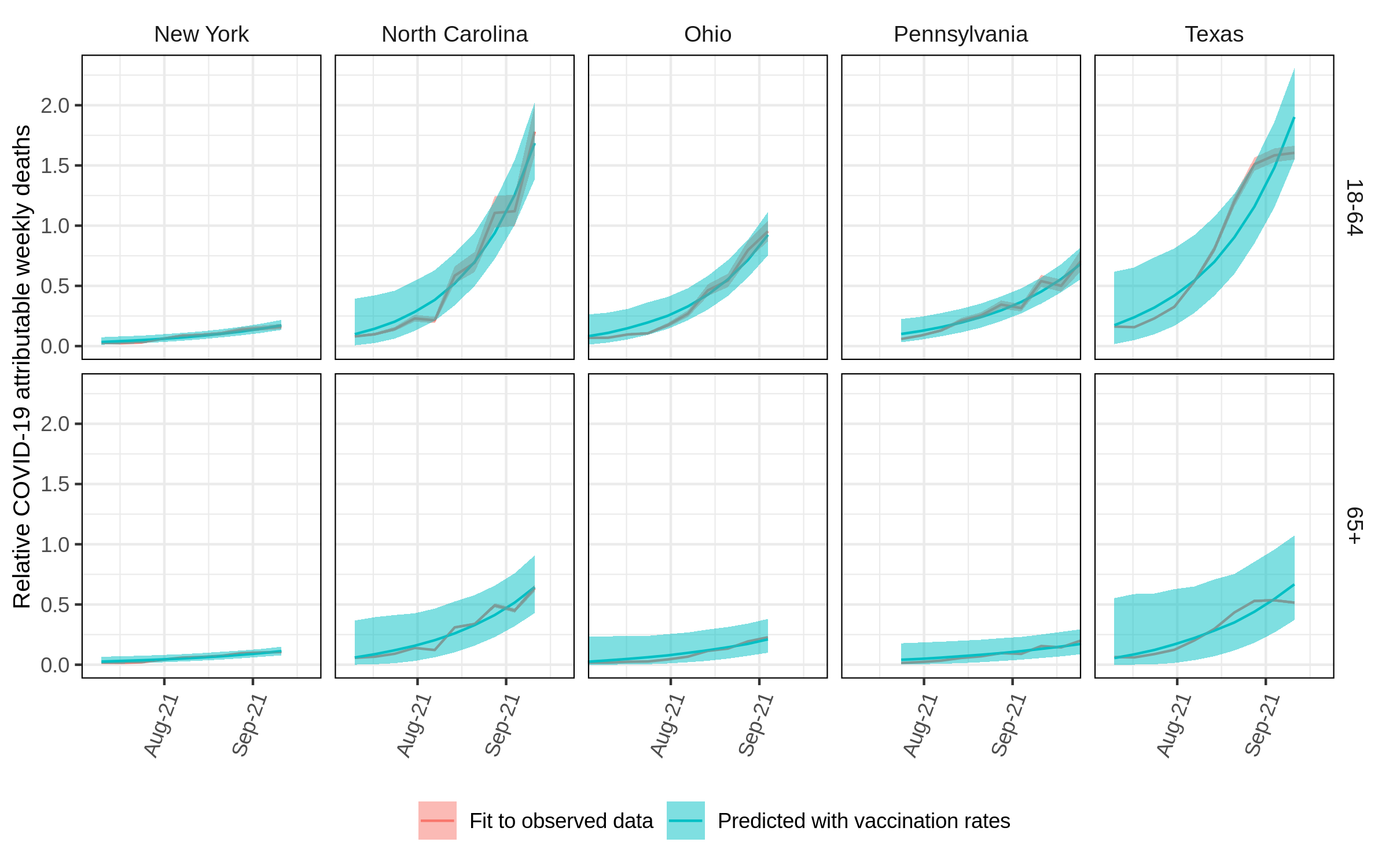}
    \caption{\textbf{Posterior predictive checks of model~\eqref{eq:model_vaccination} (part 2).} Posterior median estimates (line) and 95\% credible intervals (ribbon) of the relative COVID-19 deaths during the summer 2021 resurgence period estimated with pre-resurgence vaccination coverage~\eqref{eq:model_vaccination} are shown in blue. Posterior median estimates (line) and 95\% credible intervals (ribbon) of the relative COVID-19 deaths during the summer 2021 resurgence period estimated with weekly deaths data~\eqref{eq:rescaledeaths}, on which~\eqref{eq:model_vaccination} was fitted, are shown in red. Row facets show resurgences in relative COVID-19 deaths in individuals aged 18-64 and $65+$.}
    \label{fig:PPC_preprop_vac2}
\end{figure}

\begin{figure}[t]
    \centering
    \includegraphics[width = 0.95\textwidth]{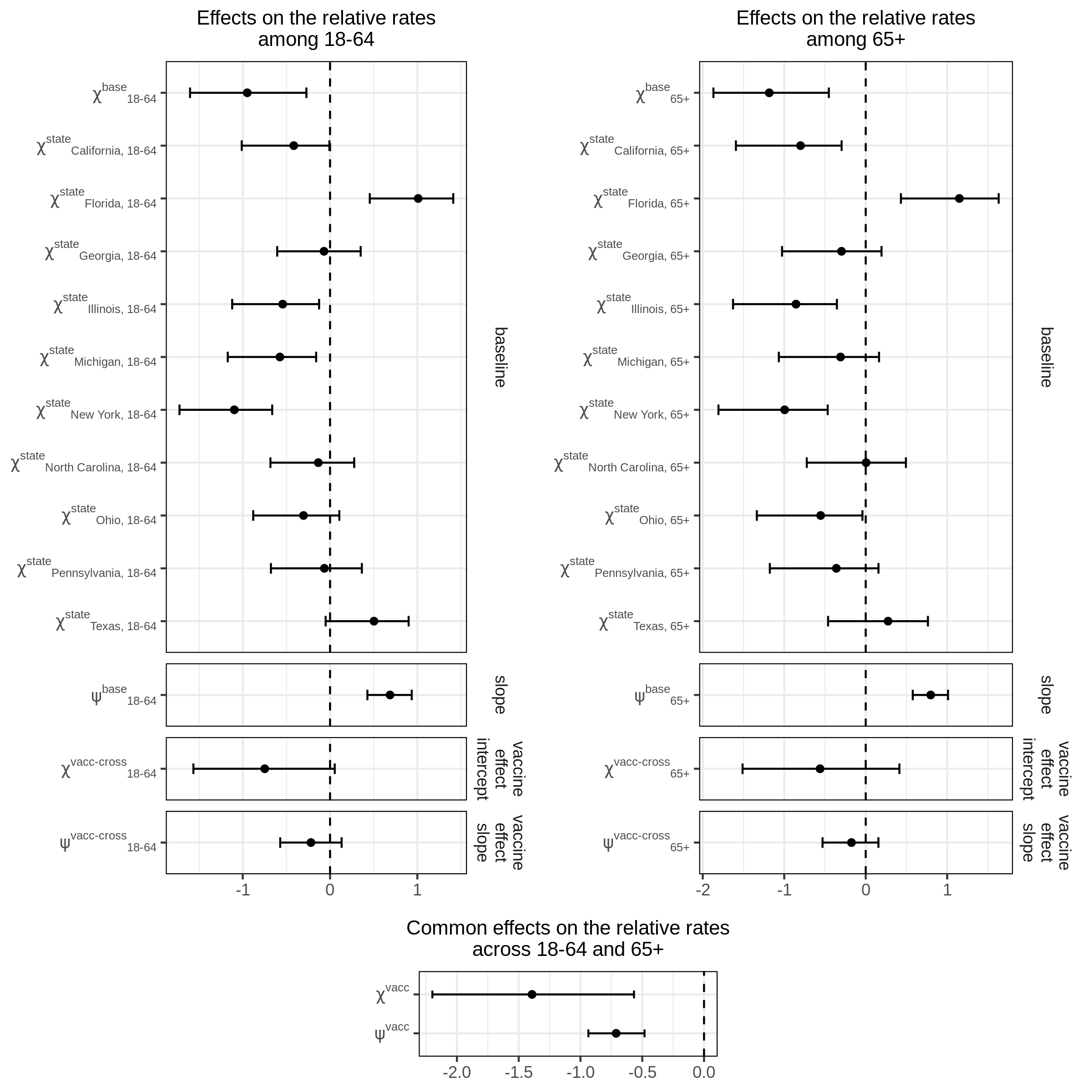}
    \caption{\textbf{Forest plot of the effects in model~\eqref{eq:model_vaccination}.} Posterior median estimates (dot) and 95\% credible intervals (error bars) of model~\eqref{eq:model_vaccination}'s parameters are shown. The parameters predicting the relative COVID-19 deaths among $18$-$64$ are shown on the left and among $65+$ on the right.}
    \label{fig:params_preprop_vac}
\end{figure}

\begin{figure}[t]
    \centering
    \includegraphics[width = 0.9\textwidth]{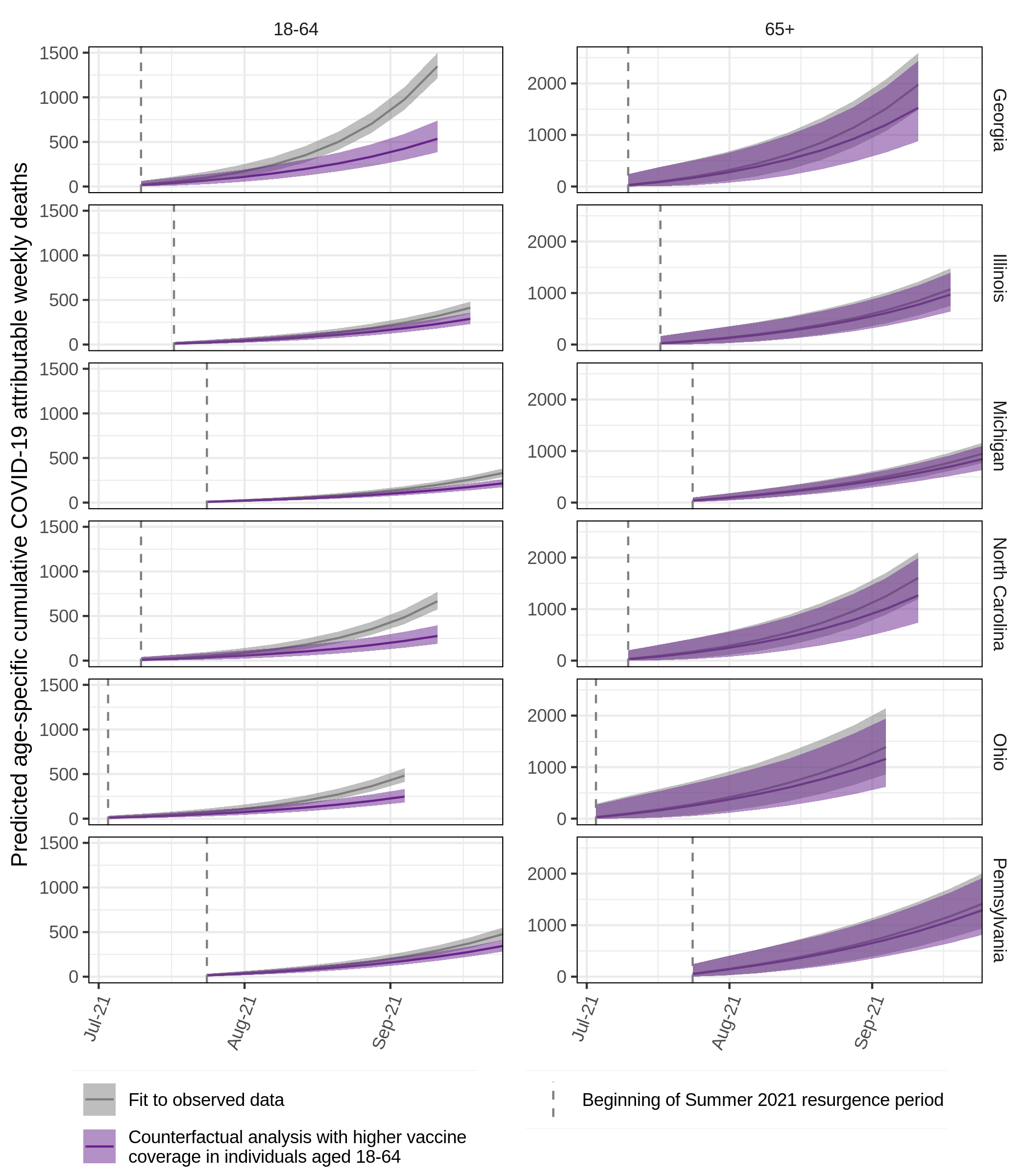}
    \caption{\textbf{Projected avoidable deaths under higher vaccine coverage in individuals aged 18-64 during the summer 2021 resurgence period in six US states.}
    In counterfactual scenarios, we predicted the number of COVID-19 deaths assuming that the vaccine coverage in individuals aged $18$-$64$ in all states had been the same as the maximum vaccine coverage in individuals aged $18$-$64$ across states ($46$\%). 
    Posterior median estimates of the predicted weekly deaths (line) and 95\% credible intervals (ribbon) are shown for the observed COVID-19 deaths (black) and the counterfactual (purple). The start of the 2021 summer resurgence is shown as dashed vertical line.}
    \label{fig:weeklydeaths_counterfactual_remaining}
\end{figure}


\clearpage
\newpage

\begin{table*}[!p]
\small
\begin{threeparttable}
 \centering
  \centering
      \begin{tabular}{l l | r r r}
      \toprule
      \hline
       \multicolumn{2}{c|}{\multirow{2}{*}{Method}}  & \multicolumn{3}{c}{Simulation scenarios} \\[0.05cm]
      && Weakly correlated & Mildly correlated & Strongly correlated \\[0.05cm]
      \hline
      &&&&\\[-0.3cm]
      && \multicolumn{3}{c}{Proportion of data inside the 95\% credible interval} \\[0.05cm]
      \cline{3-5} 
        &&&&\\[-0.3cm]

        \multicolumn{2}{l|}{\textbf{Standard 2D GP}} \\
        &&98.27\%  & 99.07\% & 98.67\% \\[0.3cm]
        
       \multicolumn{5}{l}{\textbf{Standard B-splines surface}}  \\
        &Number of knots  & & \\
        & $10$ & 98.93\%  & 98.87\%  & 98.60\% \\[0.05cm]
        & $30$ & 99.13\% & 98.53\%  & 97.87\%  \\[0.05cm]
         & $40$ &  98.73\%   & 98.00\% & 97.40\%  \\[0.2cm]
         
         \multicolumn{5}{l}{\textbf{Bayesian P-splines surface}}  \\
        &Number of knots  & & \\
        & $10$ & 99.13\% & 99.00\% & 98.27\% \\[0.05cm]
        & $30$ & 99.67\%  & 98.87\%  & 98.93\%  \\[0.05cm]
         & $40$ &  99.80\%  & 98.60\% & 99.00\%  \\[0.2cm]
        
      \multicolumn{5}{l}{\textbf{Regularised B-splines projected 2D GP}} \\
        &Number of knots & & \\
        & $10$ & 98.47\%  & 98.93\% & 98.47\%  \\[0.05cm]
        & $30$ & 98.53\%  & 99.00\% & 98.53\%  \\[0.05cm]
         & $40$ &  98.73\%  & 99.00\%  & 98.60\% \\[0.05cm]
          
        \hline
         &&&&\\[-0.35cm]

\bottomrule
 \end{tabular}
    \caption{{\bf Predictive accuracy of the four spatial models on simulated data.} Proportion of simulated observations from the test data withheld from fitting, lying inside the estimated 95\% credible interval. }
    \label{tab:CI_simu} 
    \end{threeparttable}
\end{table*}

\begin{table*}[!hbt]
\small
\begin{threeparttable}
 \centering
 \resizebox{1\textwidth}{!}{%
  \centering
      \begin{tabular}{l l | r }
      \toprule
      \hline
        \multicolumn{2}{l}{Method} & Testing MSE \\[0.05cm]
        \hline
        \multicolumn{3}{l}{\textbf{Results extracted from~\cite{mishra:2020}}} \\
        & Full rank GP with Matérn $\frac{3}{2}$ kernel & 2.47 \\
        & Low rank SPDE approximation with 1046 basis function and a Matérn $\frac{3}{2}$ kernel & 4.34 \\
        & Neural processes & 14.94 \\
        \multicolumn{3}{l}{\textbf{Results obtained by the authors}} \\
        &Standard B-splines surface & 4.45 \\
        &Bayesian P-splines surface & 2.57 \\
        &Regularised B-splines projected 2D GP & 2.96 \\

\bottomrule
 \end{tabular}
 }%

    \caption{{\bf Benchmark results obtained on 83,000 locations across East Africa with measured deviations in land surface temperatures.} }
    \label{tab:benchmarkMSE} 
    \end{threeparttable}
\end{table*}

\begin{table*}[!hbt]
\small
\begin{threeparttable}
 \centering
 \resizebox{1\textwidth}{!}{%
  \centering
      \begin{tabular}{l | r | r | r | r}
      \toprule
      \hline
         & \multicolumn{4}{c}{Method} \\[0.05cm]
       
        &&&&\\[-0.3cm]
      &  \multirow{2}{*}{Standard 2D GP} & Standard B-splines & Bayesian P-splines & Regularised B-splines \\
      & & surface && projected 2D GP \\[0.05cm]
       \cline{2-5}
      &&&&\\[-0.35cm]
      \multirow{2}{*}{State} & $\%$DoH observations  & $\%$DoH observations & $\%$DoH observations & $\%$DoH observations \\
       & inside 95\% CI  & inside 95\% CI & inside 95\% CI  & inside 95\% CI \\[0.05cm]
    \hline
        &&&&\\[-0.3cm]
 California & 90.91\% & 86.36\% & 84.09\% & 86.36\% \\
 Florida & 94.57\% & 96.52\% & 95.87\% & 94.35\% \\
 Texas & 97.14\% & 97.92\% & 97.14\% & 96.61\% \\
 &&&&\\[-0.3cm]
 \hline
 &&&&\\[-0.3cm]
  Average & 94.90\% & 95.29\% & 94.31\% & 93.82\% \\
\bottomrule
 \end{tabular}
 }%

    \caption{{\bf Predictive accuracy of the four spatial prior on age-specific weekly COVID-19 attributable deaths data.} Proportion of the empirical age-specific contribution to COVID-19 weekly deaths reported by DoH inside the 95\% credible interval estimated by the model presented in Section~\ref{sec:model} under four prior specifications. The four prior are placed on the random surface which specifies the age composition of deaths over time, a standard 2D GP, a standard B-splines surface and our regularised B-splines projected GP. Note that age-specific COVID-19 attributable deaths were not reported by the DoH of New York.}
    \label{tab:comparison_estimate_cum_deaths_age} 
    \end{threeparttable}
\end{table*}

\begin{table*}[!hbt]
\small
\begin{threeparttable}
 \centering
 \resizebox{0.85\textwidth}{!}{%
        \begin{tabular}{r | r | r | r}
      \toprule
      \hline
      \multicolumn{4}{c}{Method} \\[0.05cm]
       
        &&&\\[-0.3cm]
      \multirow{2}{*}{Standard 2D GP} & Standard B-splines & \multirow{2}{*}{Bayesian P-splines} & Regularised B-splines \\
      & surface && projected 2D GP \\[0.05cm]
       &&&\\[-0.3cm]
      \hline
       &&&\\[-0.3cm]
      $\Delta$ELPD (SD) & $\Delta$ELPD (SD) & $\Delta$ELPD (SD) & $\Delta$ELPD (SD) \\[0.05cm]
      \hline    
      &&&\\[-0.3cm]
      -22.03 (11.99) & -295.11 (54.45) & -16.44 (13.05) & 0.00 (0.00)\\

\bottomrule
 \end{tabular}
 }%

    \caption{{\bf Predictive performance of the four spatial models on age-specific weekly COVID-19 attributable deaths.} The differences in the expected log pointwise predictive density (ELPD)~\citep{Vehtari2016}, and standard error of the difference, of each model compared to the best performing model (for which the difference is null) are presented. A higher expected log pointwise predictive density suggests a better predictive capacity. Four prior placed on the random surface which specifies the age composition of deaths over time are compared, a standard 2D GP, a standard B-splines surface and our regularised B-splines projected GP.}
    \label{tab:ELPD_comp} 
    \end{threeparttable}
\end{table*}

\begin{table*}[!hbt]
\small
\begin{threeparttable}
 \centering
 \resizebox{1\textwidth}{!}{%
  \centering
      \begin{tabular}{l | r r r r }
      \toprule
      \hline
      \multirow{4}{*}{State} & \multicolumn{2}{c}{\textbf{Avoidable COVID-19}} & \multicolumn{2}{c}{\textbf{Percent of COVID-19 }} \\
       & \multicolumn{2}{c}{\textbf{attributable deaths}} & \multicolumn{2}{c}{\textbf{attributable deaths avoidable}} \\
      & \multicolumn{2}{c}{Among age group} & \multicolumn{2}{c}{Among age group} \\
         & $18$-$64$ & $65+$ & $18$-$64$ & $65+$ \\[0.05cm]
         \hline
         &&&&\\[-0.35cm]
        California & 185 [-47, 419] & 86 [-666, 862]  & 9.93\% [-2.58, 20.90] & 3.09\% [-27.64, 26.80]  \\
        
        Florida & 2218 [1531, 2887] & 1495 [-1902, 4613]  & 57.15\% [40.96, 70.47] & 20.09\% [-29.54, 54.04]  \\
        
        Georgia & 810 [597, 999] & 433 [-450, 1222]  & 60.28\% [45.68, 71.24] & 22.13\% [-25.12, 55.42]  \\
        
        Illinois & 123 [51, 195] & 101 [-329, 540]  & 29.90\% [13.42, 43.82] & 9.33\% [-35.78, 41.21]  \\
        
        Michigan & 116 [69, 165] & 101 [-150, 352]  & 34.97\% [22.15, 46.75] & 10.62\% [-17.29, 33.86]  \\
         
        North Carolina & 387 [261, 503] & 336 [-356, 958]  & 58.34\% [40.99, 71.50] & 21.10\% [-24.97, 53.94]  \\
        
        New York & - & -  & - & -  \\
        
        Ohio & 234 [145, 319] & 224 [-570, 995]  & 48.69\% [31.90, 61.81] & 16.30\% [-50.94, 56.41]  \\
        
        Pennsylvania & 134 [63, 207] & 122 [-474, 734]  & 27.87\% [13.96, 40.11] & 8.58\% [-40.71, 42.66]  \\
        
        Texas & 1631 [790, 2452] & 765 [-2037, 3500]  & 39.80\% [20.81, 54.85] & 13.41\% [-41.91, 49.61]  \\
        
        \hline
        &&&&\\[-0.35cm]
        
        Total &  5850 [4341, 7251] & 3730 [-2958, 9585] & 41.96\% [31.57, 50.89] & 14.39\% [-11.86, 35.30] \\

\bottomrule
 \end{tabular}
 }%

    \caption{{\bf Projected avoidable deaths under higher vaccine coverage in individuals aged 18-64 during the summer 2021 resurgence period.} In counterfactual scenarios, we predicted the number of COVID-19 deaths assuming that the vaccine coverage in individuals aged $18$-$64$ in all states had been the same as the maximum vaccine coverage in individuals aged $18$-$64$ across states ($46$\% reported in New York). These counterfactuals consider that the inferred associations between resurgent deaths and vaccine coverage are causal, which may be incorrect. Posterior median estimates and 95\% credible interval of the projected avoidable deaths (left) and proportion of deaths that were avoidable (right) over the summer 2021 resurgence period in the 10 most populated US states are shown.}
    \label{tab:weeklydeaths_counterfactual} 
    \end{threeparttable}
\end{table*}

\FloatBarrier
\clearpage
\section{B-splines and properties of the regularised B-splines projected Gaussian Process prior}
\subsection{Construction of a B-spline curve and a B-spline surface} \label{sup:bsplineconstruction}
We start this section by constructing a B-spline basis function, also called B-spline segment, which we refer to as B-spline. A B-spline is entirely defined by a polynomial degree, $d$, and a non-decreasing sequence of knots $(t_1 \leq \dots \leq t_K)$ spanning in $\mathcal{X} \subseteq \mathbb{R}$, where $K$ is the number of knots. 
For the B-spline to cover the entire span of knots, the sequence of knots has to be extended on the left ($d$ times) and on the right ($d$ times), $(\xi_1 \leq \dots \leq \xi_d \leq \xi_{d+1} \leq \dots \leq \xi_{d+K}\leq \xi_{d+K+1} \leq\dots\leq\xi_{2d+K+1})$, where $\xi_{d+1} := t_1, \dots, \xi_{d+K} := t_K$.
The value of the additional knots is generally set to the boundary knots~\citep{Perperoglou2019}. The order of a B-spline, denoted by $p$, is equal to $d - 1$. A B-spline of order $p$, written $B_{i,p}(x)$, is defined by the recursive formula, 
\begin{align} \label{eq:constructbspline}
\begin{split} 
    B_{i,0}(x) &= 
    \begin{cases}
    1 & \text{if } \xi_i \leq x < \xi_{i+1}, \\
    0 & \text{otherwise}
    \end{cases} \\
    B_{i,p}(x) &= \frac{x - \xi_i}{\xi_{i + p} - \xi_i} B_{i,p-1}(x) + \frac{\xi_{i+p+1} - x}{\xi_{i+p+1} - \xi_{i+1}} B_{i+1,p-1}(x)
\end{split} 
\end{align}
and $B_{i,0}(x) = 0$ if $\xi_i = \xi_{i+1}$, for $x \in \mathcal{X}$ and $i = 1, \dots, I$ with $I = K + d - 1$. The above is usually referred to as the Cox-de Boor recursion formula~\citep{deboor}.  

\noindent
Let us define $\mathcal{C}^{k}(A;B)$ as the set of $\mathcal{C}^{k}$ functions from $A$ to $B$.

\noindent
\textbf{Proposition 1. } \textit{
    Let a B-spline of degree $d$, constructed in~\eqref{eq:constructbspline} given a vector of strictly increasing sequence knots (no knot occurs more than $1$ time), spanning over $\mathcal{X}$. Then, the B-spline is a $C^{d-1}(\mathcal{X};\mathbb{R}^+)$.
    }
\begin{proof}
Recall from~\cite{goldman:2002} that a B-spline of degree $d$, constructed in~\eqref{eq:constructbspline} given a vector of strictly increasing sequence knots, spanning over $\mathcal{X}$ is of continuity $C^{\infty}(\mathcal{X};\mathbb{R}^+)$ piecewise, and of continuity $C^{d-1}(\mathcal{X};\mathbb{R}^+)$ on the knots. Therefore, the same B-spline is $C^{d-1}(\mathcal{X};\mathbb{R}^+)$ everywhere.
\end{proof}

Next, we construct a B-spline curve of degree $d$, a linear combination of coefficients and B-splines of degree $d$, given by,
\begin{align}
    \sum_{i = 1}^I \beta_{i} \: B_{i,d-1}(x)
\end{align}
where $\boldsymbol{\beta} = [\beta_{i}]_{i = 1, \dots, I}$ are the associated B-spline coefficients, also referred to as control points or de Boor points. 

The surface analogue to the B-spline curve is the B-spline surface. The surface is defined on the space $\mathcal{X}_1 \times \mathcal{X}_2$ with $\mathcal{X}_1 \subseteq \mathbb{R}$ and $\mathcal{X}_2\subseteq \mathbb{R}$. 
We start by defining two non decreasing sequences of knots, $\boldsymbol{t}^1 = (t^1_{1} \leq  \dots \leq t^1_{K_1})$ and $(t^2_{1} \leq \dots \leq t^2_{K_2})$ placed respectively on the space $\mathcal{X}_1$ and $\mathcal{X}_2$, where
$K_1$ is the number of knots along the first axis, and $K_2$ along the second axis. 
The corresponding integral of the B-spline surface evaluated at $x_1 \in \mathcal{X}_1$ and $x_2 \in \mathcal{X}_2$ is given by,
\begin{align} 
    \sum_{i =1}^I \sum_{j = 1}^J \: \beta_{(i, j)} \: B^1_{i,d_1-1}(x_1) \: B^2_{j,d_2-1}(x_2),
\end{align}
where $B^1_{i,d_1-1}(.)$ is the $i$th B-spline of degree $d_1$ defined over the space $\mathcal{X}_1$ on the knot vector $\boldsymbol{t}^1$. Similarly, $B^2_{j,d_2-1}(.)$ is the $j$th B-spline of degree $d_2$ defined over the space $\mathcal{X}_2$ on the knot vector $\boldsymbol{t}^2$. The total number of B-splines defined over the space $\mathcal{X}_1$ is $I \triangleq K_1+d_1-1$ and over the space $\mathcal{X}_2$ it is $J \triangleq K_2+d_2-1$. We denote the ensemble of pairs of B-spline's indices by 
\begin{align} \label{eq:bsplineindices}
\boldsymbol{U} \triangleq \big(\boldsymbol{u}_1, \dots, \boldsymbol{u}_M \big) =  \{1, \dots, I\} \times \{1, \dots, J\},
\end{align}
with $M = I \times J$.

\subsection{Proof that a base kernel function projected with cubic B-splines is a $\mathcal{C}^2$ function} \label{supp:cinfprop}
We follow the notations introduced in Section~\ref{sup:bsplineconstruction} and denote the $i$th and $j$th cubic B-spline by $B^1_{i}(x_1)$ and $B^2_{j}(x_2)$, with $i = 1, \dots I$ and $j = 1, \dots, J$\footnote{Notice that we dropped the polynomial degree index as it is fixed to $3$.} and where $x_1 \in \mathcal{X}_1\subseteq \mathbb{R}$ and $x_2 \in \mathcal{X}_2\subseteq \mathbb{R}$. They are respectively defined on two vectors of strictly increasing knots of lengths $K_1$ and $K_2$. For ease of notation, we denote the ensemble of B-spline indices by $\boldsymbol{U}$ defined in~\eqref{eq:bsplineindices}. 

\noindent
Let us define a low-rank covariance matrix $\boldsymbol{K}_{\beta}$ of size $IJ \times IJ$ obtained by evaluating at every pair of points in $\boldsymbol{u}, \boldsymbol{u}' \in \boldsymbol{U}$ a base kernel function $k_{\beta}$,
\begin{align} \big(\boldsymbol{K}_{\beta}\big)_{\boldsymbol{u},\boldsymbol{u}'} = k_{\beta}\big(\boldsymbol{u},\boldsymbol{u}'\big).
\end{align}
Let us define a high-rank covariance matrix projected by cubic B-splines by $\boldsymbol{W}$, 
obtained by evaluating at every pair of points a kernel function $w$, 
$\boldsymbol{W}_{(x_1,x_2), (x_1',x_2')} = w(x_1,x_2, x_1',x_2')$ with, 
\begin{multline}
\label{eq-def-w}
    w(x_{1},x_{2},x_{1}',x_{2}') =\\ \sum_{(u_{1},u_{2})\in \boldsymbol{U}}\sum_{(u_{1}',u_{2}')\in \boldsymbol{U}} D(x_{1},x_{2},u_{1},u_{2}) \: k_{\beta}(u_{1},u_{2},u_{1}',u_{2}') \: D(x_{1}',x_{2}',u_{1}',u_{2}').
\end{multline}
where $(x_1,x_2), (x_1',x_2') \in \mathcal{X}_1 \times \mathcal{X}_2$, and
\begin{equation}
\label{eq-def-D}
    D(x_{1},x_{2},u_{1},u_{2}) =B_{u_{2}}^{2}(x_{2}) \: B_{u_{1}}^{1}(x_{1}).
\end{equation}

\noindent
We repeat here the definition presented in~\ref{sup:bsplineconstruction}: we define $\mathcal{C}^{k}(A;B)$ as the set of $\mathcal{C}^{k}$ functions from $A$ to $B$.

\noindent
\textbf{Proposition 2. } \textit{
\textit{Let $w(\cdot)$ be defined as in \eqref{eq-def-w} and $\mathcal{S}=\mathcal{X}_1 \times \mathcal{X}_2 \times \mathcal{X}_1 \times \mathcal{X}_2 \subseteq \mathbb{R}^{4}$. Then, $w(\cdot)$ is a $\mathcal{C}^{2}(\mathcal{S};\mathbb{R})$ function.}
}
\begin{proof}
In order to show that $w(x_1,x_2, x_1',x_2')\in \mathcal{C}^{2}(\mathcal{S};\mathbb{R})$ it is sufficient to show that all its second partial derivatives are continuous functions on $\mathcal{S}$. We show this for the case of the partial derivative  $\frac{\partial^{2}w}{\partial x_{1}^{2}}$ and the mixed derivative $\frac{\partial^{2}w}{\partial x_{1}\partial x_{2}}$, as the proof for the remaining second partial derivatives is identical. 
The proof is constructed as follows: (1) we compute $\frac{\partial^{2}w}{\partial x_{1}^{2}}$  and $\frac{\partial^{2}w}{\partial x_{1}\partial x_{2}}$ and (2) we show that $\frac{\partial^{2}w}{\partial x_{1}^{2}}$ and $\frac{\partial^{2}w}{\partial x_{1}\partial x_{2}}$ are continuous.\par
\textit{Step (1):} A direct computation shows,
\begin{multline}
\label{eq-partial-derivative-w}
    \frac{\partial^{2}w}{\partial x_{1}^{2}} =\\ \sum_{(u_{1},u_{2})\in \boldsymbol{U}}\sum_{(u_{1}',u_{2}')\in \boldsymbol{U}} B^{2}_{u_{2}}(x_{2})\frac{\partial^{2}B^{1}_{u_{1}}}{\partial x_{1}^{2}}(x_{1})
    k_{\beta}(u_{1},u_{2},u_{1}',u_{2}') D(x_1',x_2',u_1',u_2'),
\end{multline}
\begin{multline} \label{eq-partial-derivative-w-x1x2}
    \frac{\partial w}{\partial x_1 \partial x_{2}} =\\ \sum_{(u_{1},u_{2})\in \boldsymbol{U}}\sum_{(u_{1}',u_{2}')\in \boldsymbol{U}} 
    \frac{\partial B^{2}_{u_{2}}}{\partial x_2}(x_{2}) \frac{\partial B^{1}_{u_{1}}}{\partial x_1}(x_{1})
    k_{\beta}(u_{1},u_{2},u_{1}',u_{2}') D(x_1',x_2',u_1',u_2'),
\end{multline}
where we have used the expression of $D$ as stated in \eqref{eq-def-D} and $\boldsymbol{U}$ is defined in~\eqref{eq:bsplineindices}. \par
\textit{Step (2):} We show that~\eqref{eq-partial-derivative-w} is a continuous function on $\mathcal{S}$. As stated in Proposition 1, the functions $B^{1}_{u_{1}}(\cdot)$ and $B^{2}_{u_{2}}(\cdot)$ are $\mathcal{C}^{2}(\mathcal{X}_{1};\mathbb{R}^+)$ and $\mathcal{C}^{2}(\mathcal{X}_{2};\mathbb{R}^+)$ functions, respectively. Therefore, $\frac{\partial^{2}B^{1}_{u_{1}}}{\partial x_{1}^{2}}(\cdot)$ and $B^{2}_{u_{2}}(x_{2})$ are continuous function on $\mathcal{X}_{1}$ and $\mathcal{X}_{2}$, respectively. 
Moreover, recall that $k_{\beta}$ is a kernel function, so for any fixed $(u_{1},u_{2}),(u_{1}',u_{2}')\in 
\boldsymbol{U}$, $k_{\beta}(u_{1},u_{2},u_{1}',u_{2}')\in \mathbb{R}$. By standard results of Multivariate Calculus (\cite{Lax2017} Theorem 2.5), for any fixed $u_{1},u_{2},u_{1}',u_{2}'$ the term 
$$B^{2}_{u_{2}}(x_{2})\frac{\partial^{2}B^{1}_{u_{1}}}{\partial x_{1}^{2}}(x_{1})k_{\beta}(u_{1},u_{2},u_{1}',u_{2}') B^{2}_{u_{2}'}(x_{2}')B^{1}_{u_{1}'}(x_{1}')$$ 
is a continuous function from $\mathcal{S}$ to $\mathbb{R}$. Moreover, the RHS of \eqref{eq-partial-derivative-w} is a linear combination of continuous function from $\mathcal{S}$ to $\mathbb{R}$, therefore it is continuous from $\mathcal{S}$ to $\mathbb{R}$. Hence, $\frac{\partial^{2}w}{\partial x_{1}^{2}}$ is a continuous function from $\mathcal{S}$ to $\mathbb{R}$.

Now, we show that~\eqref{eq-partial-derivative-w-x1x2} is a continuous function on $\mathcal{S}$. As stated in Proposition 1, the functions $B^{1}_{u_{1}}(\cdot)$ and $B^{2}_{u_{2}}(\cdot)$ are $\mathcal{C}^{2}(\mathcal{X}_{1};\mathbb{R}^+)$ and $\mathcal{C}^{2}(\mathcal{X}_{2};\mathbb{R}^+)$ functions, respectively. Therefore, $\frac{\partial B^{1}_{u_{1}}}{\partial x_{1}}(\cdot)$ and $\frac{\partial B^{2}_{u_{2}}}{\partial x_{2}}(\cdot)$ are continuous functions on $\mathcal{X}_{1}$ and $\mathcal{X}_{2}$, respectively. 
Moreover, recall that $k_{\beta}$ is a kernel function, so for any fixed $(u_{1},u_{2}),(u_{1}',u_{2}')\in 
\boldsymbol{U}$, $k_{\beta}(u_{1},u_{2},u_{1}',u_{2}')\in \mathbb{R}$. By standard results of Multivariate Calculus (\cite{Lax2017} Theorem 2.5), for any fixed $u_{1},u_{2},u_{1}',u_{2}'$ the term 
$$ \frac{\partial B^{2}_{u_{2}}}{\partial x_2}(x_{2}) \frac{\partial B^{1}_{u_{1}}}{\partial x_1}(x_{1}) k_{\beta}(u_{1},u_{2},u_{1}',u_{2}') B^{2}_{u_{2}'}(x_{2}')B^{1}_{u_{1}'}(x_{1}')$$ 
is a continuous function from $\mathcal{S}$ to $\mathbb{R}$. Moreover, the RHS of \eqref{eq-partial-derivative-w-x1x2} is a linear combination of continuous functions on from $\mathcal{S}$ to $\mathbb{R}$, therefore it is continuous from $\mathcal{S}$ to $\mathbb{R}$. Hence, $\frac{\partial w}{\partial x_1 \partial x_{2}}$ is a continuous function from $\mathcal{S}$ to $\mathbb{R}$.

The proof for the other terms follows from a similar argument, with the necessary replacement of sets and functions. Therefore $w(\cdot)$ is a $\mathcal{C}^{2}(\mathcal{S};\mathbb{R})$ function.
\end{proof}

\FloatBarrier
\clearpage
\section{Modelling COVID-19 weekly deaths} 
\subsection{Finding weekly death count from censored and missing cumulative death count} \label{sec:findweeklydeath}
We denote by $\mathcal{W}^{\text{cum}}= \{1, \dots, W^{\text{cum}}\}$ the set of week indices from the first to the last reported cumulative deaths by the CDC. The reports span from May 2, 2020 to October 02, 2021 such that $W^{\text{cum}} = 75$. We denote by $D_{m,b,w}$ the reported cumulative deaths by the CDC in state $m$ for age group $b$ in week $w$. The cumulative deaths count $D_{m,b,w}$ is reported if the count count is 0 or strictly greater than 9, and otherwise it is censored. For simplicity we suppress the state index in what follows, with all equations being analogous. We denote by $\mathcal{W}_b^{\text{censored}}$ the set of weeks where $D_{b,w}$ has been censored. 

For a given age group $b$, if $D_{b,w}$ is reported on weeks $w$ and $w + 1$, we find the weekly deaths among age group $b$ in week $w$, $d_{b,w}$, with the first order difference,
\begin{align}
    d_{b,w} = D_{b,(w+1)} - D_{b,w}
\end{align}
The weekly deaths that are retrievable from the first order difference are defined as ``retrievable'' and the set of week indices for which we can retrieve the weekly deaths is denoted by $\mathcal{W}_b^{\text{WR}}$ (WR := Week Retrievable). However, if $D_{b,w}$ is censored for $w$ or $w+1$, the first-order difference is not obtainable and the weekly deaths is said to be ``non-retrievable''. We denote by $\mathcal{W}_b^{\text{WNR}}$ the set of week indices for which we cannot retrieve the weekly deaths (WNR := Week non-retrievable), with $\mathcal{W}_b^{\text{WR}} \cap \mathcal{W}_b^{\text{WNR}} = \varnothing$. 
Notice that for a fixed age group $b$, weekly deaths are non-retrievable for a joint sequence of weeks over time, and importantly this period can happen only once (because the cumulative deaths are strictly increasing). Therefore, we consider the sequence of non-retrievable weekly deaths for a fixed age group together.
Because the boundaries of the censored cumulative death are known, we can obtain the boundaries of the sum of the non-retrievable weekly deaths sequence. Let us denote by $w_b^{\text{FC}}$ the first week when the cumulative death count is censored (FC := First Censored). Such that,
\begin{align}
    w_b^{\text{FC}} =
    \begin{cases}
    \text{min}(\mathcal{W}_b^{\text{censored}}) & \text{if } \mathcal{W}_b^{\text{censored}} \neq \varnothing \\
    \text{does not exist} & \text{otherwise}.
    \end{cases}
\end{align}
Similarly, we denote by $w_b^{\text{FNC}}$ the first week when the cumulative death count is observed after being censored (FNC := First Non Censored). Such that,
\begin{align}
w_b^{\text{FNC}} =
\begin{cases}
    \text{max}(\mathcal{W}_b^{\text{censored}}) + 1 & \text{if } \text{max}(\mathcal{W}_b^{\text{censored}}) < W^{\text{cum}} \\
    \text{does not exist} & \text{otherwise}.
\end{cases}
\end{align}
Four different scenarios may apply.

\begin{enumerate}
\item  If $D_{b,w}$ is censored for some weeks $w \in \mathcal{W}^{\text{cum}}$ but is observed on the first and the last weeks (i.e., $w = 1$ and $w = W^{\text{cum}}$), the non-retrievable weekly deaths must sum to the first positive cumulative deaths, such that,
\begin{align} \label{eq:scenario_1}
    \sum_{w \in \mathcal{W}_b^{\text{WNR}}} d_{b,w} = D_{b,w_b^{\text{FNC}}}.
\end{align}
where $\mathcal{W}_b^{\text{WNR}} = \{(w_b^{\text{FC}} -1), \mathcal{W}_b^{\text{censored}}\}$. In this case, $\mathcal{W}_b^{\text{WR}} = \{1, \dots, (w_b^{\text{FC}}-2), w_b^{\text{FNC}}, \dots, (W^{\text{cum}}-1)\}$.
We show an example of this scenario in Table~\ref{tab:example_2}.

\item If $D_{b,w}$ is censored for some weeks $w \in \mathcal{W}^{\text{cum}}$, including the first week (i.e., $w = 1$), but is observed for the last week (i.e., $w = W^{\text{cum}}$), the sum of non-retrievable weekly deaths must be between, 
\begin{align} \label{eq:scenario_2}
   D_{b,w_b^{\text{FNC}}} - 9 \leq \sum_{w \in \mathcal{W}_b^{\text{WNR}}} d_{b,w} \leq D_{b,w_b^{\text{FNC}}} - 1.
\end{align}
where $\mathcal{W}_b^{\text{WNR}} = \mathcal{W}_b^{\text{censored}}$. In this case, $\mathcal{W}_b^{\text{WR}} = \{w_b^{\text{FNC}}, \dots, (W^{\text{cum}}-1)\}$.
We show an example of this scenario in Table~\ref{tab:example_3}.

\item If $D_{b,w}$ is censored for some weeks $w \in \mathcal{W}^{\text{cum}}$, including the last week (i.e., $w = W^{\text{cum}}$), but is observed for the first week (i.e., $w = 1$), the sum of the non-retrievable weekly deaths must be between, 
\begin{align} \label{eq:scenario_3}
    1 \leq \sum_{w \in \mathcal{W}_b^{\text{WNR}}} d_{b,w} \leq 9
\end{align}
where $\mathcal{W}_b^{\text{WNR}} = \{(w_b^{\text{FC}} -1), \mathcal{W}_b^{\text{censored}}\setminus W^{\text{cum}}\}$. In this case, $\mathcal{W}_b^{\text{WR}} = \{1, \dots, (w_b^{\text{FC}} -2)\}$. Notice that $D_{b,w_b^{\text{FNC}}}$ does not exist.
We show an example of this scenario in Table~\ref{tab:example_4}.

\item Lastly, if $D_{b,w}$ is censored for all weeks $w \in \mathcal{W}^{\text{cum}}$, including the first week (i.e., $w = 1$) and the last week (i.e., $w = W^{\text{cum}}$), the sum of the non-retrievable weekly deaths must be between, 
\begin{align} \label{eq:scenario_4}
    0 \leq \sum_{w \in \mathcal{W}_b^{\text{WNR}}} d_{b,w} \leq 8
\end{align}
where $\mathcal{W}_b^{\text{WNR}} = \{\mathcal{W}_b^{\text{censored}} \setminus W^{\text{cum}}\}$. In this case, $\mathcal{W}_b^{\text{WR}} = \varnothing$. Notice that $D_{b,w_b^{\text{FNC}}}$ does not exist.
We show an example of this scenario in Table~\ref{tab:example_5}.
\end{enumerate}

The CDC did not publish a report on July 4 2020, such that the cumulative deaths in the corresponding week are not available. We denote this week by $w^{\text{missing}}$. The weekly deaths could not be obtained for the missing week or the week before, such that, $\mathcal{W}_b^{\text{WR}} \cup \mathcal{W}_b^{\text{WNR}} = \mathcal{W}^{\text{cum}} \setminus \{w^{\text{missing}}-1, w^{\text{missing}}, W^{\text{cum}}\}$. Weekly deaths that occurred on those $w^{\text{missing}}$ and $w^{\text{missing}}-1$ are defined to be missing. It is important to note the difference between non-retrievable weekly deaths, for which we have information through the censoring of cumulative deaths, and missing weekly deaths for which no information has been provided. 
Note that if the index of the missing week $w^{\text{missing}}$ and the week before $w^{\text{missing}}-1$, are in between the range of $\mathcal{W}_b^{\text{censored}}$, or are before or after the set's limits, they are included in $\mathcal{W}_b^{\text{censored}}$, with all inequations still holding.



\begin{table*}[!htb]
    \centering
    \begin{threeparttable}
    \begin{tabular}{c|ccccc}
        $w$ & 1 & 2 & 3 & 4 & 5 \\
        \hline
        $D_{w}$ & 0 & 0 & NA & NA & 11  \\
        $d_{w}$ & 0 &  \multicolumn{3}{c}{11} & - \\
    \end{tabular}
     \end{threeparttable}
    \caption{\textbf{Finding weekly death when $D_{w}$ is censored as in scenario 1.} In this example  $\mathcal{W}^{\text{censored}} = \{ 3, 4\}$,
    $w^{\text{FC}} = 3$,
    $w^{\text{FNC}} = 5$,  $\mathcal{W}^{\text{WNR}} = \{2, 3, 4\}$ and $\mathcal{W}^{\text{WR}} = \{1\}$. From~\eqref{eq:scenario_1}, we know that $\sum_{w \in \mathcal{W}^{\text{WNR}}} d_{w} = 11$.}
    \label{tab:example_2}
\end{table*}

\begin{table*}[!htb]
    \centering
    \begin{threeparttable}
    \begin{tabular}{c|ccccc}
        $w$ & 1 & 2 & 3 & 4 & 5 \\
        \hline
        $D_{w}$ & NA & NA & NA & 11 & 11  \\
        $d_{w}$ & \multicolumn{3}{c}{[2-10]} & 0 & - \\
    \end{tabular}
     \end{threeparttable}
    \caption{\textbf{Finding weekly death when $D_{w}$ is censored as in scenario 2.} In this example  $\mathcal{W}^{\text{censored}} = \{ 1, 2, 3 \}$,
    $w^{\text{FC}} = 1$,
    $w^{\text{FNC}} = 4$, $\mathcal{W}^{\text{WNR}} = \{1, 2, 3\}$ and $\mathcal{W}^{\text{WR}} = \{4 \}$. From~\eqref{eq:scenario_2}, we know that $ 2 \leq \sum_{w \in \mathcal{W}^{\text{WNR}}} d_{w} \leq 10$.}
    \label{tab:example_3}
\end{table*}

\begin{table*}[!htb]
    \centering
    \begin{threeparttable}
    \begin{tabular}{c|ccccc}
        $w$ & 1 & 2 & 3 & 4 & 5 \\
        \hline
        $D_{w}$ & 0 & 0 & NA & NA & NA  \\
        $d_{w}$  & 0 & \multicolumn{3}{c}{[1-9]} & - \\
    \end{tabular}
     \end{threeparttable}
    \caption{\textbf{Finding weekly death when $D_{w}$ is censored as in scenario 3.} In this example  $\mathcal{W}^{\text{censored}} = \{ 3,4,5 \}$,
    $w^{\text{FC}} = 3$,
    $w^{\text{FNC}}$ does not exist, $\mathcal{W}^{\text{WNR}} = \{2,3,4\}$ and $\mathcal{W}^{\text{WR}} = \{1 \}$. From~\eqref{eq:scenario_3}, we know that $ 1 \leq \sum_{w \in \mathcal{W}^{\text{WNR}}} d_{w} \leq 9$.}
    \label{tab:example_4}
\end{table*}

\begin{table*}[!htb]
    \centering
    \begin{threeparttable}
    \begin{tabular}{c|ccccc}
        $w$ & 1 & 2 & 3 & 4 & 5 \\
        \hline
        $D_{w}$ & NA & NA & NA & NA & NA  \\
        $d_{w}$ & \multicolumn{4}{c}{[0-8]} & - \\
    \end{tabular}
     \end{threeparttable}
    \caption{\textbf{Finding weekly death when $D_{w}$ is censored as in scenario 4.} In this example  $\mathcal{W}^{\text{censored}} = \{ 1, 2, 3,4,5 \}$,
    $w^{\text{FC}} = 1$, $w^{\text{FNC}}$ does not exist, $\mathcal{W}^{\text{WNR}} = \{1,2,3,4\}$ and $\mathcal{W}^{\text{WR}} = \varnothing$. From~\eqref{eq:scenario_4}, we know that $ 0 \leq \sum_{w \in \mathcal{W}^{\text{WNR}}} d_w  \leq 8$.}
    \label{tab:example_5}
\end{table*}

\FloatBarrier
\subsection{Likelihood form} ~\label{sup:likelihoodform}
In this section, we use the notation introduced in Section~\ref{sec:findweeklydeath}.
The log likelihood is,
\begin{subequations}
\begin{align}
\log \mathcal{L} &= \sum_{b \in \mathcal{B}} \sum_{w \in \mathcal{W}_b^{\text{WR}}} \log p(d_{b,w} \: ; \: \alpha_{b,w}, \theta) \label{eq:loglikret} \\
& + \sum_{b \in \mathcal{B}} \log p\big(\{d_{b,w}\}_{w \in \mathcal{W}_b^{\text{WNR}}} \: ; \: \{\alpha_{b,w}\}_{w \in \mathcal{W}_b^{\text{WNR}}}, \theta\big), \label{eq:logliknret}
\end{align}
\end{subequations}
The log likelihood for the retrievable weekly deaths in~\eqref{eq:loglikret} is
\begin{align}
    \log p(d_{b,w} \: ; \: \alpha_{b,w}, \theta) = \log \text{NegBin}(d_{b,w}\: ; \: \alpha_{b,w}, \theta).
\end{align}
The log likelihood for the non-retrievable weekly deaths in~\eqref{eq:logliknret} is
\begin{align} \label{eq:nonretrievablelik}
\begin{split}
& p\big(\{d_{b,w}\}_{w \in \mathcal{W}_b^{\text{WNR}}} \: ; \: \{\alpha_{b,w}\}_{w \in \mathcal{W}_b^{\text{WNR}}}, \theta\big) = 
p\Big(\sum_{w \in \mathcal{W}_b^{\text{WNR}}} d_{b,w} \: ; \: \sum_{w \in \mathcal{W}_b^{\text{WNR}}} \alpha_{b,w}, \theta\Big) \\
=
   & \begin{cases}
        \text{NegBin}\Big( D_{b,w_b^{\text{FNC}}} \: ; \: \sum_{w \in \mathcal{W}_b^{\text{WNR}}} \alpha_{b,w}, \theta \Big) & \text{if } 1, W^{\text{cum}} \notin \mathcal{W}_b^{\text{censored}} \\[0.6cm] 
        \text{Cdf NegBin}\Big( D_{bw_b^{\text{FNC}}} - 1 \: ; \: \sum_{w \in \mathcal{W}_b^{\text{WNR}}} \alpha_{b,w}, \theta\Big)  & \text{if } 1 \in \mathcal{W}_b^{\text{censored}} \\[0.2cm]
        - \text{Cdf NegBin}\Big( D_{bw_b^{\text{FNC}}} - 9 \: ; \: \sum_{w \in \mathcal{W}_b^{\text{WNR}}} \alpha_{b,w}, \theta\Big) & \text{and } W^{\text{cum}} \notin \mathcal{W}_b^{\text{censored}} \\[0.6cm] 
        \text{Cdf NegBin}\Big( 9 \: ; \: \sum_{w \in \mathcal{W}_b^{\text{WNR}}} \alpha_{b,w}, \theta\Big)  & \text{if } 1 \notin \mathcal{W}_b^{\text{censored}} \\[0.2cm]
        - \text{Cdf NegBin}\Big( 1 \: ; \: \sum_{w \in \mathcal{W}_b^{\text{WNR}}} \alpha_{b,w}, \theta\Big) & \text{and } W^{\text{cum}} \in \mathcal{W}_b^{\text{censored}} \\[0.6cm] 
        \text{Cdf NegBin}\Big( 8 \: ; \: \sum_{w \in \mathcal{W}_b^{\text{WNR}}} \alpha_{b,w}, \theta\Big)  & \text{if } 1,W^{\text{cum}} \in \mathcal{W}_b^{\text{censored}} \\[0.2cm]
        - \text{Cdf NegBin}\Big( 0 \: ; \: \sum_{w \in \mathcal{W}_b^{\text{WNR}}} \alpha_{b,w}, \theta\Big) & 
    \end{cases}
\end{split}
\end{align}
where the boundaries for the different scenarios have been discussed in Section~\ref{sec:findweeklydeath}. Note that to find~\eqref{eq:nonretrievablelik}, we used the fact that the sum of independent negative-binomially distributed random variables with shape parameter $\alpha_1$ and $\alpha_2$, and with the same scale parameter $\theta$ is negative-binomially distributed with a scale parameter $\theta$ and with a shape parameter $\alpha_1 + \alpha_2$.

\FloatBarrier
\clearpage
\section{Additional analyses}

\subsection{Comparison of random surface priors for one US state}\label{supp:realdatanalysis_one_state}
We used the non-parametric model (\ref{eq:fa}-\ref{e:age_composition_model}) to fit age-specific weekly COVID-19 attributable deaths from the CDC data. We compare results, for one state, Florida, across the four prior specifications on the random surface which specifies the age composition of deaths over time, a standard 2D GP, a standard B-splines surface and our regularised B-splines projected GP.

\begin{figure}[!bp]
    \centering
    \includegraphics[width=1\linewidth]{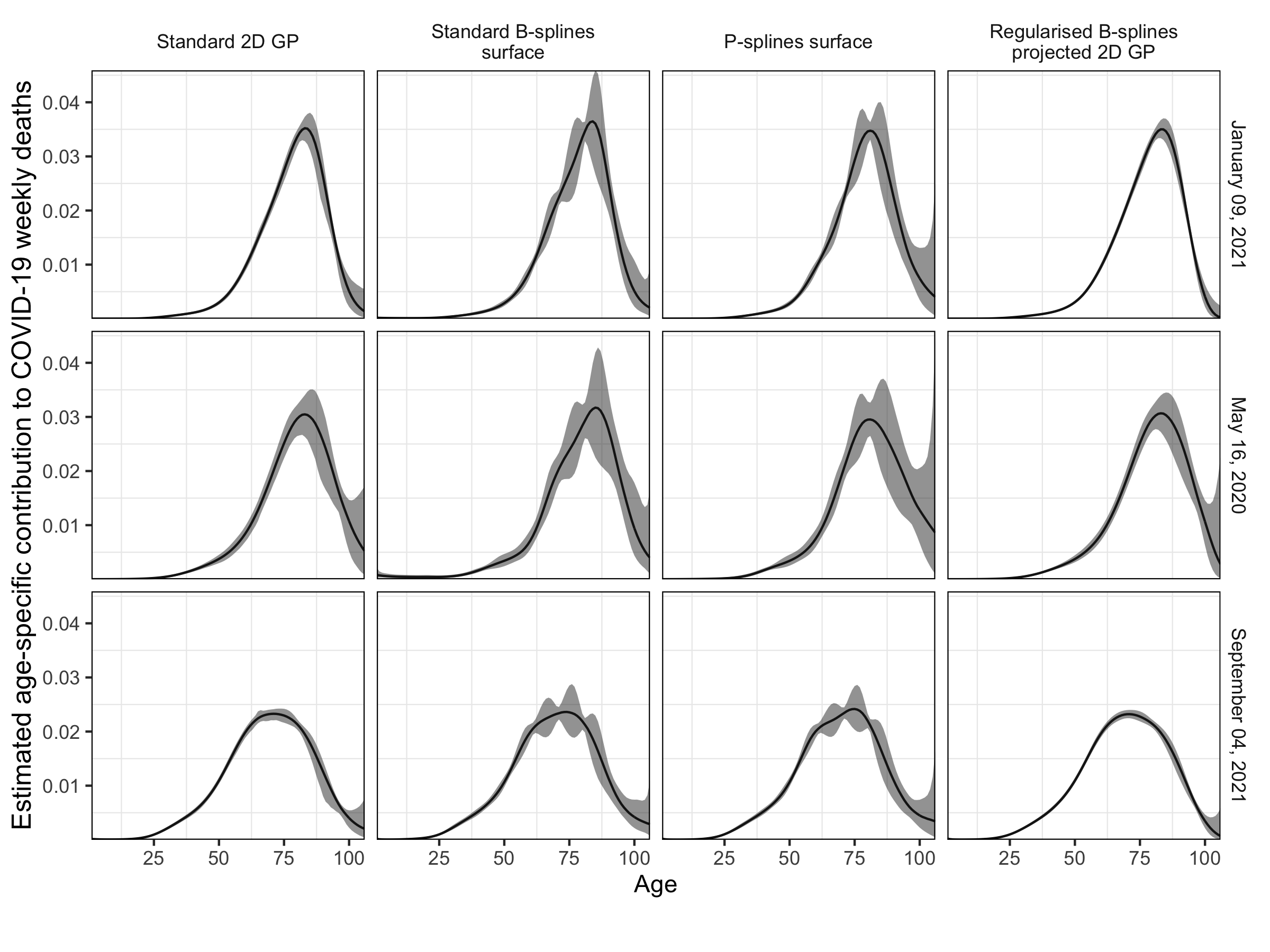}
  \caption{\textbf{Estimated age-specific contribution to weekly COVID-19 attributable deaths by four priors.} Posterior median (black line) and 95\% credible interval (ribbon) of the age-specific contribution to COVID-19 attributable deaths in three weeks. Four prior placed on the random surface which specifies the age composition of deaths over time are compared, a standard 2D GP, a standard B-splines surface and our regularised B-splines projected GP.}
  \label{fig:share_deaths} 
\end{figure}

Figure~\ref{fig:share_deaths} shows the estimated age-specific composition of weekly COVID-19 attributable deaths in three weeks for the four prior specifications.
The posterior distribution of the age profile of COVID-19 attributable deaths estimated with the standard B-splines and Bayesian P-splines prior is rough because the model's complexity is not penalized, or not enough. Such a penalty is introduced in our regularised B-splines projected GP prior, which results in a less wiggly posterior distribution with less uncertainty. 
To quantify the predictive performance of the model, we used the expected log pointwise predictive density (ELPD)~\citep{Verity2020} (Supplementary Table~\ref{tab:ELPD_comp}). The best performance was obtained by the regularised B-splines projected GP. In particular, the expected log pointwise predictive density (ELPD) difference, between a standard GP and a regularised B-splines projected GP was -22.03 (sd 11.99), suggesting significantly better predictive performance with the regularised B-splines projected GP prior. In addition, a standard GP requires a longer running time (2.34 times more). Similar results were obtained for all other states, with the regularised B-splines projected GP prior consistently outperforming the other standard methods.

\clearpage
\FloatBarrier
\subsection{Differences in the pre-vaccination age profile of COVID-19 attributable deaths across states} \label{supp:contribution_pre_vaccination}
\begin{figure}[h!]
    \centering
    \includegraphics[width = 0.9\textwidth]{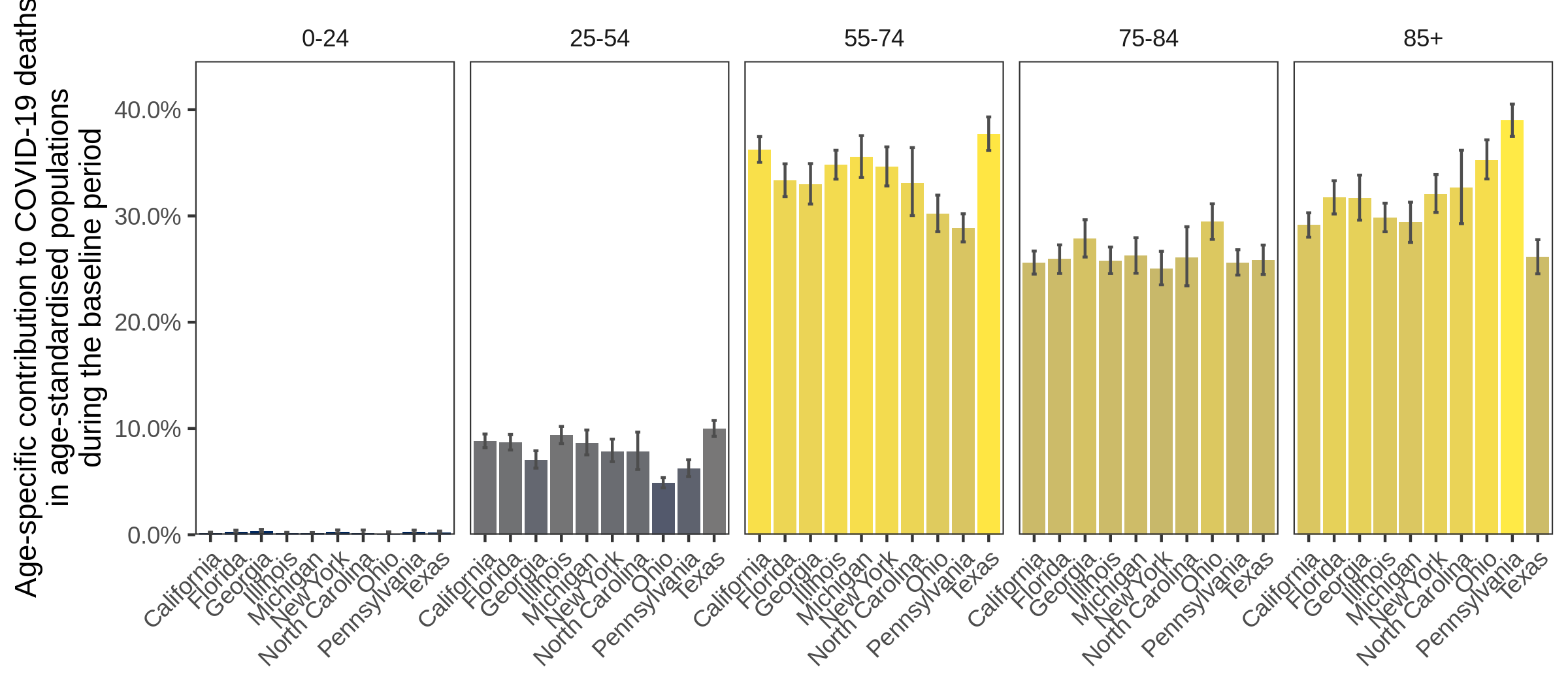}
    \caption{\textbf{Estimated age-specific contribution to COVID-19 attributable deaths adjusted for population size in the first 3 months of state epidemics.} Posterior median estimates of each age group to COVID-19 attributable deaths are shown for each state (barplot), along with 95\% credible intervals (error bars). Estimates are for the $3$ months after the $10$th cumulative death in each state, and are adjusted for differences in the age composition of state populations (see text). Brighter colors indicate higher values of the baseline contribution and darker colors indicate lower values. }
    \label{fig:baselinecontribution_adj}
\end{figure}

We investigate whether the age composition of COVID-19 attributable deaths differs across states. To address this question, we focus on the baseline period of the first $3$ months after the $10$th cumulated death in each state, and aggregate the predictions of the share of $1$-year age bands from our non-parametric model into the age strata $0$-$24$, $25$-$54$, $55$-$74$, $75$-$84$, $85+$ years for each state during the baseline period. In these calculations, we adjusted for age differences in state populations with the age composition of the US population. Pre-pandemic age-specific population counts were drawn from the U.S. Census Bureau in 2018~\citep{censuspopsize}. Strikingly, the resulting posterior estimates of the share of age groups among COVID-19 attributable deaths were statistically significantly different across states even after standardising the age composition of state populations (Figure~\ref{fig:baselinecontribution_adj}).  
For example, the contributions of individuals aged $85+$ to COVID-19 attributable deaths ranged from 26.17\% (24.56\%- 27.77\%) in Texas to 38.99\% (37.49\%- 40.53\%) in Pennsylvania. 
High heterogeneity was also found in the contributions from individuals age $55$-$74$, which varied from 28.85\% (27.56\%- 30.2\%) in Pennsylvania to
37.71\% (36.16\%- 39.31\%) in Texas.

\begin{figure}[h!]
    \centering
    \includegraphics[width = 0.9\textwidth]{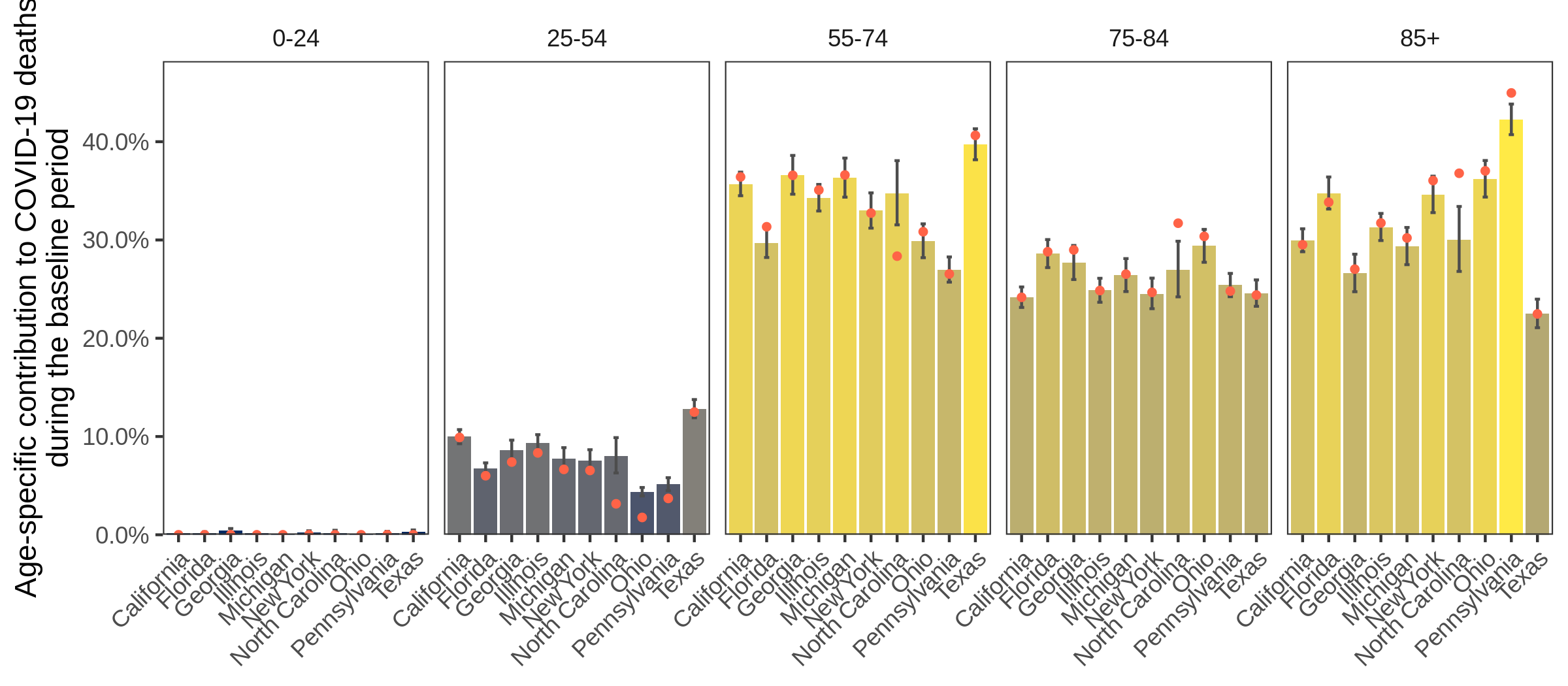}
    \caption{\textbf{Estimated and empirical age-specific contribution to COVID-19 attributable deaths in the first 3 months of state epidemics.} Posterior median estimates of each age group to COVID-19 attributable deaths are shown for each state (barplot), along with 95\% credible intervals (error bars). Estimates are for the $3$ months after the $10$th cumulative death in each state, and are not adjusted for differences in the age composition of state populations. Brighter colors indicate higher values of the baseline contribution and darker colors indicate lower values. The empirical estimates (red points) of contribution are calculated by setting the non-retrievable weekly deaths to zero. }
    \label{fig:baselinecontribution}
\end{figure}

These discrepancies across states are clearly supported in the empirical data, albeit without an assessment of their statistical significance (Figure~\ref{fig:baselinecontribution}).
Notice that the contribution from individuals aged $55+$ estimated by the model is always lower or equal to the empirical contribution while the one in $25$-$54$ is always greater or equal. This is because the empirical contributions were computed considering that unretrievable weekly deaths were null while they could, in fact, be positive.

\FloatBarrier
\clearpage

\FloatBarrier
\clearpage
\section{Templated Stan model files} \label{supp:stan_code}
\subsection{Stan model to fit regularised B-splines projected Gaussian Process on the mean surface of 2D count data}

\begin{lstlisting}[caption=Stan model to fit regularised B-splines projected Gaussian Process on the mean surface of 2D count data]
functions {
  matrix kron_mvprod(matrix A, matrix B, matrix V) 
  {
    return transpose(A*transpose(B*V));
  }
    
  matrix gp(int N_rows, int N_columns, real[] rows_idx, real[] columns_index,
            real delta0,
            real alpha_gp, 
            real rho_gp1, real rho_gp2,
            matrix z1)
  {
    
    matrix[N_rows,N_columns] GP;
    
    matrix[N_rows, N_rows] K1;
    matrix[N_rows, N_rows] L_K1;
    
    matrix[N_columns, N_columns] K2;
    matrix[N_columns, N_columns] L_K2;
    
    K1 = cov_exp_quad(rows_idx, alpha_gp, rho_gp1) + diag_matrix(rep_vector(delta0, N_rows));
    K2 = cov_exp_quad(columns_index, alpha_gp, rho_gp2) + diag_matrix(rep_vector(delta0, N_columns));

    L_K1 = cholesky_decompose(K1);
    L_K2 = cholesky_decompose(K2);
    
    GP = kron_mvprod(L_K2, L_K1, z1);

    return(GP);
  }
}

data {
  int<lower=1> n; // number of rows
  int<lower=1> m; // number of columns
  int<lower=1> N; // number of entries observed
  int coordinates[N,2]; // coordinate of entries observed
  int y[N]; // data on entries observed
  
  //splines
  int num_basis_rows; // number of B-Splines basis functions rows 
  int num_basis_columns; // number of B-Splines basis functions columns 
  matrix[num_basis_rows, n] BASIS_ROWS; // B-splines basis functions on the rows
  matrix[num_basis_columns, m] BASIS_COLUMNS; // B-splines basis functions on the columns
  
  // GP
  real IDX_BASIS_ROWS[num_basis_rows]; // index of the B-splines basis functions rows
  real IDX_BASIS_COLUMNS[num_basis_columns]; // index of the B-splines basis functions columns
}

transformed data {
  real delta = 1e-9;
}

parameters {
  real<lower=0> rho_1; // length scale rows
  real<lower=0> rho_2; // length scale columns
  real<lower=0> alpha_gp; // output variance
  matrix[num_basis_rows,num_basis_columns] eta; // GP variables
  real<lower=0> nu_unscaled; // overdispersion parameter
}

transformed parameters {
  real<lower=0> nu = (1/nu_unscaled)^2;
  real<lower=0> theta = (1 / nu);
  matrix[num_basis_rows,num_basis_columns] beta = gp(num_basis_rows, 
                                                     num_basis_columns, 
                                                     IDX_BASIS_ROWS, 
                                                     IDX_BASIS_COLUMNS,
                                                     delta,
                                                     alpha_gp, 
                                                     rho_1,  rho_2,
                                                     eta); 
                                                     
  matrix[n,m] f = exp( (BASIS_ROWS') * beta * BASIS_COLUMNS );
  matrix[n,m] alpha = f / nu;
}

model {
  nu_unscaled ~ normal(0,1);
  
  rho_1 ~ inv_gamma(5, 5);
  rho_2 ~ inv_gamma(5, 5);
  alpha_gp ~ cauchy(0,1);
  
  for(i in 1:num_basis_rows){
    for(j in 1:num_basis_columns){
        eta[i,j] ~ std_normal();
    }
  }
  
  for(k in 1:N){
      y[k] ~ neg_binomial(alpha[coordinates[k,1],coordinates[k,2]], theta);
  }

}
\end{lstlisting}

\subsection{Stan model to fit regularised B-splines projected Gaussian Process on the mean surface of 2D continuous data}
\begin{lstlisting}[caption=Stan model to fit regularised B-splines projected Gaussian Process on the mean surface of 2D continuous data]
functions {
  matrix kron_mvprod(matrix A, matrix B, matrix V) 
  {
    return transpose(A*transpose(B*V));
  }
    
  matrix gp(int N_rows, int N_columns, real[] rows_idx, real[] columns_index,
            real delta0,
            real alpha_gp,
            real rho_gp1, real rho_gp2,
            matrix z1)
  {
    
    matrix[N_rows,N_columns] GP;
    
    matrix[N_rows, N_rows] K1;
    matrix[N_rows, N_rows] L_K1;
    
    matrix[N_columns, N_columns] K2;
    matrix[N_columns, N_columns] L_K2;
    
    K1 = cov_exp_quad(rows_idx, alpha_gp, rho_gp1) + diag_matrix(rep_vector(delta0, N_rows));
    K2 = cov_exp_quad(columns_index, alpha_gp, rho_gp2) + diag_matrix(rep_vector(delta0, N_columns));

    L_K1 = cholesky_decompose(K1);
    L_K2 = cholesky_decompose(K2);
    
    GP = kron_mvprod(L_K2, L_K1, z1);

    return(GP);
  }
}

data {
  int<lower=1> n; // number of rows
  int<lower=1> m; // number of columns
  int<lower=1> N; // number of entries observed
  int coordinates[N,2]; // coordinate of entries observed
  vector[N] y; // data on entries observed
  
  //splines
  int num_basis_rows; // number of B-Splines basis functions rows 
  int num_basis_columns; // number of B-Splines basis functions columns 
  matrix[num_basis_rows, n] BASIS_ROWS; // B-splines basis functions on the rows
  matrix[num_basis_columns, m] BASIS_COLUMNS; // B-splines basis functions on the columns
  
  // GP
  real IDX_BASIS_ROWS[num_basis_rows]; // index of the B-splines basis functions rows
  real IDX_BASIS_COLUMNS[num_basis_columns]; // index of the B-splines basis functions columns
}

transformed data {
  real delta = 1e-9;
}

parameters {
  real<lower=0> rho_1; // length scale rows
  real<lower=0> rho_2; // length scale columns
  real<lower=0> alpha_gp; // output variance
  matrix[num_basis_rows,num_basis_columns] eta; // GP variables
  real<lower=0> sigma; // observational noise 
}

transformed parameters {
  matrix[num_basis_rows,num_basis_columns] beta = gp(num_basis_rows, 
                                                     num_basis_columns, 
                                                     IDX_BASIS_ROWS, 
                                                     IDX_BASIS_COLUMNS,
                                                     delta,
                                                     alpha_gp,
                                                     rho_1,  rho_2,
                                                     eta); 
                                                     
  matrix[n,m] f = (BASIS_ROWS') * beta * BASIS_COLUMNS;
}

model {
  rho_1 ~ inv_gamma(5, 5);
  rho_2 ~ inv_gamma(5, 5);
  alpha_gp ~ cauchy(0,1);
  
  sigma ~ cauchy(0,1);
  
  for(i in 1:num_basis_rows){
    for(j in 1:num_basis_columns){
        eta[i,j] ~ std_normal();
    }
  }
  
  for(k in 1:N){
        y[k] ~ normal(f[coordinates[k,1],coordinates[k,2]], sigma);
  }

}
\end{lstlisting}

\subsection{R code to obtain B-splines}
\begin{lstlisting}[caption=R code to obtain B-splines]
bsplines = function(data, knots, degree)
 {
  # data: support 
  # knots: value of the knots
  # spline_degree: spline degree

  K = length(knots)
  num_basis = K + degree - 1
  
  intervals = find_intervals(knots, degree)
  
  m = matrix(nrow = num_basis, ncol = length(data), 0)
  
  for(k in 1:num_basis)
  {
    m[k,] = bspline(data, k, degree + 1, intervals) 
  }
  
  m[num_basis,length(data)] = 1
  
  return(m)
}

bspline = function(x, k, order, intervals)
  {
  
  if(order == 1){
    return(x >= intervals[k] & x < intervals[k+1])
  }
  
  w1 = 0; w2 = 0
  
  if(intervals[k] != intervals[k+order-1])
    w1 = (x - intervals[k]) / (intervals[k+order-1] - intervals[k])
  if(intervals[k+1] != intervals[k+order])
    w2 = 1 - (x - intervals[k+1]) / (intervals[k+order] - intervals[k+1])
  
  spline = w1 * bspline(x, k, order - 1, intervals) +
    w2 * bspline(x, k+1, order - 1, intervals)
  
  return(spline)
}

find_intervals = function(knots, degree)
  {
  
  K = length(knots)
  
  intervals = vector(mode = 'double', length = 2*degree + K)
  
  # support of knots
  intervals[(degree+1):(degree+K)] = knots
  
  # extreme
  intervals[1:degree] = min(knots)
  intervals[(degree+K+1):(2*degree+K)] = max(knots)

  return(intervals)
}



\end{lstlisting}

\FloatBarrier
\clearpage
\section{State level summary} \label{sec:statesummary}
\includeCountrySummaryIfexists{1}{CA}{California}
\includeCountrySummaryIfexists{1}{FL}{Florida}
\includeCountrySummaryIfexists{1}{GA}{Georgia}
\includeCountrySummaryIfexists{1}{IL}{Illinois}
\includeCountrySummaryIfexists{1}{MI}{Michigan}
\includeCountrySummaryIfexists{1}{NC}{North Carolina}
\includeCountrySummaryIfexists{1}{NY}{New York}
\includeCountrySummaryIfexists{1}{OH}{Ohio}
\includeCountrySummaryIfexists{1}{PA}{Pennsylvania}
\includeCountrySummaryIfexists{1}{TX}{Texas}

\end{document}